%% file: LC21.tex
             \newif\ifboyscout\boyscouttrue          
             \newif\ifsubmission\submissionfalse     
             \newif\ifblog\blogfalse 
  \boyscoutfalse                         

\documentclass[12pt]{iopart}
\usepackage{iopams} 

\usepackage[pdftex]{graphicx}
\graphicspath{{../figs/}{../Fig/}}  

\ifsubmission
\else 
    \usepackage{color} 
    \usepackage[colorlinks]{hyperref} 
\fi

    \ifsubmission\else
\input{biblatex}    
    \fi
\input{defsReversal}      

\begin{document}

\title[Chaotic lattice field theory]
      {A chaotic lattice field theory in one dimension}

    \author{
H Liang
         and
P Cvitanovi{\'c}
    }\address{
            School of Physics,
            Georgia Institute of Technology,
            Atlanta, GA 30332-0430, USA
    } \ead{predrag.cvitanovic@physics.gatech.edu}
    \vspace{10pt}
    \begin{indented}
    \item[]
    May 20, 2022   
    \end{indented}

    \begin{abstract}
Motivated by Gutzwiller's semiclassical quantization, in which
unstable periodic orbits of low-dimensional deterministic dynamics serve
as a WKB `skeleton' for chaotic quantum mechanics, we
construct the corresponding deterministic skeleton for
infinite-dimensional lattice-discretized scalar field theories. In the
field-theoretical formulation, there is no evolution in time, and there
is no `Lyapunov horizon'; there is only an enumeration of lattice states
that contribute to the theory's partition sum, each a global spatiotemporal
solution of system's deterministic {\ELe}s.

The reformulation aligns `chaos theory' with the standard solid state, field
theory, and statistical mechanics. In a spatiotemporal, crystallographer
formulation, the time-periodic orbits of dynamical systems theory are replaced by
periodic $d$-dimensional Bravais cell tilings of spacetime, each weighted
by the inverse of its instability, its Hill determinant. Hyperbolic shadowing
of large cells by smaller ones ensures that the predictions of the theory are
dominated by the smallest Bravais cells.

The form of the partition function of a given field theory is determined
by the group of its spatiotemporal symmetries, that is, by the space group of its
lattice discretization, best studied on its reciprocal lattice. Already
1-dimensional lattice discretization is of sufficient interest to be the focus
of this paper. In particular, from a spatiotemporal field theory perspective,
`time'-reversal is a purely crystallographic notion, a reflection point
group, leading to a novel, symmetry quotienting perspective of
time-reversible theories and associated topological zeta functions.
    \end{abstract}

\pacs{02.20.-a, 05.45.-a, 05.45.Jn, 47.27.ed
      }

\vspace{2pc}
\noindent{\it Keywords}:
chaotic field theory, many-particle systems, coupled map lattices,
periodic orbits, symbolic dynamics, cat maps

\submitto{\jpa} 
    \ifsubmission
\maketitle 
    \fi

\begin{quote}
     Dedicated to Fritz Haake 1941--2019\rf{GGSZ21}.
\end{quote}

\begin{figure}
\begin{center}
            \begin{minipage}[c]{0.35\textwidth}\begin{center}
{harmonic} field theory
\bigskip

\includegraphics[width=0.85\textwidth]{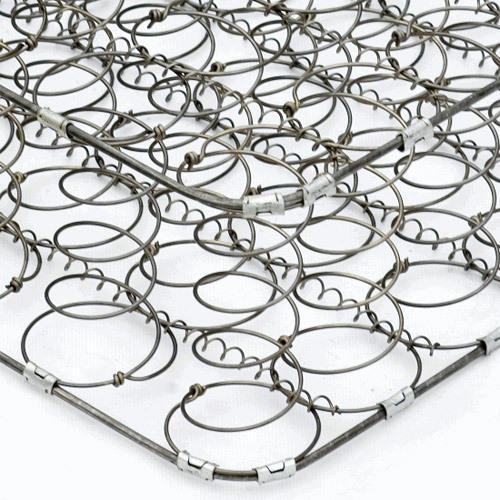}\\
{tight-binding} model \\ ({Helmholtz})
            \end{center}\end{minipage}
            \hspace{2ex}
            \begin{minipage}[c]{0.36\textwidth}\begin{center}
{chaotic} field theory
\medskip

\includegraphics[width=1.1\textwidth]{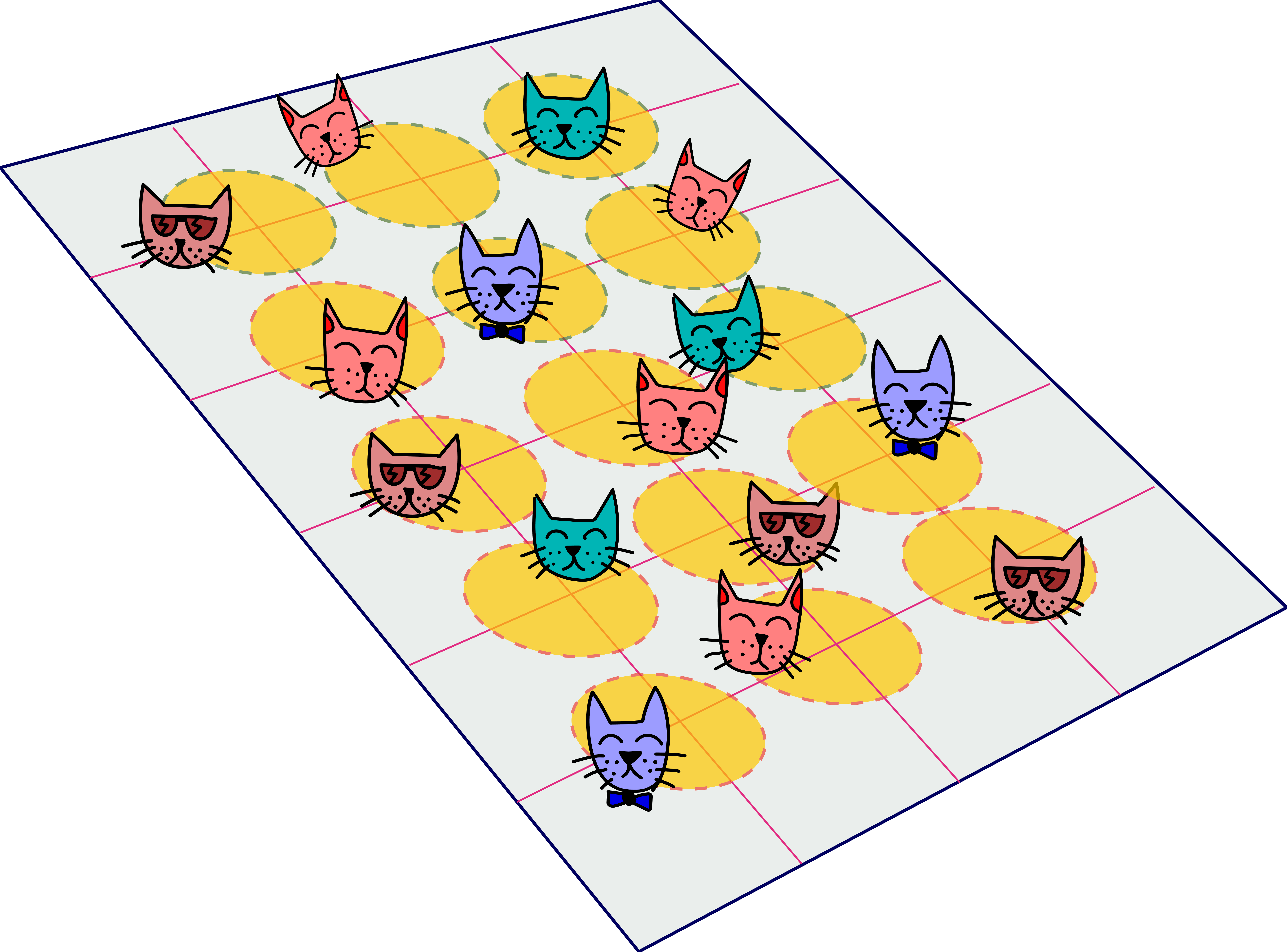}\\
\medskip

Euclidean {Klein-Gordon} \\ (damped {Poisson})
            \end{center}\end{minipage}
\end{center}
    \caption{\label{GraffitiCats}
(Color online)~~~
The simplest of all chaotic field theories
is the `\catlatt', a deterministic Klein-Gordon field theory on a
hypercubic lattice, with an unstable, ``anti-harmonic" rotor
at each lattice site, a cat that runs away rather than pushes back.
In contrast to its elliptic sibling, the Helmholtz equation and its
oscillatory solutions, {\catlatt}'s {\lsts} are hyperbolic and
unstable.
    }
\end{figure}

The year was 1988.
Roberto Artuso, Erik Aurell and P.C. had just worked out the cycle
expansions formulation of the deterministic and semiclassical chaotic
systems\rf{AACI,AACII}, and a Niels Bohr Institute ``Quantum Chaos
Symposium'' was organized to introduce the newfangled theory to
unbelievers (for a history, see \toChaosBook{section.A.4} {Append.~A.4
{\em Periodic orbit theory}}). In the first row (here \reffig{GraffitiCats})
of the famed auditorium where long ago Niels Bohr and his colleagues used
to nod off, sat a man with an impressive butterfly bow tie and a big
smile. At the end of our presentation, Fritz --for that was Fritz Haake--
stood up and exclaimed

\begin{quote}
     ``Amazing! I did not understand a single word!''
\end{quote}

And indeed, there is a problem of understanding what is `chaos' as
encountered in different disciplines, so we start this offering to Fritz
Haake's memory by {`a fair coin toss'} theory of chaos
(\refsect{s:coinToss}), as was presented in the 1988 symposium, but in a
modern, field theorist's vision. In those days `chaos' was a phenomenon
exhibited by low-dimensional systems. In this and companion papers%
\rf{CL18,GuBuCv17} we develop a theory of `chaotic' or
`turbulent' infinite\dmn\ deterministic field theories. Deterministic
chaotic field theory is of interest on its merits, as a method of
describing turbulence in strongly nonlinear deterministic field theories,
such as \NS\ or \KS\rf{GudorfThesis,GuBuCv17}, or as a Gutzwiller WKB
`skeleton' for a chaotic quantum field theory\rf{gutbook,CFTsketch}
or a stochastic field theory%
\rf{noisy_Fred,conjug_Fred,diag_Fred,LipCvi08,CviLip12}.
Lattice reformulation aligns `chaos' with standard solid state, field
theory and statistical mechanics, but the claims are radical: we've been
doing `turbulence' all wrong.
In ``explaining'' chaos we talk the talk as though we have never moved
beyond Newton; here is an initial state of a system, at an instant in
time, and here are the differential equations that evolve it forward in
time. But when we -all of us- walk the walk, we do something altogether
different (see the references preceding eq.~\refeq{nXdCycle}), much
closer to the 20th century spacetime physics.
Our papers realign the theory to what we
actually {\em do} when solving `chaos equations', using not much more
than linear algebra.
In the field-theoretical formulation, there is no evolution in time, and
there is no `Lyapunov horizon'; every contributing {\em \lst} is a
robust global solution of a \spt\ fixed point condition, and there is no
dynamicist's exponential blowup of initial state perturbations.

To a field theorist, the fundamental object is \emph{global}, the
partition function sum over probabilities of all possible spacetime field
configurations.
To a dynamicist, the fundamental notion is \emph{local}, an ordinary or
partial differential time-evolution equation. From the field-theoretic
perspective, the spacetime formulation is fundamental, elegant and
computationally powerful, while moving in step-lock with time is only one
of the methods, a `transfer matrix' for construction of the
partition sum, a method at times awkward and computationally unstable.

We start our introduction to chaotic field theory (\refsect{s:FT}) by
rewriting the two most elementary examples of deterministic chaos,
the for\-ward-in-time first order difference equation for the Bernoulli
map (\refsect{s:coinToss}), and
the for\-ward-in-time second order difference equation for a 1\dmn\
lattice of coupled rotors (\refsect{s:kickRot}) as, respectively,
the `{temporal Bernoulli}' two-term discrete lattice recurrence relation,
and
the `\templatt'  three-term discrete lattice recurrence relation.
We then apply the approach to the simplest
nonlinear field theories, the 1\dmn\ discretized
scalar {$\phi^3$} and {$\phi^4$} theories
(\refsects{s:henlatt}{s:phi4latt}).

Their spacetime generalization, the simplest of all chaotic field theories,
is the `\catlatt'\rf{GutOsi15,GHJSC16,CL18}, a discretization of the
Klein-Gordon equation, a deterministic scalar field theory on a $d$\dmn\
hypercubic lattice, with an unstable ``anti-harmonic" rotor $\ssp_{z}$
at each lattice site $z$, a rotor that gives rather than pushes back,
coupled to its nearest neighbors. Dear Fritz, if you lack inclination
to plunge into what follows, please take home \reffig{GraffitiCats}.
In contrast to its elliptic sibling, the Helmholtz equation and its
oscillatory solutions, {\catlatt}'s {\lsts} are hyperbolic and
`turbulent', just as in contrast to oscillations of a harmonic
oscillator, Bernoulli coin flips are unstable and chaotic.

The key to constructing partition sums for deterministic field theories
(\refsect{s:FT}) are the {\HillDet}s of  the
`{\jacobianOrbs}' (\refsect{s:JacobianOrb}) that describe the global
stability of linearized deterministic equations.
How is this global, high-dimensional orbit stability related
to the stability of the conventional low-dimensional, forward-in-time evolution
(\refsect{s:timeJacob})?
The two notions of stability are related by Hill's
formulas (\refsect{s:Hill}, \refappe{s:Hill2step}), relations that rely on higher-order
derivative equations being rewritten as sets of first order ODEs,
relations equally applicable to mechanical, energy conserving
systems, as to viscous, dissipative systems.

In order to explain the
key ideas,
we focus in this paper on 1\dmn\ field theories, postponing the
2\dmn\ Bravais lattices' subtleties to the sequel\rf{CL18}.
In \refsect{s:latt1d} we show that the partition function of a
given field theory is determined by the group of its symmetries, \ie,
by the space group of its lattice discretization.
Lattice discretization of the theory enables us to apply
solid state computational methods, such as the reciprocal lattice and
space group crystallography, to what are conventionally dynamical
system problems (\refsect{s:recip1d}).
On the level of counting {\lsts}, their {\tzeta}s are purely
group-theoretic Lind zeta functions (\refsect{s:Lind1d}).
As long as the only symmetry is time translation, we recover the
conventional \po\ theory\rf{ChaosBook}
(\refsect{sect:ArtinMazur}).
However, from a \spt\ field theory perspective, `time'-reversal is a
purely crystallographic notion, leading to --to us very surprising--
dihedral space group zeta function
for the  `time-reversible' theories
(\refsect{sect:KiLePa}).

Our results are summarized and open problems discussed in
\refsect{s:summary}.
The work that forms the basis of our formulation of chaotic field
theory is reviewed in \refappe{s:HillHistory}.
Icon
\raisebox{-0.4ex}[0pt][0pt]{\includegraphics[height=1em]{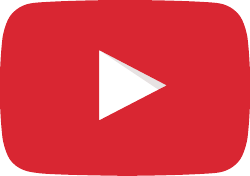}}
on the margin links the block of text to a supplementary online video.


\section{Deterministic lattice field theory}
\label{s:FT}   

\begin{figure}
  \centering
\includegraphics[width=0.50\textwidth]{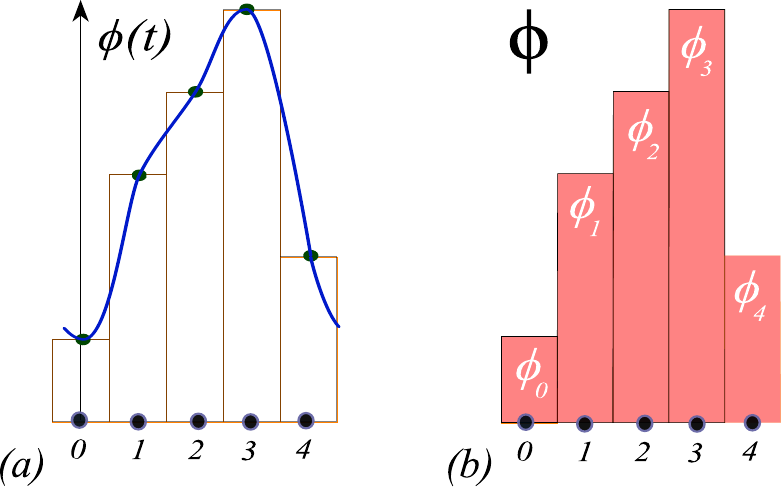} 
    \caption{\label{FieldConfig}
(Color online)~~~
Discretization of a 1\dmn\ field theory.
Horizontal: $\zeit$ coordinate, with lattice sites marked by dots and
labelled by $\zeit\in\integers$.
(a)
A periodic field $\ssp(\zeit)$, plotted as a function of continuous
coordinate $\zeit$.
(b)
A corresponding discretized period-$5$ {Bravais cell} {\lst}
$\Xx=\cycle{\ssp_0 \ssp_1 \ssp_2 \ssp_3 \ssp_4}$,
with discretized field $\ssp_\zeit$ plotted as a bar
centred at lattice site $\zeit$.
In what follows we use `lattice units' \(a=1\).
Continued in \reffig{fig:1dLattRefl}.
    }
\end{figure}

A scalar field $\ssp(x)$ over $d$ Euclidean coordinates can be
discretized by
replacing the continuous space by a $d$\dmn\ hypercubic {integer lattice}
$\integers^d$,
    %
\toVideo{youtube.com/embed/EWLQJ6ZpUWQ}
with lattice spacing $a$, and
evaluating the {field} only on the
lattice points\rf{MonMun94,MunWal00}
\beq
\ssp_z
=
\ssp(x)
    \,,\qquad \qquad x = az= \mbox{lattice point}
    \,,\quad z \in \integers^d
\,,
\ee{LattField}
see \reffig{FieldConfig}.
A {\em field configuration} (here in one {\spt} dimension)
\beq
\Xx =
\cdots {\ssp}_{-3} {\ssp}_{-2}\,{\ssp}_{-1}\,
       {\ssp}_0\,
      {\ssp}_{1} {\ssp}_{2} {\ssp}_{3} {\ssp}_{4}  \cdots
\,,
\ee{stateSp}
takes any set of values $\ssp_{z}\in\reals$ in system's $\infty$\dmn\
\emph{\statesp}.
A {periodic} {field configuration} $\Xx$ satisfies
\beq
\Xx({x} + {R}) = \Xx({x})
\ee{1DprimePO}
for any discrete translation
$R$ in the \emph{Bravais lattice}
\beq
\lattice = \Big\{\sum_{i=1}^d \cl{i} {\bf a}_i
                 \;\vert\; \cl{i} \in\mathbb{Z} \Big\}
\ee{BravLatt}
where the $d$ independent integer lattice vectors
\(
\{{\bf a}_1,{\bf a}_2,\cdots,{\bf a}_d\}
\) 
define a \emph{Bravais cell}.

A {field configuration} occurs with probability density
\beq
p(\Xx)\,=\, \frac{1}{Z}\,\e^{-\action[\Xx]}
\,,\qquad Z=Z[0]
\,.
\label{ProbConf}
\eeq
Here $Z$ is a normalization factor, given by the \emph{partition
function}, the integral over probabilities of all
configurations,
\beq
Z[\Source]	
    \,=\, \int d\Xx\,e^{-\action[\Xx] + \Xx \cdot \Source}
    \,,\qquad
d\Xx = \prod_{z}^{\lattice} d\ssp_z
\,,
\ee{partFunct}
where $\Source=\{\source_{z}\}$ is an external source $\source_{z}$ that
one can vary site by site, and $\action[\Xx]$ is the action that defines
the theory (discussed in more detail in \refsect{s:nonlinFT}).
The dimension of the partition function integral equals the number of lattice
sites $N_\lattice$.

Motivated by WKB  semi-classical, saddle-point
approximations\rf{gutbook} to the partition function \refeq{partFunct},
in this paper we describe their deterministic underpinning, the corresponding
\emph{deterministic} field theory, with partition function built from
solutions to the variational saddle-point condition
\beq
F[\Xx_c]_z =
\frac{\delta{\action[\Xx_c]}}{\delta\ssp_z~} = 0
\,,
\ee{eqMotion}   
with a global deterministic solution $\Xx_c$ satisfying this local
extremal condition on every lattice site. We shall refer to the defining
condition \refeq{eqMotion} as system's {\em `\ELe'},
keeping in mind that the field theories studied here might have a
Lagrangian formulation (for example, scalar $\phi^k$ field theories of
\refsect{s:nonlinFT}), or be dissipative (for example, temporal
Bernoulli, \spt\ \Henon\ for non-area preserving parameter values, \KS\
or \NS\ equations).

In order to distinguish a \emph{solution} to the {\ELe}s \refeq{eqMotion}
from an {arbitrary} \emph{field configuration} \refeq{stateSp}, we refer
to the solutions as \emph{{\lsts}}, each a set of lattice site field
values
\beq
\Xx_c = \{\ssp_{c,z}\}
\,,
\label{1dLattStat}
\eeq
that satisfies the {condition} \refeq{eqMotion} globally,
over all lattice sites.
For a finite lattice \lattice\ one needs to specify the boundary
conditions ({\bcs}).
The companion article \refref{GHJSC16} tackles the Dirichlet {\bcs}, a
difficult, time-translation symmetry breaking, and from the \po\ theory
perspective, a wholly unnecessary, self-inflicted pain. All that one
needs to solve a field theory are the $\cl{}$-periodic,
time-translation enforced {\bcs} that we shall use here.

An example is the 1 {\spt} dimension \brick\ of fields of period
$\cl{}=5$ {Bravais cell} sketched in \reffig{FieldConfig}\,(b),
\beq
\Xx_c = \ssp_0 \ssp_1 \ssp_2 \ssp_3 \ssp_{4}
\,,
\ee{1dLattStatC_n}
with its infinite repetition sketched in
\reffig{fig:1dLatStatC_5}\,(1).
The first field value $\ssp_0$ in the \brick\ is evaluated on the lattice site 0,
the second $\ssp_1$ on the lattice site 1,
the $(\cl{}+1)$th $\ssp_{\cl{}}=\ssp_0$ on the lattice site $\cl{}$,
with $k$th lattice site field value $\ssp_{k}=\ssp_{\ell}$,
where $\ell=k$~(mod $\cl{}$).

What we call here a chaotic `field' at a discretized spacetime lattice
site $z$, a solid state physicist would call a `particle' at
crystal site $z$, coupled to its nearest neighbors. A solid state
physicist endeavours to understand $N$-particle chaotic systems in
many-body or `large $N$' settings, where in practice $N$ not much larger than 2
can be `large'. Chaotic field theory is {\em ab initio} formulated for infinite
time and infinite space lattice, but its periodic theory description is
-thanks to hyperbolicity-- often accurate already for $N=2, 3,
\cdots$, where $N$ is the number of sites in a Bravais cell that tiles the
spacetime.

Each {\lst} is a distinct deterministic solution $\Xx_c$ to the
discretized {\ELe}s \refeq{eqMotion}, so its
probability density is a $N_\lattice$\dmn\ Dirac delta function
(that's what we mean by the system being \emph{deterministic}),
a delta function per site ensuring that {\ELe}
\refeq{eqMotion} is satisfied everywhere,
\beq
p_c(\Xx)\,=\, \frac{1}{Z}\,\delta(F[\Xx])
\,, \qquad \Xx \in \pS_c
\,,
\label{DiracDeltaExp}
\eeq
where $\pS_c$ is a small neighborhood of \lst\ $\Xx_c$.
In \refsect{s:Hill} we  verify that this
definition agrees with the for\-ward-in-time \FP\ probability
density evolution\rf{CBmeasure}.
However, we find field-theoretical formulation vastly preferable to the
for\-ward-in-time formulation, especially when it comes to higher \spt\
dimensions\rf{CL18}.

As is case for a WKB approximation\rf{gutbook}, the {deterministic} field
theory partition sum has support only on lattice field values that are
solutions to the {\ELe}s \refeq{eqMotion},
and the partition function \refeq{ProbConf} is now a sum over
configuration \statesp\ \refeq{stateSp} \emph{points}, what in theory
of dynamical systems is called the `deterministic trace formula'\rf{ChaosBook},
\beq
Z[0] =
        \sum_c \int_{\pS_c}\!\!\!\!\!d\Xx\,\delta(F[\Xx])
    =
        \sum_c \frac{1}{\left|\Det\jMorb_c\right|}
\,,
\label{ClassPartitF}
\eeq
and we refer to
the $[N_\lattice\!\times\!N_\lattice]$ matrix of second derivatives
\beq
(\jMorb_c)_{z'z} = \frac{\delta F_{z'}[\Xx_c]}{\delta \ssp_{z}}
             = \action[\Xx_c]_{z'z}
\ee{jacobianOrb}
as the \emph{\jacobianOrb}, and to its determinant $\Det\jMorb_c$ as the
\emph{\HillDet}. Support being on  \statesp\ \emph{points} means that we
do not need to worry about potentials being even or odd (thus unbounded),
or the system being energy conserving or dissipative, as long as its
nonwandering {\lsts} $\Xx_c$ set is bounded in \statesp.
In what follows, we shall deal only with deterministic field
theory and mostly omit the subscript `$c$' in $\Xx_c$.

How is a deterministic chaotic field theory different from a conventional
field theory?
By ``spontaneous breaking of the symmetry'' in a conventional theory one
means that a solution does not satisfy a symmetry such as
$\phi\rightarrow-\phi$; we always work in the ``broken-symmetry'' regime,
as almost every `turbulent', {\spt}ly chaotic deterministic solution
breaks all symmetries.
We work `beyond perturbation theory', in the anti-integrable, strong
coupling regime, in contrast to much of the literature that focuses on
weak coupling expansions around a `ground state'.
And, in contrast to \refref{Wyld61,MaSiRo73,GurMig96,BFKL97,RosSmo22},
our `far from equilibrium' field theory has no added dissipation, and is
not driven by external noise.
All chaoticity is due to the intrinsic unstable deterministic dynamics,
and our trace formulas \refeq{ClassPartitF} are exact, not merely saddle points
approximations to the exact theory.

\subsection{Lattice Laplacian}
\label{s:lattLap}

Lattice free field theory is defined by action\rf{Rothe05}
    %
\toVideo{youtube.com/embed/V4pyM2vuXL0}
\beq
\action_0[\Xx]=
          \frac{1}{2}\transp{\Xx}\left(-\Box + {\mu}^2\mathsf{1} \right)\Xx
\,,
\ee{freeAction}
where the `discrete Laplace operator', `central difference operator', or
the `graph Laplacian'%
\rf{PerViv,Pollicott01,Cimasoni12,MraRin12,GodRoy13,Pozrikidis14}
\beq
\Box\,\ssp_z =
    \sum_{||z'-z||=1} \!\! (\ssp_{z'} - \ssp_z)
 \quad \mbox{for all} \ z,z' \in \lattice 
\ee{LapOp} 
is the average of the lattice field variation $\ssp_{z'}-\ssp_z$
over the sites nearest to the site $z$.
For example, for a hypercubic lattice in one and two dimensions this
discretized Laplacian is given by
\bea
\Box\,\ssp_\zeit &=& \ssp_{\zeit+1} - 2\,\ssp_{\zeit} + \ssp_{\zeit-1}
    \label{LapTime}\\
\Box\,\ssp_{j,\zeit}
     &=&
\ssp_{j,\zeit+1} + \ssp_{j+1,\zeit} - 4\,\ssp_{j,\zeit}
                 + \ssp_{j,\zeit-1} + \ssp_{j-1, \zeit}
\,.
\label{LapSpTime}  
\eea

For the free field theory action \refeq{freeAction} the {\ELe}
\refeq{eqMotion} is the discretized {\sPe}\rf{FetWal03}, also known
as the {Yukawa} or Klein–Gordon equation, where  ${\mu}^2>0$ is the
Klein–\-Gordon mass-squared.

\subsection{1\dmn\ lattice field theories}
\label{s:FT1d}

    %
\toVideo{youtube.com/embed/qOkt0X7AZPo}
Discrete time evolution is frequently recast into a 1\dmn\ temporal
lattice field theory form, by anyone who rewrites a dynamical systems
discrete time evolution problem as a $k$-term recurrence, for example in
\refrefs{FeHa82,noisy_Fred,conjug_Fred,diag_Fred}. As already in one
\spt\ dimension there is much to be learned about the role symmetries
play in solving lattice field theories, that is what we will focus on in
this paper (time-reversal \refsects{s:latt1d}{s:Lind1d}), with $2$\dmn\
\spt\ field theories studied in the sequel\rf{CL18}.

We start with the first order difference equation that we call
`{temporal Bernoulli}' (\refsect{s:coinToss}),
\bea
- \ssp_{\zeit+1} + ({s}\ssp_{\zeit} - \Ssym{\zeit})
    \;\;
    &=&
0
\label{1dBernLatt}
\eea
in order to motivate the 2-component field formulation
\refeq{1stOrderDiffEqs} of second-order difference {\ELe}s \refeq{1dTempFT}
that we call, in the cases considered here,
the `{\templatt}' (\refsect{s:catLagrange}),
`temporal {$\phi^3$} theory'
(\refsect{s:henlatt}), and `temporal {$\phi^4$} theory'
(\refsect{s:phi4latt}), respectively:
\bea
- \ssp_{\zeit+1} + 2\,\ssp_{\zeit} - \ssp_{\zeit-1}
\;+\;  {\mu^2}\ssp_{\zeit} - \Ssym{\zeit}
    &=&
0
\label{1dTemplatt}\\ 
- \ssp_{\zeit+1} + 2\,\ssp_{\zeit} - \ssp_{\zeit-1}
\;+\;  \mu^2\,({1}/{4}-\ssp_{\zeit}^2)
    &=&
0
\label{1dHenlatt}\\    
- \ssp_{\zeit+1} + 2\,\ssp_{\zeit} - \ssp_{\zeit-1}
\;+\; \mu^2(\ssp_{\zeit}-\,\ssp_{\zeit}^3)
    &=&
0
\,.
\label{1dPhi4}     
\eea
Qualifier `temporal' is used here to emphasize that we view 1\dmn\
examples as special cases of `\spt' field theories; much of our
methodology for $d$\dmn\ deterministic field theories can be profitably
explained by working out $1$\dmn\ field theories.
Lurking here is the totality of the map-iteration dynamical systems
theory, but the reader will find it more profitable, and less confusing,
to think of these simply as lattice problems, and forget that the subscript
$\zeit$ often stands for `time'.

So, what is a `chaotic', or `turbulent' field theory?
    %
\toVideo{youtube.com/embed/CQ21AITMy84} 
As we shall see in \refsect{s:recip1d}, all of the above, as well as their
higher\dmn\ \spt\ siblings are `chaotic' for sufficiently large `stretching
parameter' ${s}$, or `Klein-Gordon mass'  ${\mu^2}$.
Our goal here is to make this `{\spt} chaos' tangible and precise, by
acquainting the reader what we believe are some of the simplest examples
of chaotic field theories.

\section{A fair coin toss}
\label{s:coinToss}
\renewcommand{\ssp}{\ensuremath{x}}               

The very simplest example of a deterministic law of evolution that gives
rise to `chaos' is the {\em Bernoulli} map, \reffig{fig:BernPart}\,(a),
which models a
\HREF{https://www.random.org/coins/?num=2&cur=40-antique.aurelian} {coin
toss}. Starting with a random initial state, the map generates,
deterministically,  a sequence of tails and heads with 50-50\%
probability.

We introduce the model in its conventional, time-evolution dynamical
formulation, than reformulate it as a lattice field theory, solved by
enumeration of all admissible \emph{{\lsts}}, field configurations that
satisfy a  global fixed point condition, and use this simple setting to
motivate
(1) the \emph{fundamental fact}: for a given lattice period, the {\em
\HillDet} of stabilities of global solutions counts their number
(\refsect{sect:fundFact}), and
(2) the {\tzeta} counts their translational symmetry group orbits
(\refsect{s:PoThe}).

\subsection{Bernoulli map} 
\label{s:Bernoulli}

\begin{figure}
  \centering
{(a)}
\includegraphics[width=0.35\textwidth]{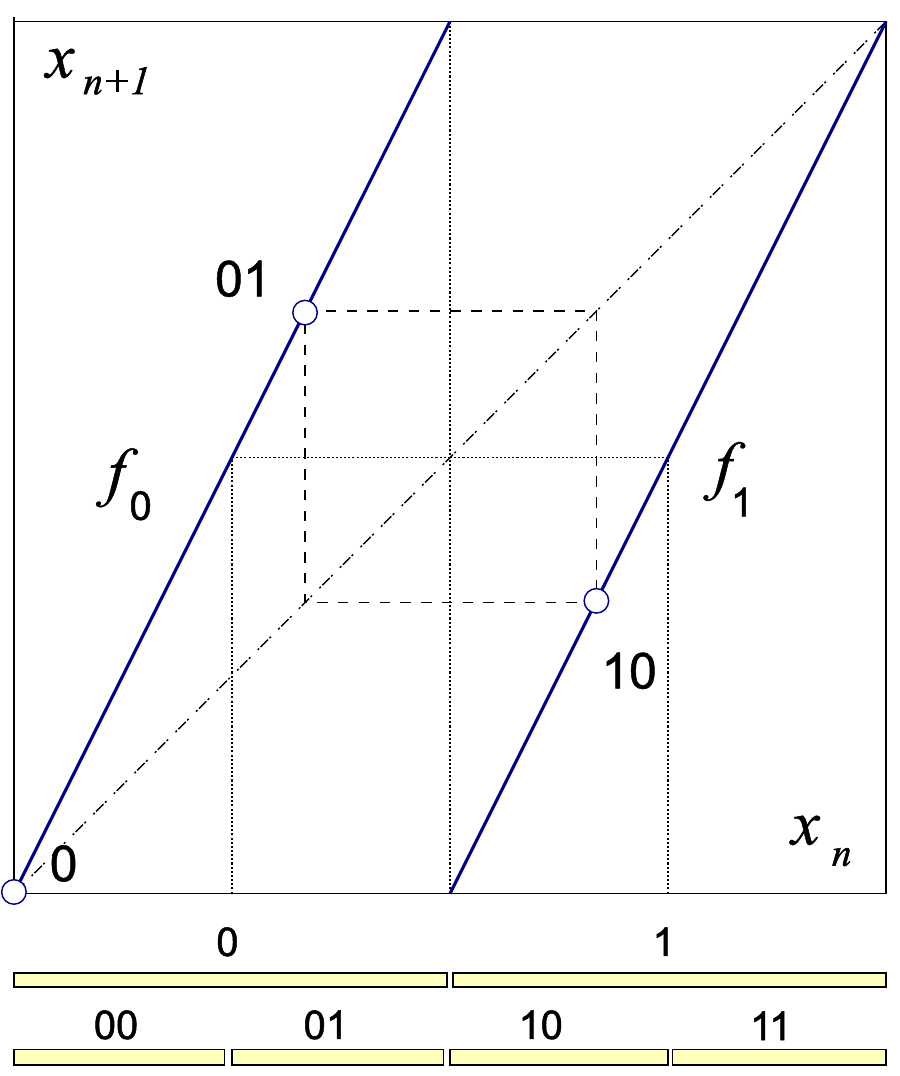} 
~~~
{(b)}$\!\!\!\!$
\includegraphics[width=0.40\textwidth]{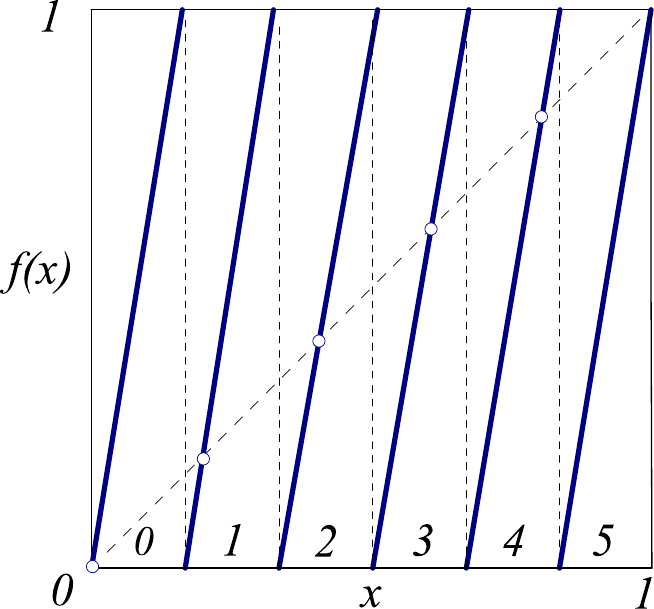} 

  \caption{\label{fig:BernPart}
(Color online)~~~
(a)
The `coin toss' map \refeq{BerShift}, together with the
$\cycle{0}$ fixed point, and the \cycle{01} 2-cycle. Preimages
of the critical point $\ssp_c=1/2$ partition the unit interval into
$\{\pS_0,\pS_1\}$, $\{\pS_{00},\pS_{01},\pS_{10},\pS_{11}\}$, $\dots$,
subintervals.
(b)
The base-${s}$ Bernoulli map, here with the `dice throw' stretching parameter ${s}=6$,
partitions the unit interval into $6$ subintervals $\{\pS_{\Ssym{}}\}$,
labeled by the ${6}$-letter alphabet \refeq{base-sAlph}. As the map is a
circle map, $\ssp_{5}=1=0=\ssp_{0} \quad(\mbox{mod}\;1)$.
          }
\end{figure}
%

The base-2 {\em Bernoulli} shift map,
\index{Bernoulli!shift}
\index{shift!Bernoulli}
    %
\toVideo{youtube.com/embed/Twbe1bAH678} 
\beq
\ssp_{\zeit+1} =
\left\{ \begin{array}{ll}
        f_0(\ssp_{\zeit}) =  2 \ssp_{\zeit} \,, \quad
                                    & \ssp_{\zeit} \in \pS_0=[0,1/2) \\
        f_1(\ssp_{\zeit}) =  2 \ssp_{\zeit} \;\; (\mbox{mod}\;1)\,, \quad
                                    & \ssp_{\zeit} \in \pS_1 =[1/2,1)
         \end{array}\right.
\,,
\ee{BerShift}
is shown in \reffig{fig:BernPart}\,(a).
If the linear part of such map has an integer-valued slope,
or `stretching' parameter $s\geq2$,
\beq
\ssp_{\zeit+1} \,=\, {s} \ssp_{\zeit}
\ee{BerStretch}
that maps state $\ssp_{\zeit}$ into a state in the `extended \statesp',
outside the unit interval,
the $(\mbox{mod}\;1)$ operation results in the base-${s}$ Bernoulli
circle map,
\renewcommand{\ssp}{\ensuremath{\phi}}             
\beq
\ssp_{\zeit+1}
= {s} \ssp_{\zeit}
\;\; (\mbox{mod}\;1)
\,,
\ee{n-tuplingMap}
sketched as a \HREF{https://www.random.org/dice/}{dice throw} in
\reffig{fig:BernPart}\,(b).
The $(\mbox{mod}\;1)$ operation subtracts
$\Ssym{\zeit}=\left\lfloor{s}\ssp_{\zeit}\right\rfloor$, the integer part of ${s}
\ssp_{\zeit}$, or the circle map \emph{winding number}, to keep
$\ssp_{\zeit+1}$ in the unit interval $[0,1)$, and partitions the unit
interval into ${s}$ subintervals $\{\pS_\Ssym{}\}$,
\beq
\ssp_{\zeit+1}
= {s} \ssp_{\zeit} - \Ssym{\zeit}
\,,\qquad  \ssp_{\zeit}\in\pS_{\Ssym{\zeit}}
\,,
\ee{circ-m}
where $\Ssym{\zeit}$ takes values in the ${s}$-letter alphabet
\beq
\Ssym{} \in \A=\{0,1,2,\cdots,s-1\}
\,.
\ee{base-sAlph}

The Bernoulli map is a highly instructive example of a
hyperbolic dynamical system. Its symbolic dynamics is simple:
the base-${s}$ expansion of the initial point $\ssp_0$ is also its
temporal itinerary, with symbols from alphabet \refeq{base-sAlph}
indicating that at time $\zeit$ the orbit visits the subinterval
$\pS_{\Ssym{\zeit}}$. The map is a `shift':
a multiplication by ${s}$ acts on the base-${s}$
representation of $\ssp_{0}=.\Ssym{1}\Ssym{2}\Ssym{3}\cdots $ (for
example, binary, if ${s}=2$) by shifting its digits,
\bea
\ssp_{1}
    &=& \map(\ssp_{0})
    =.\Ssym{2}\Ssym{3}\cdots
\,.
\label{shiftBern}
\eea

Periodic points can be counted by observing that the preimages of
critical points $\{\ssp_{c1},\ssp_{c2},\cdots\ssp_{c,s-1}\}$ =
$\{{1}/s,{2}/s,\cdots,(s-1)/s\}$ partition the unit interval into
$s^\cl{}$ subintervals
$\{\pS_0,\pS_1,\cdots,\pS_{s-1}\}$,
$s^2$ subintervals
$\{\pS_{\Ssym{1}\Ssym{2}}\}$, $\dots$,
$s^\cl{}$ subintervals, each containing {\em one}  unstable
period-$\cl{}$ periodic point
$\ssp_{\Ssym{1}\Ssym{2}\cdots\Ssym{\cl{}}}$, with stability multiplier
${s}^\cl{}$, see \reffig{fig:BernPart}. The Bernoulli map is a
full shift, in the sense that every itinerary is \admissible, with one
exception: on the circle, the rightmost fixed point is the same as the
fixed point at the origin, $\ssp_{s-1}=\ssp_{0}\quad(\mbox{mod}\;1)$,
so these fixed points are identified and counted as one, see
\reffig{fig:BernPart}. The total number of periodic points of period
$\cl{}$ is thus
\beq
N_{\cl{}} = s^{\cl{}} - 1
\,.
\ee{noPerPtsBm}

\subsection{Temporal Bernoulli}
\label{s:1D1dLatt}

To motivate our formulation of a \spt\ chaotic field theory to be
developed below,
    %
\toVideo{youtube.com/embed/2Rl-KKsiXFw} 
we now recast the local initial value, time-evolution
Bernoulli map problem as a \emph{temporal lattice} fixed point condition,
the problem of enumerating and determining all global solutions.

`Temporal' here refers to the lattice site field  $\ssp_\zeit$ and the
winding number $\Ssym{\zeit}$ taking their values on the lattice
sites of a 1\dmn\ \emph{temporal} integer lattice $\zeit\in\integers$.
    %
\toVideo{youtube.com/embed/wjQ1DmwhkEM} 
Over a finite lattice segment, these can be written compactly  as a
\emph{{\lst}} and the corresponding \emph{symbol \brick}
\beq
\transp{\Xx} 
             = (\ssp_{\zeit+1},\cdots,\ssp_{\zeit+\cl{}})
\,,\quad
\transp{\Mm} 
             = (\Ssym{{\zeit+1}},\cdots,\Ssym{{\zeit+\cl{}}})
\,,
\ee{pathBern}
where $\transp{(\cdots)}$ denotes a transpose.
The Bernoulli equation \refeq{circ-m}, rewritten as a first-order
difference equation
\beq
-\ssp_{\zeit+1} + ({s}\ssp_{\zeit} - \Ssym{\zeit}) =0
\,,\qquad  \ssp_{\zeit} \in [0,1)
\,,
\ee{1stepDiffEq}  
takes the matrix form
\beq
\jMorb\,\Xx - \Mm =0
\,,\qquad
\jMorb =  - {\shift} + {s}\id
\,,
\ee{tempBern}
where the $[\cl{}\!\times\!\cl{}]$ matrix
\beq
\shift_{jk}=\delta_{j+1,k}
\,,\qquad
\shift
=  \left(\begin{array}{ccccc}
             0    &  1    &        &   &  \cr
                  &  0    &   1    &   &  \cr
                  &       &        & \ddots &  \cr
                  &       &        & 0 & 1 \cr
             1    &       &        &   & 0
          \end{array} \right)
\,,
\ee{hopMatrix}
implements the shift operation \refeq{shiftBern}, a cyclic permutation
that translates for\-ward-in-time {\lst} $\Xx$ by one site,
$\transp{(\shift \Xx)}=(\ssp_2,\ssp_3,\cdots,\ssp_\cl{},\ssp_1)$. The
time evolution law \refeq{circ-m} must be of the same form for all times,
so the operator $\shift$ has to be time-translation invariant, with
$\shift_{\cl{}+1,\cl{}}=\shift_{1\cl{}}=1$ matrix element enforcing its
periodicity.

As the {temporal Bernoulli} condition \refeq{tempBern} is a linear
relation, a given \brick\ $\Mm$, or `code' in terms of alphabet
\refeq{base-sAlph}, corresponds to a unique temporal {\lst} $\Xx_\Mm$.
That is why Percival and Vivaldi\rf{PerViv} refer to such symbol \brick\
$\Mm$ as a {\em linear code}.
The temporal Bernoulli, however, is {\em not} a linear dynamical system:
as illustrated by \reffig{fig:BernPart}, it is a set of piecewise-linear
${s}$-stretching maps and their compositions, one for each \statesp\
region $\pS_{\Mm}$.

\subsection{Bernoulli as a continuous time dynamical system}
\label{s:bernODE}

The discrete time derivative of a lattice configuration \Xx\ evaluated at the
lattice site \zeit\ is given by the {difference operator}\rf{Elaydi05}
    \index{lattice!derivative}\index{derivative, lattice}
    \index{lattice!derivative, forward}\index{difference operator}
\beq
\dot{\ssp}_\zeit =
\left[\frac{\partial\Xx}{\partial\zeit}\right]_\zeit
        =
    \frac{\ssp_{\zeit+1} - \ssp_{\zeit}}{\Delta\zeit}
\,.
\ee{lattTimeDer}
The {temporal Bernoulli} condition \refeq{tempBern} 
can be thus viewed as forward Euler method,
a time-discretized, first-order ODE dynamical
system
\beq
   \dot{\Xx} \,=\, \vel(\Xx,\Mm) \,,
\ee{1stepVecEq}
where the `velocity' vector field $\vel$ is given by
\[
\vel(\Xx,\Mm) \,=\,
(s-1)\,\Xx-\Mm
\,,
\]
with the time increment set to $\Delta\zeit=1$, and perturbations that
grow (or decay) with rate $({s}-1)$. By inspection of
\reffig{fig:BernPart}\,(a), it is clear that for \emph{shrinking},
${s}<1$  parameter values the orbit is stable for\-ward-in-time, with a
single linear branch, 1-letter alphabet $\A=\{0\}$, and the only
{\lsts} being the single fixed point  $\ssp_0=0$, and its repeats
$\Xx=(0,0,\cdots,0)$. However, for \emph{stretching},  ${s}>1$  parameter
values, the Bernoulli system (more generally, R{\'e}nyi's beta
transformations\rf{Renyi57}) that we study here, every {\lst}
$\Xx_\Mm$ is unstable, and there is a {\lst} for each admissible
symbol \brick\ \Mm.

\paragraph{A fair coin toss, summarized.}
We refer to the \emph{global} temporal lattice condition \refeq{tempBern}
as the `\emph{temporal} Bernoulli', in order to distinguish it from the
1-time step Bernoulli evolution \emph{map} \refeq{n-tuplingMap}, in
preparation for the study of \emph{\spt} systems to be undertaken in
\refref{CL18}. In the lattice formulation, a \emph{global} {temporal
{\lst}} $\Xx_\Mm$ is determined by the requirement that the
\emph{local} temporal lattice condition \refeq{1stepDiffEq} is satisfied
at every lattice site. In \spt\ formulation there is no need for
for\-ward-in-time, close recurrence searches for the returning periodic
points. Instead, one determines each global {temporal {\lst}}
$\Xx_\Mm$ at one go, by solving the fixed point condition
\refeq{eqMotion}. The most importantly for what follows, the \spt\
field theory of \refref{CL18}, this calculation requires no recourse to any
\emph{explicit coordinatization and partitioning of system's state
space}.


\section{A kicked rotor}
\label{s:kickRot}

Temporal Bernoulli is the simplest example of a chaotic lattice field
theory. Our next task is to formulate a deterministic {\spt}ly chaotic
field theory, Hamiltonian and energy conserving, because (a) that is
physics, and (b) one cannot do quantum theory without it. We need a
system as simple as the Bernoulli map, but mechanical. So, we move on
from running in circles, to a mechanical rotor to kick.

The 1-degree of freedom maps that describe kicked rotors
subject to discrete time sequences of angle-dependent force pulses
    %
\toVideo{youtube.com/embed/kWrvgeqYaBU} 
$P(\coord_{\zeit})$, $\zeit\in\integers$,
\bea
\coord_{\zeit+1} &=& \coord_{\zeit} + p_{\zeit+1} \qquad  (\mbox{mod}\;1),
    \label{PerViv2.1b}\\    
p_{\zeit+1}       &=& p_{\zeit} + P(\coord_{\zeit})
\,,
    \label{PerViv2.1a}      
\eea
with $2\pi \coord$ the  angle of the rotor, $p$ the momentum conjugate to
the angular coordinate $\coord$, and the angular pulse
$P(\coord_{\zeit})=P(\coord_{\zeit+1})=-V'(\coord_{\zeit})$ lattice
periodic with period $1$, play a key role in the theory of deterministic
and quantum chaos in  atomic physics, from the Taylor, Chirikov and
Greene  standard map\rf{Lichtenberg92,Chirikov79}, to the cat maps that
we turn to now. The equations are of the Hamiltonian form:
eq.~\refeq{PerViv2.1b} is $\dot{\coord}=p/m$ in terms of discrete
time derivative \refeq{lattTimeDer}, \ie, the configuration trajectory
starting at $\coord_{\zeit}$ reaches
$\coord_{\zeit+1}=\coord_{\zeit}+p_{\zeit+1}\Delta{\zeit}/m$ in one time
step $\Delta{\zeit}$. Eq.~\refeq{PerViv2.1a} is the time-discretized
$\dot{p}=-\partial V(\coord)/\partial \coord$: at each kick the angular
momentum $p_{\zeit}$ is accelerated to $p_{\zeit+1}$ by the force pulse
$P(\coord_{\zeit})\Delta{\zeit}$, with the time step and the rotor mass
set to $\Delta{\zeit}=1$,  $m=1$.

\subsection{Cat map}
\label{s:catPV}

The simplest kicked rotor is subject to force pulses
$P(\coord)={\mu}^2\coord$ proportional to the angular displacement
$\coord$: in that case, the map
(\ref{PerViv2.1b},\ref{PerViv2.1a}) is of form
 \beq
 \left(\begin{array}{c}
 \coord_{\zeit+1}  \\
   p_{\zeit+1}
  \end{array} \right )=
  \jMat \left(\begin{array}{c}
 \coord_{\zeit}  \\
   p_{\zeit}
  \end{array} \right )\quad (\mbox{mod}\;1)
    \,,  \qquad
 {\jMat} =\left(\begin{array}{cc}
 {\mu}^2+1 & 1 \\
  {\mu}^2 & 1
  \end{array} \right)
\,.
\ee{catMap}
The $(\mbox{mod}\;1)$ makes the map a
discontinuous `sawtooth,' unless ${\mu}^2$ is a positive integer.
The map is then a Continuous Automorphism of the Torus
known as the Thom-Anosov-Arnol'd-Sinai
{\em `cat map'}\rf{ArnAve,deva87,StOtWt06}, extensively studied as the
simplest example of a chaotic Hamiltonian system.

The determinant of the one-time-step Jacobian is
$\det \jMat=1$, \ie, the forward-in-time mapping is area-preserving.
Let ${s}=\tr{\jMat}={\mu}^2+2$ be the trace of the Jacobian.
For $|s|>2$ the $\jMat$ {characteristic equation}
\beq
\ExpaEig^{2} - {s}\ExpaEig + 1 = 0
\,,
\ee{catCharEq} 
has real roots
$(\ExpaEig\,,\;\ExpaEig^{-1})$  and a positive Lyapunov exponent
$\Lyap >0$,
\beq
\ExpaEig=e^{\Lyap} = \frac{1}{2}(s+\sqrt{(s-2)(s+2)})
\,,\qquad
{s}=\tr{\jMat}=\ExpaEig+\ExpaEig^{-1}
\,.
\ee{StabMtlpr}
The eigenvalues are functions of the stretching parameter $s$, and
for $|s| > 2$ the cat map \refeq{catMap} is a fully chaotic
Hamiltonian dynamical system.

\subsection{\tempLatt}
\label{s:catLagrange}
\renewcommand{\period}[1]{{\ensuremath{n_{#1}}}}

In order to motivate our formulation of $d$-dimensional \spt\ chaotic
field theories, to be developed in \refref{CL18}, we now recast the
\emph{local} initial value, Hamiltonian time-evolution  as a
\emph{global} solution to the {\ELe}s.

The 2-component field at the
temporal lattice site \zeit,
\(
\ssp_{\zeit} =(\coord_{\zeit},p_{\zeit}) \in  (0,1]\times(0,1]
\)
is kicked rotor's the angular position and momentum.
Hamilton's equations (\ref{PerViv2.1b},\ref{PerViv2.1a}) induce
for\-ward-in-time evolution on a 2-torus  $(\coord_{\zeit},p_\zeit)$ {\em
phase space}.
Eliminating the momentum by Hamilton's discrete time velocity
eq.~\refeq{PerViv2.1b},
\beq
(\coord_\zeit,p_\zeit) =
\left(
    \coord_{\zeit},\frac{\coord_{\zeit} - \coord_{\zeit-1}}{\Delta\zeit}
\right)
\,,
\ee{Ham2Lagr}
setting the time step to $\Delta\zeit=1$, and forgetting for a moment
the $(\mbox{mod}\;1)$ condition, the
for\-ward-in-time Hamilton's first order difference equations are brought
to the second order difference, 3-term recurrence {\ELe}s
for scalar field $\ssp_{\zeit}=q_\zeit$,
\beq
\ssp_{\zeit+1} - 2\,\ssp_{\zeit} + \ssp_{\zeit-1} + V'(\ssp_{\zeit})
    \,=\, 0
\,.
\ee{kittyMap}
But that is Newton's Second Law: ``acceleration equals
force,'' so Percival and Vivaldi\rf{PerViv} refer to this formulation as
`Newtonian'. Here we follow Allroth\rf{Allroth83}, Mackay, Meiss,
Percival, Kook \& Dullin\rf{MacMei83,meiss92,MKMP84,DulMei98,kooknewt},
and Li and Tomsovic\rf{LiTom17b} in referring  to it as `Lagrangian'.

For the cat map \refeq{catMap}, the Lagrangian passage
\refeq{Ham2Lagr} to the  scalar field  $\ssp_{\zeit}$ leads to the \PV\
`two-configuration representation'\rf{PerViv}
\beq
 \left(\begin{array}{c}
 \ssp_{\zeit}  \\
 \ssp_{\zeit+1}
 \end{array} \right )=
 \jMat_{PV} \left(\begin{array}{c}
 \ssp_{\zeit-1}  \\
 \ssp_{\zeit}
 \end{array} \right ) 
 - \left(\begin{array}{c}
 0  \\
 \Ssym{\zeit}
 \end{array} \right )
 \,,  \qquad
 {\jMat_{PV}} =\left(\begin{array}{cc}
 0 & 1 \\
 -1 & s
 \end{array} \right ),
\ee{PerViv}     
with matrix $\jMat_{PV}$ acting on the 2\dmn\ space of successive
configuration points $\transp{(\ssp_{\zeit-1},\ssp_{\zeit})}$. As was
case for the Bernoulli map \refeq{1stepDiffEq}, the cat map
$(\mbox{mod}\;1)$ condition \refeq{catMap} is enforced by integers
$\Ssym{\zeit}\in  \A$, where for a given integer stretching parameter $s$
the alphabet \A\ ranges over $|\A|={s}\!+\!1$ possible values for
$\Ssym{\zeit}$,
\beq
\A=\{\underline{1},0,\dots s\!-\!1\}
\,,
\ee{catAlphabet}
necessary  to keep $\ssp_{\zeit}$ for all times $t$ within the unit
interval $[0,1)$. (We find it convenient to have symbol
$\underline{\Ssym{}}{}_{\zeit}$ denote $\Ssym{\zeit}$ with the negative
sign, \ie, `$\underline{1}$' stands for symbol `$-1$'.)

Written out as a second-order difference equation, the \PV\ map
\refeq{PerViv} takes a particularly elegant form,
    %
\toVideo{youtube.com/embed/RjV30zx_Pp0} 
that we shall
refer to as the {\em \templatt} \refeq{1dTemplatt},
\beq
-\ssp_{\zeit+1}  +  ({s}\,\ssp_{\zeit} - \Ssym{\zeit}) - \ssp_{\zeit-1}
    =
0
\,,
\ee{catMapNewt}
or,
in terms of a {{\lst}} $\Xx$, the corresponding {symbol \brick}
$\Mm$ \refeq{pathBern}, and the $[\cl{}\!\times\!\cl{}]$ time translation operator
$\shift$ \refeq{hopMatrix},
\beq
(-\shift + s\id - \shift^{-1})\,\Xx =  \Mm
\,,
\ee{catTempLatt}
very much like the {temporal Bernoulli} condition \refeq{tempBern}, with
the winding numbers $\Mm$ taking their values on the lattice
sites of a 1\dmn\ \emph{temporal} lattice $\zeit\in\integers$.

As was the case for {temporal Bernoulli} \refeq{tempBern}, the condition
\refeq{PerViv} is a linear relation: a given `code'
$\{\Ssym{\zeit}\}$ in terms of alphabet \refeq{catAlphabet} corresponds
to a unique temporal sequence $\{\ssp_\zeit\}$. That is why Percival and
Vivaldi\rf{PerViv} refer to such symbol \brick\ $\Mm$ as a {\em linear
code}. As for the Bernoulli system, $\Ssym{\zeit}$ can also be
interpreted as `winding numbers'\rf{Keating91}, or, as they shepherd
stray points back into the unit torus, as `stabilising
impulses'\rf{PerViv}.
The cat map, however, is {\em not} a linear dynamical system:
it is a set of piecewise-linear maps and their
convolutions, one for each \statesp\ region $\pS_{\Mm}$.

The lattice formulation \refeq{catMapNewt} lends itself immediately to
$d$\dmn\ generalizations. An example is the Gutkin and
Osipov\rf{GutOsi15} \catlatt\ in  $d=2$ dimensions\rf{CL18},
    %
\toVideo{youtube.com/embed/rTh_I0KOasY} 
an Arnold
cat map-inspired Euclidean scalar field theory  of form
\refeq{freeAction} for which the {\ELe}
\refeq{eqMotion} is a 5-term recurrence relation
\beq
      -\ssp_{j,\zeit+1} - \ssp_{j,\zeit-1}
+ (2{s}\,\ssp_{j\zeit}-\Ssym{j\zeit})
     - \ssp_{j+1,\zeit} - \ssp_{j-1, \zeit}
     \,=\,  0
\,,
\ee{CatMap2d}
where we refer to parameter ${s}$, related to the Klein-Gordon mass in
\refeq{freeAction} by ${\mu}^2=d({s}-2)$, as the `stretching
parameter'.

\subsection{\tempLatt\ as a continuous time dynamical system}
\label{s:tempCatODE}

Recall that the Bernoulli first-order difference equation could be viewed as
a time-discretization of the first-order linear ODE \refeq{1stepVecEq}. The
second-order difference equation \refeq{catMapNewt} can be interpreted as the
second order discrete time derivative ${d^2}/{dt^2}$, or the temporal
lattice Laplacian \refeq{LapTime},
\beq
\Box\,\ssp_\zeit \equiv
\ssp_{\zeit+1} - 2\ssp_{\zeit} + \ssp_{\zeit-1}
= (s-2)\ssp_{\zeit} -\Ssym{\zeit}
\,,
\ee{PerViv2.2}
 with the time step set to $\Delta\zeit=1$.
In other words, if we include the cat map forcing pulse
\refeq{PerViv2.1a}
\(
P(\ssp_\zeit)= - V'(\ssp_\zeit) = (s-2)\,\ssp - \Ssym{\zeit}
\)
into the definition of
the on-site potential, 
the force
is linear in the angular displacement $\ssp$, so
the \templatt\ {\ELe} takes form (see free action
\refeq{freeAction})
\beq
(-\Box + {\mu}^2\id)\,\Xx = \Mm
\,,
\ee{OneCat}
where the Klein-Gordon mass ${\mu}$
    %
\toVideo{youtube.com/embed/uyu1O3tZgFM} 
is related to the cat-map
stretching parameter ${s}$ by ${\mu}^2={s}-2$.

Here we study the strong stretching, $s>2$ case, known as the discrete
\sPe\rf{Dorr70,GoVanLo96,HuCon96,HuRyCo98,FetWal03,Pozrikidis14},
whose solutions are hyperbolic. We refer to the
Euler–\-Lagrange equation
\refeq{OneCat} as the `{\em \templatt}', both to distinguish it from
the for\-ward-in-time Hamiltonian cat \emph{map} \refeq{catMap}, and in the
anticipation of the \emph{\catlatt} to be discussed in the sequel
\refref{CL18}.
The field $\ssp_\zeit$ compactification to unit circle makes the
{\catlatt} a strongly \emph{nonlinear} deterministic field theory, with
nontrivial symbolic dynamics.

\paragraph{\tempLatt, summarized.}
In the \spt\ formulation a \emph{global} {temporal {\lst}}
\beq
\transp{\Xx} 
             = (\ssp_\zeit,\ssp_{\zeit+1},\cdots,\ssp_{\zeit+k})
\ee{path}
is not determined by a for\-ward-in-time `cat map' evolution
\refeq{catMap}, but rather by the fixed point condition
\refeq{eqMotion}
that the \emph{local}, 3-term discrete temporal lattice {\ELe}s \refeq{catMapNewt} are satisfied at every lattice point. This
temporal 1\dmn\ lattice reformulation is the bridge that takes us from
the single cat map \refeq{catMap} to the higher-\dmn\ coupled
``multi-cat'' \spt\ lattices\rf{GutOsi15,GHJSC16,CL18}.

\renewcommand{\period}[1]{{\ensuremath{T_{#1}}}}         

\section{Nonlinear field theories}
\label{s:nonlinFT}

The `mod~1' in the definition of the `linear' kicked rotor, makes the cat
map \refeq{catMap} a highly nonlinear, discontinuous map.
In contrast, discretized scalar
$d$\dmn\ Euclidean $\phi^k$ theories\rf{Munster10}
are defined by smooth, polynomial  actions \refeq{partFunct}
given as lattice sums over the Lagrangian density
\beq
S[\Xx] = \sum_z \left\{
    \frac{1}{2} \sum_{\mu =1}^d
(\partial_{\mu}\ssp)_z^2
+
V(\ssp_z)
            \right\}
\,,
\ee{actionDscr}
    %
\toVideo{youtube.com/embed/F-iOrF-G-1M} 
with nonlinear self-interaction\rf{Wolff14}
\beq
V(\ssp) =    \frac{\;\mu^2}{2}\,\ssp^2
           - \frac{g}{k!}\,\ssp^k
    \,,\quad k\geq3
\,,
\ee{polynPotent}
where $V(\ssp)$ is a {local} nonlinear potential%
\rf{FriMil89,DulMei00,LiMal04,AnBoBa17,AnBoBa18,Anastassiou21}, the same
for every lattice site $z$, $\mu^2\geq0$ is the Klein-Gordon
mass-squared, $g\geq0$ is the strength of the self-coupling, and we set
lattice constant to unity, $a=1$, throughout. The signs of the terms of
\refeq{polynPotent} reflect our focus on deterministic \spt\ chaos, \ie,
we shall study systems for whom all solutions are unstable.
    %
\toVideo{youtube.com/embed/cKuPh3sfW5c} 

The discrete {\ELe}s \refeq{eqMotion} now take form of 3-term recurrence,
second-order difference equations
\bea
- \Box\,\ssp_z + V'(\ssp_z)
    &=&
0  
\,.  
\label{1dTempFT} 
\eea

\subsection{A $\ssp^3$ field theory}
\label{s:henlatt}

 The simplest such nonlinear action turns out to correspond to the
paradigmatic dynamicist's model of a 2\dmn\ nonlinear dynamical system,
the {\HenonMap}\rf{henon}
\bea
    x_{\zeit+1} &=& 1-a\,x_{\zeit}^2 + b\,y_{\zeit}
        \continue
    y_{\zeit+1} &=& x_{\zeit}
\,.
\label{HenMap}
\eea
For the contraction parameter value $b=-1$ this is a Hamiltonian map
(see \refeq{HamHenDet} below).

The {\HenonMap} is the simplest map that captures chaos that arises from
the smooth stretch \& fold dynamics of nonlinear {\PoincMap}s of flows
such as R\"ossler\rf{ross}.
Written as a  2nd-order inhomogeneous difference equation\rf{DulMei00},
\refeq{HenMap} takes the
{\em \henlatt} 3-term recurrence form, time-translation
and time-reversal invariant {\ELe} \refeq{1dHenlatt},
\beq
-\ssp_{\zeit+1} + ({a}\,\ssp_{\zeit}^2 - 1) - \ssp_{\zeit-1} = 0
\,.
\ee{Hen3term} 
Just as the kicked rotor (\ref{PerViv2.1b},\ref{PerViv2.1a}), the map can
be interpreted as a kicked driven anaharmonic oscillator\rf{Heagy92},
with the nonlinear, cubic Biham-Wenzel\rf{afind}
    %
\toVideo{youtube.com/embed/PxV7q8R-NOc} 
lattice site potential \refeq{polynPotent}
\beq
V(\ssp) =  
   \frac{1}{2}\,\mu^2\ssp^2  - \frac{1}{3!}\,g\,\ssp^3
\,,
\ee{BWcubic} 
giving rise to kicking pulse \refeq{PerViv2.1a}, so we refer to this
field theory as $\ssp^3$ theory. As discussed in detail in
\refref{WWLAFC22}, one of the parameters can be rescaled away by
translations and rescalings of the field $\ssp$, and the {\ELe} of the
system can brought to various equivalent forms, such as the \Henon\
form \refeq{Hen3term}, or the anti-integrable form \refeq{1dHenlatt},
\bea
\,-\,\frac{1}{d}\sum_{||z'-z||=1}\!\!\! (\ssp_{z'} - \ssp_z)
\;+\;
\mu^2\,({1}/{4}-\ssp_{\zeit}^2)
    &=&
0
\,.
\label{1dPhi3}
\eea
For a sufficiently large `stretching parameter' $a$,
or `mass parameter'  ${\mu^2}$, the
{\lsts} of this $\phi^3$ theory are in
one-to-one correspondence to the unimodal \HenonMap\
Smale horseshoe repeller, cleanly split into the `left', positive stretching and
`right', negative stretching lattice site field values.
A plot of this horseshoe, given in, for example,
\toChaosBook{exmple.15.4} {Example 15.4} is helpfull in understanding
that \statesp\ of deterministic solutions of strongly nonlinear
field theories has fractal support.
Devaney, Nitecki, Sterling and Meiss\rf{Devaney79,StMeiss98,SteDuMei99}
have shown that the Hamiltonian {\HenonMap} has a complete Smale
horseshoe for `stretching parameter' $a$ values above
\beq
      a > 5.699310786700\cdots
\;.
\ee{SterlHen}
In numerical\rf{ChaosBook} and analytic\rf{EndGal06} calculations we fix
(arbitrarily) the stretching parameter value to $a=6$, in order to guarantee
that all $2^\cl{}$ periodic points  $\ssp=\flow{\cl{}}{\ssp}$ of the
{\HenonMap} \refeq{HenMap} exist, see \reftab{tab:HamHenon}. The symbolic
dynamics is binary, as simple as the Bernoulli map of
\reffig{fig:BernPart}\,(a), in contrast to the \templatt\ which has
nontrivial pruning, see \reftab{tab:lattstateCountCat}.

\subsection{A {$\phi^4$} field theory}
\label{s:phi4latt}

If a symmetry forbids the odd-power potentials such as \refeq{BWcubic},
one starts instead with the
Klein-Gordon\rf{BCKRBW00,BeBoJu00,BeBoVr02,BouSko12,AnBoBa17} quartic
potential \refeq{polynPotent}
\beq
V(\ssp) = \frac{1}{2}\,\mu^2\ssp^2  - \frac{1}{4!}\,g\,\ssp^4
\,,
\ee{phi4pot}
leading, after some translations and rescalings\rf{WWLAFC22}, to the \ELe\
for the lattice scalar {$\phi^4$} field theory \refeq{1dPhi4},
\bea
\,-\,\frac{1}{d}\sum_{||z'-z||=1}\!\!\! (\ssp_{z'} - \ssp_z)
\;+\;
\mu^2(\ssp_{\zeit}-\,\ssp_{\zeit}^3)
    &=&
0
\,.
\label{1dPhi4a}
\eea
Topology of the \statesp\ of $\phi^4$ theory is very much like
what we had learned for the unimodal \HenonMap\ $\phi^3$ theory,
except that the repeller set is now bimodal. As long as $\mu^2$
is sufficiently large, the repeller is a full 3-letter shift.
Indeed, while Smale's first horseshoe\rf{smale}, his fig.~1, was unimodal, he
also sketched the $\phi^4$ bimodal repeller, his fig.~5.

\subsection{Computing {\lsts} for nonlinear theories}
\label{s:nonlinLstsComp}

Unlike the {temporal Bernoulli} \refeq{1dBernLatt} and the
{\templatt} \refeq{1dTemplatt}, for which the {\lst} fixed
point condition \refeq{eqMotion}
is linear and easily solved, for nonlinear lattice field theories the
{\lsts} are roots of polynomials of arbitrarily high order.
    %
\toVideo{youtube.com/embed/r7DtAeQON5c}
While Friedland and Milnor\rf{FriMil89},
Endler and Gallas\rf{EndGal06,EG05a,EndGal06}
and others\rf{Brown81,Stephen1992A}
have developed a powerful theory that yields {\HenonMap} {\po}s in
analytic form, it would be unrealistic to demand such explicit solutions for
general field theories on multi-dimensional lattices.
    %
\toVideo{youtube.com/embed/JYP6sqcxhh0}
We take a
pragmatic, numerical route\rf{WWLAFC22,orbithunter}, and search for the fixed-point solutions
\refeq{eqMotion}
starting with the deviation of an approximate trajectory from the 3-term
    %
\toVideo{youtube.com/embed/vOaIjt44lCM} 
recurrence \refeq{1dTempFT}, given by the lattice deviation vector
\beq
v_{\zeit} = - \Box\,\ssp_\zeit + V'(\ssp_\zeit)
\,,
\ee{BWdeviate}
and minimizing this error term by any suitable variational or
\toVideo{youtube.com/embed/4bJAcix9pE0}
optimization method,
possibly in conjunction with a high-dimensional
variant of the Newton method%
\rf{CvitLanCrete02,lanVar1,orbithunter,WanLan22,ParSch22}.
\toVideo{youtube.com/embed/JAvOcKjGTVM}

\section{For\-ward-in-time stability}
\label{s:timeJacob}

Consider a temporal lattice with a $d$-component field on each lattice site
$\zeit$, with time evolution given by a $d$\dmn\ map
(1st order difference equation)
\beq
\ssp_{\zeit}-\map(\ssp_{\zeit-1}) = 0
    \,,\quad
\ssp_{\zeit}=\{\ssp_{\zeit,1},\ssp_{\zeit,2},\dots,\ssp_{\zeit,d}\}
\,.
\ee{dComponField}
A small deviation $\Delta\ssp_{\zeit}$ from $\ssp_{\zeit}$ satisfies the
linearized equation
\beq
\Delta\ssp_{\zeit} - \jMat_{\zeit-1}\,\Delta\ssp_{\zeit-1} = 0
\,,\qquad
(\jMat_{\zeit})_{ij}
=
        \frac{\partial \map(\ssp_\zeit)_i }
             {\partial \ssp_{\zeit,j}                }
\,,
\ee{d-1stepJac}
where $\jMat_{\zeit}=\jMat(\ssp_{\zeit})$ is the 1-time step
$[d\!\times\!{d}]$ \jacobianM, evaluated on lattice site $\zeit$.
The formula for the linearization of $\cl{}$th iterate of the map
\beq
\jMat^\cl{} = \jMat_{\cl{}-1}\jMat_{\cl{}-2}\cdots\jMat_1\jMat_0
\ee{jMat_n}
in terms of 1-time step  \jacobianM\ \refeq{d-1stepJac} follows by chain
rule for iterated functions. For a period-\cl{} {Bravais cell} \lst\
$\Xx_c$ we refer to this for\-ward-in-time $[d\!\times\!{d}]$ matrix as
the Floquet (or monodromy) matrix.

If the only symmetry of the system is time translation (the map
\refeq{dComponField} is the same at all temporal lattice sites, but not
invariant under a
lattice reflection), a \lst\ $\Xx_p$ is \emph{prime} if it is not a
repeat of a shorter period \lst. Evaluated at lattice site $\zeit_0$,
its  {\FloquetM} is
\beq
\jMat_p = \jMat_{\cl{p}-1}\jMat_{\cl{p}-2}\cdots\jMat_1\jMat_0
\,.
\ee{jMat_p}

Consider a {\lst} $\Xx$ which is $m$th repeat of a
period-$\cl{}$ prime {\lst} $\Xx_p$ (for a sketch, see
\reffig{fig:1dLatStatC_5}\,(1)).
Due to the multiplicative structure
\refeq{jMat_n} of \jacobianMs, the {\FloquetM} for the $m$th
repeat of a prime period-$\cl{}$ \lst\ $\Xx_p$  is
\beq
\jMat^{m\cl{}}(\ssp_0)
        = \jMat^\cl{}(\ssp_{(m-1)\cl{}}) \cdots
    \jMat^\cl{}(\ssp_\cl{})
    \jMat^\cl{}(\ssp_0)
    = \jMat_{p}^{m} \,.
\ee{Jrepeat}
Hence it suffices to restrict our considerations to the
{\FloquetM} of prime {\lsts}.

For example, for the Hamiltonian, $b=-1$ {\HenonMap} \refeq{HenMap}, the
1-time step \jacobianM\ \refeq{d-1stepJac} is
\beq
\jMat_\zeit =
            \left(\begin{array}{cc}
                -2a\,\ssp_\zeit & -1 \\
                         1 & 0
            \end{array}\right)
\,,\qquad (\ssp_\zeit,\varphi_\zeit) = \map^{\zeit}(\ssp_0,\varphi_0)
\,.
\ee{Hen-1stepJac}
So, once we have a determined a \henlatt\ \lst\ $\Xx_p$, we have its
\FloquetM\ $\jMat_p$. When $\jMat_p$ is hyperbolic, only the expanding
eigen\-value $\ExpaEig_1=1/\ExpaEig_2$ needs to be determined, as the
determinant of the {\Henon} 1-time step \jacobianM\ \refeq{Hen-1stepJac} is
unity,
\beq
\det\jMat_p = \ExpaEig_1 \ExpaEig_2 = 1
\,.
\ee{HamHenDet}    
The map is Hamiltonian in the sense that it preserves areas in the
$[\ssp,\varphi]$ plane.

\section{Orbit stability}
\label{s:JacobianOrb}

The discretized Euler–\-Lagrange $F[\Xx_c]=0$ fixed point condition
\refeq{eqMotion} is central to the theory of robust {global methods} for finding
{\po}s. In global multi-shooting, collocation\rf{auto,GM00aut,ChoGuck99}, and
Lindstedt-Poincar{\'e}\rf{DV02,DV03,DV04} searches for \po s, one discretizes
a \po\ into $\cl{}$ sites  temporal lattice
configuration\rf{CvitLanCrete02,lanVar1,DingCvit14,DCTSCD14}, and lists the
field value at a point of each segment
\beq
\transp{\Xx}=(\ssp_0,\ssp_{1},\cdots,\ssp_{\cl{}-1})
\,.
\ee{nXdCycle}
Starting with an initial guess for \Xx, a zero of function
$F[\Xx_c]$ can then be found by Newton iteration,
    %
\toVideo{youtube.com/embed/RvDGNXUMX_c} 
which requires
an evaluation of the $[\cl{}\!\times\!\cl{}]$ \emph{\jacobianOrb}
\beq
\jMorb_{\zeit\zeit'} =\frac{\delta F[\Xx_c]_\zeit}{\delta \ssp_{\zeit'}}
\,.
\ee{jacobianOrb}

The {temporal Bernoulli} condition \refeq{tempBern}
and
the {\templatt} discretized Euler–\-Lagrange equation \refeq{catTempLatt}
can be viewed as searches for zeros of the vector of
$\cl{}$ functions
\bea
F[\Xx_\Mm] &=& \jMorb\Xx-\Mm = 0
                \label{tempFixPoint}\\
&& \mbox{temporal Bernoulli: } \qquad \jMorb =  - {\shift} + {s}\id
                \label{bernFixPoint}\\
&& \mbox{\templatt: } \qquad\qquad\;\;\,    \jMorb =  -\shift + s\id - \shift^{-1}
                \label{tempCatFix}
\,,
\eea
with the entire periodic \emph{{\lst}} ${\Xx}_{\Mm}$ treated as a
single fixed \emph{point} $(\ssp_0,\ssp_{1},\cdots,\ssp_{\cl{}-1})$ in the
\cl{}\dmn\ \statesp\ unit hypercube $\Xx\in[0,1)^\cl{}$.

For uniform stretching systems, such as the
{temporal Bernoulli}
and
the {\templatt},
 $[\cl{}\!\times\!\cl{}]$ {\jacobianOrb}  $\jMorb$ is a
circulant, time-translation invariant matrix.
Written out as matrices they are, respectively,
{temporal Bernoulli} \refeq{bernFixPoint}
\beq
\jMorb 
  \;\;=\;\;
\left(\begin{array}{ccccccc}
{s}&-1 & 0 & 0 &\dots &0        & 0 \\
0 & {s}&-1 & 0 &\dots &0        & 0 \\
0 & 0 & {s}&-1 &\dots &0        & 0 \\
\vdots & \vdots&\vdots & \vdots & \ddots &\vdots &\vdots\\
 0 & 0 & 0     & 0     &\dots   & {s}&-1 \\
-1 & 0 & 0     & 0     &\dots& 0& {s}
        \end{array} \right)
\,,
\ee{bernJacOrb}
and \templatt\ \refeq{tempCatFix}
\beq
\jMorb 
  \;\;=\;\;
\left(\begin{array}{ccccccc}
{s}&-1 & 0 & 0 &\dots &0&-1 \\
-1 & {s}&-1 & 0 &\dots &0&0 \\
0 &-1 & {s}&-1 &\dots &0 & 0 \\
\vdots & \vdots &\vdots & \vdots & \ddots &\vdots &\vdots\\
 0 & 0 & 0     & 0      &\dots  & {s}&-1 \\
-1 & 0 & 0     & 0      &\dots&-1 & {s}
        \end{array} \right)
\,.
\ee{Hessian}
While in Lagrangian mechanics matrices such as \refeq{Hessian} are often
called ``Hessian'', here we refer to them collectively as `{\jacobianOrbs}',
to emphasize that they describe the stability of any dynamical system, be it
energy-conserving, or a dissipative system without a Lagrangian formulation.

\medskip

Solutions of a {nonlinear}  field theory, such as the \lst\ sketched in
\reffig{FieldConfig}\,(b), are in general not translation invariant, so
the {\jacobianOrb} \refeq{jacobianOrb} (or the `discrete
Schr{\"o}dinger operator'\rf{Bount81,Simon82})
\beq
\jMorb_c =
\left(\begin{array}{ccccccc}
 {s}_{0} &-1 & 0 & 0 & \cdots & 0 &-1 \\
-1 & {s}_{1} &-1 & 0 & \cdots & 0 & 0 \\
0 &-1 & {s}_{2}  &-1 & \cdots & 0 & 0 \\
\vdots & \vdots & \vdots & \vdots & \ddots & \vdots & \vdots \\
0 & 0 & 0 & 0 & \cdots & {s}_{\cl{}-2} &-1 \\
-1 & 0 & 0 & 0 & \cdots & -1 & {s}_{\cl{}-1}
          \end{array} \right)
\ee{jMorb1dFT} 
is not a circulant matrix: each {\lst} $\Xx_c$ has its own {\jacobianOrb}
$\jMorb_c=\jMorb[\Xx_c]$, with the `stretching factor'
${s}_{\zeit}=V''(\ssp_\zeit)+2$
at the lattice site $\zeit$ a function of the
site field $\ssp_\zeit$.

The \jacobianOrb\ of a period-$(m\cl{})$ {\lst} $\Xx$, which is a $m$th
\emph{repeat} of a period-$\cl{}$ prime {\lst} $\Xx_p$, has a tri-diagonal
block circulant matrix form that follows by inspection from
\refeq{jMorb1dFT}:
\bea
\jMorb=
\left(
\begin{array}{ccccc}
 \mathbf{s}_p & -\mathbf{\shift} &  &  & -\transp{\mathbf{\shift}} \\
 -\transp{\mathbf{\shift}} & \mathbf{s}_p & -\mathbf{\shift} &  &  \\
  & \ddots & \ddots & \ddots &  \\
  &  & -\transp{\mathbf{\shift}} & \mathbf{s}_p & -\mathbf{\shift} \\
 -\mathbf{\shift} &  &  & -\transp{\mathbf{\shift}} & \mathbf{s}_p \\
\end{array}
\right) \,,
\label{orbJprimeRpt}
\eea
where block matrix $\mathbf{s}_p$ is a
$[\cl{}\!\times\!\cl{}]$ symmetric Toeplitz matrix
\bea
\mathbf{s}_p &=&
\left(
\begin{array}{ccccc}
 {s}_0 &-1 &  &  & 0 \\
-1 & {s}_1 &-1 &  &  \\
  & \ddots & \ddots & \ddots &  \\
  &  &-1 & {s}_{\cl{}-2} &-1 \\
 0 &  &  &-1 & {s}_{\cl{}-1} \\
\end{array}
\right) \,,\quad
\mathbf{\shift} =
\left(
\begin{array}{ccccc}
  0 &  & \cdots &  & 0 \\
  &  &  &  &  \\
  &  & \ddots &  & \vdots \\
  &  &  &  &  \\
 1 &  &  &  & 0 \\
\end{array}
\right)
\,,
\label{orbJprimBlck}
\eea
and $\mathbf{\shift}$ and its transpose enforce the periodic \bcs.
This period-$(m\cl{})$ {\lst} $\Xx$ \jacobianOrb\ is as translation-invariant
as the \templatt\ \refeq{Hessian}, but now under Bravais lattice translations
by multiples of \cl{}.
As discussed in \refsect{s:latt1d}, one can visualize this {\lst} as a
tiling of the integer lattice $\integers$ by a generic {\lst} field
decorating a tile of length \cl{}.
The \jacobianOrb\ $\jMorb$ is now a block circulant matrix which can be
brought into a block diagonal form by a unitary transformation, with a
repeating $[\cl{}\!\times\!\cl{}]$ block along the diagonal, see
\refsect{s:jacOrbSpectr}.

\renewcommand{\statesp}{state space}
\renewcommand{\Statesp}{State space}
\renewcommand{\stateDsp}{state-space}
\renewcommand{\StateDsp}{State-space}

\section{Stability of an orbit vs. its time-evolution stability}
\label{s:Hill}          

The {\jacobianOrb} $\jMorb[\Xx_c]_{z'z}$ is a
high-\dmn\ linear stability matrix
for the extremum condition
$F[\Xx_c]=0$, evaluated on the {\lst} $\Xx_c$. How is the stability
so computed related to the dynamical systems'
for\-ward-in-time stability?

As we shall show now,
the two notions of stability are related by \emph{Hill's formula}
    %
\toVideo{youtube.com/embed/y4SdeDFCLYk} 
\beq
\left|\Det\jMorb_c \right|= \left|\det (\id - \jMat_c)\right|
\ee{detDet}
which relates the characteristic polynomial of the for\-ward-in-time
evolution {\po} {\FloquetM} (monodromy matrix) $\jMat_c$ to the
determinant of the global {\jacobianOrb} $\jMorb_c$.

While first discovered in a Lagrangian setting, Hill's formulas apply
equally well to dissipative dynamical systems, from the Bernoulli map of
\refsect{s:coinToss} to \NS\ and \KS\ systems\rf{GudorfThesis,GuBuCv17},
with the Lagrangian formalism of
\refrefs{MacMei83,TreZub09,BolTre10,kooknewt} mostly getting in the way
of understanding them.
We find the discrete spacetime derivations given below a good starting
point to grasp their simplicity.

Why do we need them?
Let's say that an \cl{}-periodic $\ssp_{\zeit+\cl{}}=\ssp_\zeit$ {\lst}
$\Xx_c$ is known `numerically exactly', that is to say, to a high (but not
infinite) precision. One way to present the solution is to list the field
value $\ssp_0$ at a single temporal lattice site $\zeit=0$, and instruct the reader
to reconstruct the rest by stepping forward in time,
$\ssp_{\zeit}=\map^\zeit(\ssp_0)$.
However, for a linearly unstable orbit a single  field value $\ssp_0$ does
not suffice to present the solution, because there is always a finite
`Lyapunov time' horizon $\zeit_{Lyap}$  beyond which
$\map^\zeit(\ssp_0)$ has lost all memory of the entire {\lst} $\Xx_c$. This
problem is particularly vexing in searches for {`\ecs s'} embedded in
turbulence, where even the shortest period solutions have to be computed to
the (for a working fluid dynamicist excessive) machine
precision\rf{GHCW07,channelflow,openpipeflow}, in order to complete the first
return to the initial state. And so the ``$\cdots$ sensitivity to $\cdots$''
incantations of introductory chaos
courses\rf{strogb,ASY96,ottbook,ChaosBook,haake01} bear no relation to what we
actually \emph{do} in practice.

In practice, instead of relaying on for\-ward-in-time numerical
integration, \emph{global methods} for finding {\po}s\rf{ChoGuck99} view
them as equations for the vector fields $\dot{\ssp}$ on spaces of closed
curves, or, as we shall
see\rf{CvitLanCrete02,lanVar1,GHJSC16,CL18,GuBuCv17}, on $D$-tori
spacetime tilings. In numerical implementations  \refeq{nXdCycle} one
discretizes a \po\ into sufficiently many short
segments\rf{auto,GM00aut,ChoGuck99,DingCvit14,DCTSCD14}, and lists one
field value for each segment
\(
(\ssp_1,\ssp_2,\cdots,\ssp_\cl{})
\,.
\) 
For a $\cl{}$\dmn\ discrete time map $\map$ obtained by cutting the
flow by $\cl{}$ local {\PoincSec s}, with the \po\ now of discrete period
$\cl{}$, every trajectory segment can be reconstructed by short time
integration, and satisfies
\beq
\ssp_{\zeit+1}=\map_\zeit(\ssp_\zeit)
\,,
\ee{CyclePntErr}
to high accuracy, as for sufficiently short times the exponential
instabilities are numerically controllable. That is why a very rough, but
topologically correct global guess can robustly lead to a solution that
forward-in-time methods fail to find.

\subsection{Hill's formula for a 1st order difference equation}
\label{s:notHill}   

As Hill's formula is fundamental to our formulation of the \spt\ chaotic field
theory, we rederive it now in three ways, relying on nothing more than
elementary linear algebra.
Here is its first, `multi-shooting' derivation (where the reader is invited
to take care of the convergence of the formal series used).

Consider a temporal lattice with a $d$-component field \refeq{dComponField}.
It suffices to work out a temporal period $\cl{}=3$ example to understand
the calculation for any period. In terms of the $[3d\!\times\!3d]$
generalized \refeq{hopMatrix} block shift matrix $\shift$
\beq
\shift =
\left(
\begin{array}{ccc}
0     & \id_d& 0   \\
0     & 0     & \id_d\\
\id_d& 0     & 0
\end{array}
\right)
\,,
\ee{3shift}
where $\id_d$ is the $d$\dmn\ identity matrix,
the \jacobianOrb\
\refeq{jacobianOrb} has a block matrix form
\beq
\jMorb_c \,=\,
\id-\shift^{-1}\jMat
\,,\quad
\jMat =
\left(
\begin{array}{ccc}
\jMat_0 & 0 & 0 \\
0 & \jMat_1 & 0  \\
0 & 0 & \jMat_2
\end{array}
\right)
\,,
\ee{jacobianOrbC3}
where $\jMat_{\zeit}=\jMat(\ssp_{\zeit})$ is the 1-time step
$[d\!\times\!{d}]$ \jacobianM\ \refeq{d-1stepJac}.
Next, consider
\beq
\shift^{-1}\jMat =
\left(
\begin{array}{ccc}
0       & 0       & \jMat_2   \\
\jMat_0 & 0       & 0  \\
0       & \jMat_1 & 0
\end{array}
\right)
,\;\;
(\shift^{-1}\jMat)^2 \,=\,
\left(
\begin{array}{ccc}
0 & \jMat_2\jMat_1 & 0 \\
0 & 0 & \jMat_0\jMat_2  \\
\jMat_1\jMat_0 & 0 & 0
\end{array}
\right)
\,,
\ee{stabShift}
and note that the $\cl{}=3$ repeat of $\shift^{-1}\jMat$ is block-diagonal
\bea
(\shift^{-1}\jMat)^3  =
\left(
\begin{array}{ccc}
\jMat_2\jMat_1\jMat_0 & 0 & 0 \\
0 & \jMat_0\jMat_2\jMat_1 & 0  \\
0 & 0 & \jMat_1\jMat_0\jMat_2
\end{array}
\right)
\,,
\label{stabCube}
\eea
with $[d\!\times\!{d}]$ blocks along the diagonal cyclic permutations of each other.
The trace of the
$[\cl{}d\!\times\!\cl{}d]$ matrix for a period $\cl{}$ {\lst} $\Xx_c$
\beq
\Tr(\shift^{-1}\jMat)^k=
\sum_{m=1}^{\infty} \delta_{k,m\cl{}}\,\cl{}\,\tr\jMat_c^m
\,,\quad
\jMat_c = \jMat_{\cl{}-1}\jMat_{\cl{}-2}\cdots\jMat_1\jMat_0
\ee{FloqMat}
is non-vanishing only if $k$ is a multiple of $\cl{}$, where $\jMat_c$ is the
for\-ward-in-time $[d\!\times\!{d}]$ {\FloquetM} of the  period-\cl{} {\lst} $\Xx_c$,
evaluated at lattice site 0.
Evaluate the {\HillDet}
$\Det\jMorb_c$ by expanding
\bea
\ln\Det\jMorb_c &=&
\Tr\ln(\id-{\shift}^{-1}\jMat)
                \,=\,
-\sum_{k=1}^\infty\frac{1}{k}\,\Tr({\shift}^{-1}\jMat)^k
    \continue
                 &=&
-\tr\sum_{m=1}^\infty\frac{1}{m} \jMat_c^{m}
  =
\ln\det(\id_d-\jMat_c)
\,,
\label{LnDet=TrLn2}
\eea
where `$\Tr$, $\Det$' refer to the big, $[\cl{}d\!\times\!\cl{}d]$ global matrices,
while `$\tr$, $\det$' refer to the small, $[d\!\times\!{d}]$ time-stepping matrices.

The {\jacobianOrb} $\jMorb_c$ evaluated on a {\lst} $\Xx_c$
that is a solution of the temporal lattice first-order difference equation
\refeq{dComponField}, and the dynamical, for\-ward-in-time \jacobianM\
$\jMat_c$ are thus connected by \emph{Hill's formula}
\refeq{detDet}
which relates the global orbit stability to the Floquet, for\-ward-in-time
evolution stability. This version of Hill's formula applies to all
first-order difference equations, \ie, systems whose evolution laws are first
order in time.

Perhaps the simplest example of Hill's formula is afforded by the temporal
{Bernoulli} lattice \refeq{tempBern}.
The site field $\ssp_\zeit$ is a scalar, so $d=1$, the 1-time step
$[1\!\times\!1]$ time-evolution \jacobianM\ \refeq{d-1stepJac}  is the same
at every lattice point $\zeit$, $\jMat_{\zeit}={s}$, the {\jacobianOrb}
\refeq{tempBern} is the same for all {\lsts} of period $\cl{}$, and (see
\refsect{sect:fundFact}) in this case the Hill's formula \refeq{detDet}
counts the numbers of {\lsts}
\beq
\mbox{temporal {Bernoulli}: }\quad
N_\cl{} = |\Det\jMorb| = {s}^{\cl{}} - 1
\,,
\ee{detBern}
in agreement with the time-evolution count \refeq{noPerPtsBm}; all
itineraries are allowed, except that the periodicity of
$\shift^\cl{}=\id$ accounts for $\cycle{0}$ and
$\cycle{s\!-\!1}$ fixed points (see \reffig{fig:BernPart}) being a
single periodic point.

\subsection{Hill's formula for the trace of an evolution operator}
\label{s:forwardHill}

Our second derivation redoes the first, but now in the
\toChaosBook{chapter.19} {evolution operator formulation} of the
deterministic transport of \statesp\ orbits densities\rf{CBmeasure}, setting
up the generalization of the time-\po\ theory to the
spacetime-{\po} theory\rf{CL18}.

For a $d$\dmn\ deterministic map \refeq{dComponField} the {\FPoper}
\beq
     \Lop\,\msr(\ssp_{\zeit+1})
= \int_\pS\!\! d^d\!\ssp_{\zeit}\,
            \Lop(\ssp_{\zeit+1},\ssp_{\zeit})
            \,\msr(\ssp_{\zeit})
\ee{PerronFrobenius}
maps a \statesp\ density distribution $\msr(\ssp_{\zeit})$ one step
for\-ward-in-time. Applied repeatedly, its kernel, the $d$\dmn\ Dirac
delta function
\bea
\Lop(\ssp_{\zeit+1},\ssp_{\zeit})
    = \delta(\ssp_{\zeit+1} - \map(\ssp_{\zeit}))
\,,
\eea
satisfies the semigroup property
\beq
\Lop^2(\ssp_{\zeit+2},\ssp_{\zeit})
    = \int_\pS\!\! d^d\!\ssp_{\zeit+1}\,
            \Lop(\ssp_{\zeit+2},\ssp_{\zeit+1})\,
            \Lop(\ssp_{\zeit+1},\ssp_{\zeit})
    = \delta(\ssp_{\zeit+2}-\map{^2}(\ssp_{\zeit}))
\,.
\ee{FPsemiGroup}
The time-evolution {\po} theory\rf{ChaosBook} relates the
long time chaotic averages to the traces of {\FPoper}s
\beq
\tr\Lop^\cl{}
     = \int_\pS\!\!d^d\!\ssp\,\Lop^\cl{}(\ssp,\ssp)
     = \int_\pS\!\!d^d\!\ssp\,\delta(\ssp -\map^{\cl{}}(\ssp))
\,,
\ee{EvOp-n}
and their weighted evolution operator generalizations, with support on all
deterministic period-$\cl{}$ temporal {\lsts}
$\ssp_{c}=\map^\cl{}(\ssp_c)$.
Usually one
evaluates this trace by restricting the $d$\dmn\ integral
over $\pS$ to an infinitesimal open neighborhood
$c$ around a lattice site field $\ssp_{c,0}$,
\beq
\tr_c\Lop^\cl{} =
       \int_{c} d^d\!\ssp_0\,\delta(\ssp_{0}-\map^\cl{}(\ssp_0))
       =\frac{1}{\left|\det(\id - \jMat_c)\right|}
\,,
\ee{ForwardInTimeTr}
where $\jMat_c$ is the for\-ward-in-time $[d\!\times\!d]$ {\FloquetM}
\refeq{FloqMat} evaluated at the period-$\cl{}$ {Bravais cell} temporal site field
$\ssp_{c,0}$.

Alternatively, one can use the group property \refeq{FPsemiGroup} to insert
integrations over all $\cl{}$ temporal lattice site fields, and rewrite
$\Lop^\cl{}$ as a product of one-time-step operators $\Lop$:
\bea
\tr \Lop^\cl{} &=&
\int  d\Xx \prod_{\zeit=0}^{\cl{}-1} \delta(\ssp_{\zeit+1}-\map(\ssp_{\zeit}))
    \,,\quad
           d\Xx= \prod_{\zeit=0}^{\cl{}-1} d^d\!\ssp_{\zeit}
\,,
\label{FPtrace}
\eea
where $\ssp_{\cl{}} = \ssp_0$.
The lattice site field $\ssp_{\zeit}$ is a $d$-component field
\refeq{dComponField}, so a period-\cl{} {Bravais cell} {\lst} \Xx\ is
$(\cl{}d)$\dmn, with the $(\cl{}d)$\dmn\ Dirac delta function of the
deterministic field theory form \refeq{ClassPartitF},
\beq
\tr \Lop^\cl{} = \int d\Xx\,\delta(\shift\,\Xx - \map(\Xx))
\,,
\ee{orbitDiracDelt}
where $\shift$ is the cyclic $[\cl{}d\!\times\!\cl{}d]$ version of the
time translation operator \refeq{3shift}, and $\map(\Xx)$ acts within $d$\dmn\ blocks
\refeq{dComponField} along the diagonal.
We recognize the argument \refeq{dComponField} of this
$(\cl{}d)$\dmn\ Dirac delta function as the {\ELe} \refeq{eqMotion}
of the system,
\[
F[\Xx_c] = \shift \Xx_c - \map(\Xx_c) = 0
\,,
\]
with \lst\ $\Xx_c$ satisfying the local {\ELe}
\refeq{dComponField} lattice site by site.
Now evaluate the trace by integrating over the $d$ components of the
$\cl{}$ lattice site fields,
\beq
\tr_c \Lop^\cl{} =
            \int_{\pS_c}\!\!\!d\Xx\,\delta(F[\Xx])
= \frac{1}{\left|\Det\jMorb_c\right|}
\,,
\ee{GlobalTr}
where $\jMorb_c=\jMorb[\Xx_c]$ is the $[\cl{}d\!\times\!\cl{}d]$
{\jacobianOrb} \refeq{jacobianOrb} of a period-$\cl{}$ {\lst} $\Xx_c$, and
$\pS_c$ is an $(\cl{}d)$\dmn\ infinitesimal open neighborhood of $\Xx_c$.
By comparing the trace evaluations \refeq{ForwardInTimeTr} and \refeq{GlobalTr},
we see that we have again proved  Hill's formula \refeq{detDet} for first-order,
forward-in-time difference equations, this time without writing down any
explicit matrices such as (\ref{jacobianOrbC3}-\ref{stabCube}).

In dynamical systems theory, one often replaces higher order derivatives (for
example, {\ELe}s) by multi-component fields satisfying first
order equations (for example, Hamilton's equations), and the same is true for
discrete time systems, where a $k$th order difference equation is the
discrete-time analogue of a $k$th order differential equation\rf{Elaydi05}.
For example, the cat map and \HenonMap\ are usually presented as discrete
time evolution over a 2-component phase space \refeq{catMap} and \refeq{HenMap},
rather than the 3-term scalar field recurrence conditions \refeq{catMapNewt} and
\refeq{Hen3term}.

One could compute a \HillDet\ for such system
using the for\-ward-in-time Hill's formula for the $k$-component lattice site
field, with the corresponding $[k\cl{}\!\times\!k\cl{}]$ {\jacobianOrb}
determinant \refeq{GlobalTr}, or use the recurrence relation to reduce the
dimension of the {\jacobianOrb}. For example, in \refsect{s:catLagrange},
in passage from the Hamiltonian to the Lagrangian formulation, the $k=2$
component phase space field $(\coord_{\zeit},p_{\zeit})$ is
replaced by 1 component scalar field $\ssp_{\zeit}$. And using the 1
component scalar field one can compute the $[\cl{}\!\times\!\cl{}]$ {\jacobianOrbs}
such as (\ref{bernJacOrb}-\ref{jMorb1dFT}), whose \HillDet\ equals the
forward-in-time $[2\!\times\!2]$ phase space $\left|\det(\id - \jMat_c)\right|$.
\refAppe{s:Hill2step}, our third derivation of a
Hill's formula, is an example of such relation.

\section{\HillDet s}
\label{s:HillDet}

Having shown that the inverse of {\HillDet} $1/|\Det\jMorb_c|$ gives us the
\lst's probability \refeq{DiracDeltaExp} in the deterministic partition
function, our next task is to compute it. As we shall see in
\refsect{sect:Hillrecip1d}, that is often best done on the reciprocal
lattice. But first we show that on hypercubic lattices we can visualize
a {\HillDet} geometrically, as the volume of the associated {\fundPip}.

\subsection{Fundamental fact} 
\label{sect:fundFact}

Consider {temporal Bernoulli} and
{\templatt}.
    %
\toVideo{youtube.com/embed/Ztt1v8uGCUE} 
The {\jacobianOrb} \jMorb\ stretches the \statesp\ unit hypercube
$\Xx\in[0,1)^\cl{}$ into the \cl{}\dmn\ {\em \fundPip}, and maps each
periodic {\lst} ${\Xx}_{\Mm}$ into an integer lattice $\integers^\cl{}$
site, which is then translated by the winding numbers $\Mm$ into the
origin, in order to satisfy the fixed point condition
\refeq{tempFixPoint}. Hence $N_\cl{}$, the total number of the solutions
of the fixed point condition equals the number of integer lattice points
within the {\fundPip}, a number given by what Baake \etal\rf{BaHePl97}
call the \emph{`fundamental fact'},
\beq
N_\cl{} = |\Det\jMorb|
\,,
\ee{fundFact}
\ie, fact that the number of integer points in the {\fundPip} is equal to
its volume, or, what we refer to as its {\HillDet}.

\begin{figure}
  \centering
(a)~\includegraphics[width=0.29\textwidth]{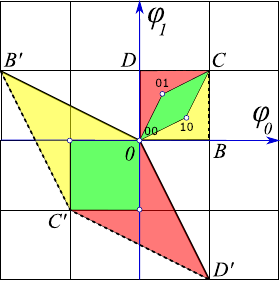}
~~~~~~                          
(b)~\includegraphics[width=0.21\textwidth]{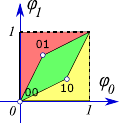}
  \caption{\label{fig:BernC2Jacob}
(Color online)~~~
  (a)
The Bernoulli map \refeq{BerShift} periodic points
$\Xx_\Mm=(\ssp_0,\ssp_1)$ of period 2 are the $\cycle{0}=(0,0)$ fixed
point, and the 2-cycle $\Xx_{01}=({1}/{3},{2}/{3})$, see
\reffig{fig:BernPart}\,(a). They all lie within the unit square $[0BCD]$,
which is mapped by the {\jacobianOrb} $-\jMorb$ \refeq{bernFundPar} into
the {\fundPip} $[0B'C'D']$. Periodic points $\Xx_\Mm$ are mapped by
$\jMorb$ onto the integer lattice, $\jMorb\Xx_\Mm\in\integers^\cl{}$, and
are sent back into the origin by integer translations $\Mm$, in order to
satisfy the fixed point condition \refeq{tempFixPoint}. Note that this
{\fundPip} is covered by  3 unit area quadrilaterals, hence
$|\Det\jMorb|=3$.
    (b)
Conversely, in the flow conservation sum rule \refeq{H-OdeA_mapsOrb2} sum
over all {\lsts} $\Mm$ of period $\cl{}$, the inverse of the
{\HillDet} defines the `neighborhood' of a lattices state as the
corresponding fraction of the unit hypercube volume.
          }
\end{figure}
%
The action of the {\jacobianOrb}
$\jMorb$ for period-2 {\lsts} (periodic points) of the Bernoulli map of
\reffig{fig:BernPart}\,(a), suffices to convey the idea. In this
case, the $[2\!\times\!2]$ {\jacobianOrb} \refeq{tempBern}, the unit
square basis vectors, and their images are
\bea
\jMorb &=&
 \left(\begin{array}{cc}
  2 & -1 \\
 -1 &  2
 \end{array} \right) \,,
    \continue
\Xx^{(B)} &=&
 \left(\begin{array}{c}
 1  \\
 0
 \end{array} \right)
\;\to\;
\Xx^{(B')} = \jMorb\,\Xx^{(B)} =
 \left(\begin{array}{c}
  2  \\
 -1
 \end{array} \right) \,,
 \continue
\Xx^{(D)} &=&
 \left(\begin{array}{c}
 0  \\
 1
 \end{array} \right)
\;\to\;
\Xx^{(D')} = \jMorb\,\Xx^{(D)} =
 \left(\begin{array}{c}
 -1  \\
 2
 \end{array} \right) \,,
\eea
\ie, the columns of the {\jacobianOrb} are the edges of the {\fundPip},
\beq
\jMorb = \left(\Xx^{(B')}\Xx^{(D')}\right)
\,,
\ee{bernFundPar}
see \reffig{fig:BernC2Jacob}\,(a), and $N_2=|\Det\jMorb|=3$,
in agreement with the {\po} count \refeq{noPerPtsBm}.

\begin{figure}
  \centering
(a)~\includegraphics[width=0.34\textwidth]{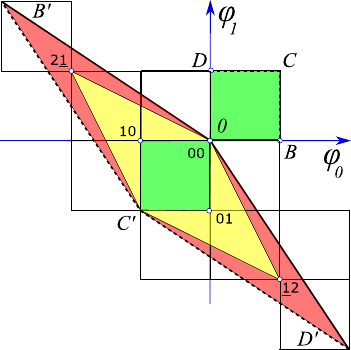}
~~~
(b)~\includegraphics[width=0.32\textwidth]{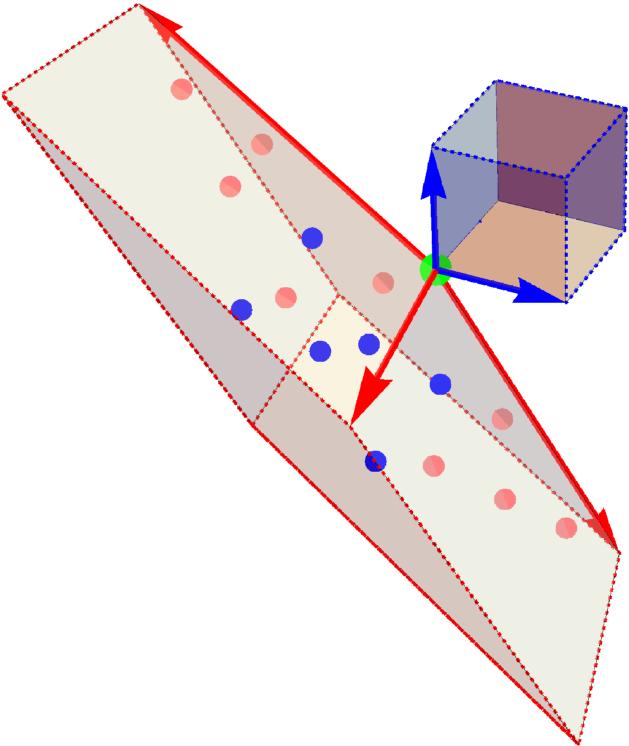} 
  \caption{\label{fig:catCycJacob}
    (Color online)~~~
(a)
    For $s=3$, the \templatt\ \refeq{catTempLatt} has 5 period-2
    {\lsts} $\Xx_\Mm=(\ssp_0,\ssp_1)$: $\Xx_{00}$ fixed point and
    period-2 {\lsts} $\{\Xx_{01},\Xx_{10}\}$,
    $\{\Xx_{\underline{1}2},\Xx_{2\underline{1}}\}$. They lie
    within the unit square $[0BCD]$, and are mapped by the
    $[2\!\times\!2]$ {\jacobianOrb} $-\jMorb$ \refeq{catFundPar2} into the
    {\fundPip} $[0B'C'D']$, as in, for example, Bernoulli
    \reffig{fig:BernC2Jacob}. The images of periodic points $\Xx_\Mm$
    land on the integer lattice, and are sent back into the origin by
    integer translations $\Mm= \Ssym{0}\Ssym{1}$, in order to satisfy the
    fixed point condition
    $\jMorb\Xx_\Mm+\Mm=0$.
(b) A 3-dimensional [{\color{blue} blue} basis vectors] unit-cube stretched by
    $-\jMorb$ \refeq{catFundPar3} into the [{\color{red} red} basis vectors]
    {\fundPip}. For $s=3$, the \templatt\
    \refeq{catTempLatt} has 16 period-3 {\lsts}: a $\Xx_{000}$
    fixed point at the vertex at the origin, [{\color{red} pink dots}] 3
    period-3 orbits on the faces of the {\fundPip}, and
    [{\color{blue} blue dots}] 2 period-3 orbits in its interior.
    An \cl{}\dmn\ \statesp\ unit hypercube $\Xx\in[0,1)^\cl{}$ and the
    corresponding {\fundPip} are half-open, as indicated
    by dashed lines, so the integer lattice points on the far corners, edges
    and faces do not belong to it.
}
\end{figure}

In general, the unit vectors of the \statesp\ unit hypercube
$\Xx\in[0,1)^\cl{}$ point along the \cl{} axes; {\jacobianOrb} \jMorb\
stretches them into a {\fundPip} basis vectors $\Xx^{(j)}$, each one a
column of the $[\cl{}\!\times\!\cl{}]$ matrix
\beq
\jMorb = \left(\Xx^{(1)}\Xx^{(2)}\cdots\Xx^{(\cl{})}\right)
\,.
\ee{lattJac}
The {\HillDet}
\beq
\Det \jMorb = \Det\!\left(\Xx^{(1)}\Xx^{(2)}\cdots\Xx^{(\cl{})}\right)
\,,
\ee{lattVol}
is then the volume of the {\fundPip} whose edges are basis vectors
$\Xx^{(j)}$. Note that the unit hypercubes and {\fundPip}s are half-open,
as indicated by dashed lines in \reffig{fig:BernC2Jacob}\,(a), so that
their translates form a partition of the extended \statesp\
\refeq{BerStretch}. For another example of {\fundPip}s, see
\reffig{fig:catCycJacob}.

For \templatt\ the total number of {\lsts} is again, as for the Bernoulli
system, given by the fundamental fact \refeq{fundFact}.
However, while for the temporal Bernoulli every sequence of alphabet
letters \refeq{base-sAlph}  but one is admissible, for \templatt\ the
condition \refeq{catMapNewt} constrains admissible winding numbers {\brick}s \Mm.

For period-1, constant field  {\lsts}
$\ssp_{\zeit+1}=\ssp_{\zeit}=\ssp_{\zeit-1}$
it follows from
\refeq{catMapNewt} that
\[
        ({s}-2)\ssp_\zeit = \Ssym{\zeit}
\,,
\]
so the {\jacobianOrb} is a $[1\times1]$ matrix, and there are
\beq
N_1={s}-2
\ee{catFundPar1}
period-1 {\lsts}. This is easy to verify by counting the admissible
$\Ssym{\zeit}$ values. Since $\ssp_\zeit \in [0,1)$, the range of
\Ssym{\zeit} is $\Ssym{\zeit} \in [0,s-2)$. So three of the
\refeq{catAlphabet}  \templatt\  letters are not admissible:
$\underline{1}$ is below the range, and $s-2$ and $s-1$ are above it.

The action of the \templatt\ {\jacobianOrb} can be hard to visualize,
as a period-2 {lattice field} is a 2-torus,
period-3 {lattice field} a 3-torus, \etc. Still, the {\fundPip} for the period-2
and period-3 {\lsts}, \reffig{fig:catCycJacob}, should suffice to
convey the idea. The {\fundPip} basis vectors 
are the
columns of $\jMorb$. The $[2\!\times\!2]$ {\jacobianOrb} \refeq{Hessian}
and its {\HillDet} are
\beq
\jMorb =
 \left(\begin{array}{cc}
  s & -2 \\
 -2 &  s
 \end{array} \right)
\,,\qquad
N_2=\Det\jMorb=({s}-2)({s}+2)
\,,
\ee{catFundPar2}
(compare with the {\lsts} count
\refeq{1stChebGenF}),
with the resulting {\fundPip} shown in \reffig{fig:catCycJacob}\,(a).
Period-3
{\lsts} for $s=3$ are contained in the half-open {\fundPip} of
\reffig{fig:catCycJacob}\,(b), defined by the columns of $[3\!\times\!3]$
{\jacobianOrb}
\beq
\jMorb =
\left(
\begin{array}{ccc}
 {s}& -1 & -1 \\
 -1 & {s}& -1 \\
 -1 & -1 & {s}
\end{array}
\right)
\,,
\qquad
N_3 = \Det \jMorb
    = ({s}-2)({s}+1)^2
\,,
\label{catFundPar3}
\eeq
again in agreement with the {\po} count \refeq{1stChebGenF}.
The 16 period-3, ${s}=3$ {\lsts} $\Xx_\Mm=(\ssp_0,\ssp_1,\ssp_2)$
are the $\Xx_{000}$ fixed point at the vertex at the origin, 3 period-3
orbits on the faces of the {\fundPip}, and 2 period-3 orbits in its
interior.

    In this example there is no need to go further with the fundamental
fact \HillDet\ evaluations, as the explicit formula for the numbers
of periodic {\lsts} is well known\rf{Isola90,Keating91}.
The \templatt\ equation \refeq{catMapNewt} is
a linear {$2$nd-order inhomogeneous difference} equation
($3$-term recurrence relation) with constant coefficients
that can be solved by standard methods\rf{Elaydi05} that
parallel the theory of linear differential equations.
Inserting a solution of form $\ssp_{\zeit}=\ExpaEig^\zeit$ into the
\Ssym{\zeit}=0 homogenous {$2$nd-order \templatt\ condition}
\refeq{catMapNewt}
yields the {characteristic equation} \refeq{catCharEq}
with roots
$\{\ExpaEig\,,\;\ExpaEig^{-1}\}$.
The result is that the number
of temporal {\lsts} of period $\cl{}$ is
\beq
N_{\cl{}}  = |\Det\jMorb| =
    \ExpaEig^{\cl{}} + \ExpaEig^{-\cl{}} - 2
\,,
\ee{1stepDiffSolu}
often written as
\beq
N_\cl{}
 = 2\,T_{\cl{}}(s/2) -2
\,,
\label{POsChebyshev}
\eeq
where $T_{\cl{}}(s/2)$ is the Chebyshev polynomial of the first kind
(this discussion continues in \refsect{sect:Hillrecip1d}).

Note that in the {temporal lattice} reformulation, both {temporal
Bernoulli} and \templatt\ happen to involve two distinct lattices:
\begin{itemize}
  \item[(i)]
In the latticization \refeq{LattField} of a time continuum,
one replaces a time-dependent
field $\ssp(\zeit)$ at time $\zeit\in\reals$ of \emph{any} dynamical system by a
discrete set of its values $\ssp_\zeit=\ssp(a\,\zeit)$, at time instants
$\zeit\in\integers$. Here the subscript `$\zeit$' indicates a \emph{coordinate} over
which the field $\ssp$ lives.
  \item[(ii)]
A peculiarity of the {temporal Bernoulli} and \templatt\ is that the
\emph{field} $\ssp_{\zeit}$, \refeq{n-tuplingMap} and \refeq{PerViv}, is
confined to the unit interval $[0,1)$, imparting an integer lattice structure
onto the intermediate calculational steps in the extended \statesp\
\refeq{BerStretch} on which the {\jacobianOrb} \jMorb\ acts.
Nothing like that applies to general nonlinear field theories of
\refsect{s:nonlinFT}.
\end{itemize}

\subsection{\Po\ theory}
\label{s:PoThe}

How come that a \HillDet\ \refeq{fundFact} counts {\lsts}?

For a general, nonlinear fixed point condition $F[\Xx]=0$, expansion
\refeq{LnDet=TrLn2} in terms of traces is a cycle
expansion\rf{inv,AACI,ChaosBook}, with support on \emph{{\po}s}.
Ozorio de Almeida and Hannay\rf{OzoHan84} were the first to relate the
number of periodic points to a \JacobianM\ generated volume; in 1984 they
used such relation as an illustration of their `principle of uniformity':
``periodic points of an ergodic system, counted with their natural
weighting, are uniformly dense in phase space.''
    %
\toVideo{youtube.com/embed/Al7iFxJkMMo} 
In \po\
theory\rf{inv,CBgetused} this principle is stated as a
\HREF{http://chaosbook.org/chapters/ChaosBook.pdf\#section.27.4} {flow
conservation} sum rule, where the sum is over all {\lsts} $\Mm$ of
period $\cl{}$,
\beq
\sum_{|\Mm|=\cl{}}
    \frac{1}{|\det (\id - \jMat_\Mm)|}
    \;= 1
\,,
\label{H-OdeA_mapsOrb}
\eeq
or, by Hill's formula \refeq{detDet},
\beq
\sum_{|\Mm|=\cl{}}
    \frac{1}{|\Det\jMorb_\Mm|}
    \;=1
\,.
\label{Det(jMorb)eights}
\eeq
For the Bernoulli and \templatt\ systems the `natural weighting' takes a
particularly simple form, as the {\HillDet} of the {\jacobianOrb} is the
same for all periodic points of period $\cl{}$,
$\Det\jMorb_\Mm=\Det\jMorb$, whose number is thus given by
\refeq{detBern}.
For example, the sum over the $\cl{}=2$ {\lsts} is,
\beq
      \frac{1}{|\Det{\color{green}\jMorb_{00}}|}
   +    \frac{1}{|\Det{\color{red}\jMorb_{01}}|}
   + \frac{1}{|\Det{\color{yellow}\jMorb_{10}}|}
    =1
\,,
\ee{H-OdeA_mapsOrb2}
see \reffig{fig:BernC2Jacob}\,(b).
Furthermore, for any piece-wise
linear system all curvature corrections\rf{CBcount} for orbits of
periods $k>\cl{}$ vanish, leading to explicit {\lst}-counting
formulas of kind reported in this paper.

In the case of temporal Bernoulli or \templatt, the hyperbolicity is the same
everywhere and does not depend on a particular solution $\Xx_c$, counting
\po s is all that is needed to solve a cat-map dynamical system
completely; once \po s are counted, all {\cycForm s}\rf{CBtrace} follow.

Fritz, this is the `\po\ theory'. And if you don't know,
\HREF{https://www.youtube.com/watch?v=_JZom_gVfuw} {now you know}.

\section{Translations and reflections} 
\label{s:latt1d}

Though this exposition is nominally about `evolution in time', `time' is
such a loaded notion, a straightjacket hard to escape, that it is best to
forget about `time' for time being, and think instead like a
crystallographer, about lattices and the space groups that describe their
symmetries.

Of necessity, there are many group-theoretic notions a crystallographer must
juggle (see \toChaosBook{section.11.2}{sect.~11.2}), but only a few key
things to understand.
    %
\toVideo{youtube.com/embed/QSPGT0XK2PI} 
For a 1\dmn\ lattice, there are only two kinds of qualitatively different
symmetry transformations,
\begin{itemize}
  \item[(i)]
translations \refeq{C_infty}
and
reflections \refeq{shiftRefl}, which reverse the direction of translation.
  \item[(ii)]
There are two kinds of reflections \refeq{DinftyClassRefl},
across a lattice site,
and
across a mid-point between lattice sites, \reffig{fig:1dLattRefl}.
  \item[(iii)]
While the lattice \lattice\ and its space group $\Group$ are both
infinite, \emph{orbits} of {\lsts} are finite and described
by finite cyclic and dihedral groups, \reffig{fig:1dLatStatC_5}.
  \item[(iv)]
A {\lst} has one of the 4 possible symmetries,
\reffig{fig:symmLattStates}. They are the building blocks of zeta
functions of \refsect{s:Lind1d}.
\end{itemize}

Should the reader find symmetries of infinite lattices too obstruse:
to understand all that one needs to know about translations  and
reflections, it suffices to understand the symmetries of a
triangle and a square, \reffig{fig:1dLatStatD3D4}.

\subsection{Internal symmetries}
\label{s:InternalSymm}

In addition to the spacetime symmetries, a field theory might have an
{\em internal} symmetry, a group of transformations that leaves the
{\ELe}s invariant, but acts only on a lattice site field, not on the
spacetime lattice.

The {$\phi^4$} action \refeq{phi4pot} is invariant
under the  $\Dn{1}$ reflection $\ssp_z\to-\ssp_z$.
The temporal Bernoulli \refeq{1stepDiffEq} and
\templatt\ \refeq{catMapNewt} {\ELe}s are invariant under
$\Dn{1}$ inversion of the field though the center of the
$0\leq\ssp_{z}< 1$ unit interval:
\beq
\bar{\ssp}_z = 1 - \ssp_z \quad \mbox{mod}\;1
\,.
\ee{InterInverBern}
If $\Xx = \{\ssp_z\}$ is a {\lst} of the system, its inversion
$\bar{\Xx}=\{\bar{\ssp}_z\}$ is also a \lst. So every lattice state of
the temporal Bernoulli and the {\templatt} either belongs to a pair of
asymmetric \lst s $\{\Xx, \bar{\Xx}\}$, or is symmetric under the
inversion. \refFigs{fig:BernC3}{fig:catD3} illustrate such symmetries.

In principle, the internal symmetries should also be quotiented, but to
keep things as simple as possible, they are not quotiented in this paper.

\subsection{Symmetries of 1\dmn\ lattices, sublattices}
\label{s:1dLatt}

A space group $\Group$ is the set of all translations and rotations that
puts a crystallographic structure \lattice\ in coincidence with itself.
To make the exposition as simple as possible, here we focus on 1\dmn\
crystals, with sites labeled by integer lattice $ \lattice=\integers$.
Their space groups crystallographers\rf{Dresselhaus07} call \emph{line
groups}.
There are only two \HREF{https://en.wikipedia.org/wiki/Line_group}
{1\dmn\ space groups}  $\Group$: $p1$, or the \emph{infinite cyclic
group}  \Cn{\infty} of all lattice translations,
and
$p1m$, the \emph{infinite dihedral group} $\Dn{\infty}$  of all
translations and reflections\rf{KiLePa03},
\beq
  \Dn{\infty} = \{
\cdots, \shift_{-2},\Refl_{-2}, \shift_{-1},\Refl_{-1},
        1,\Refl,
        \shift_{1},\Refl_{1}, \shift_{2},\Refl_{2}, \cdots
             \}
\,.
\ee{D_infty}
A half of the elements are translations (`shifts'; for finite period
lattices, `rotations').
$\shift_{0}=1$ denotes the identity,
and
the
$\shift_{1}=\shift$, $\shift_{2}=\shift^2$, $\cdots$,
$\shift_{k}=\shift^k$, $\cdots$,
denote translations by $1,2,\cdots,k,\cdots$ lattice points. They form
the \emph{infinite cyclic group}
\beq
\Cn{\infty}
    =       \{
\cdots, \shift_{-2}, \shift_{-1},
        1,
        \shift_{1}, \shift_{2}, \shift_{3}, \cdots
             \}
\,,
\label{C_infty}
\eeq
a subgroup of $\Dn{\infty}$,
in crystallography called the translation group.

The other half of elements are reflections $\Refl_{k}^2=1$
(`inversions', `time reversals', `flips'), defined by first translating
by $k$ steps, and then reflecting over the 0th lattice point,
resulting in a `translate-reflect' operation
\beq
\Refl_{k}=\Refl\shift_k
\,.
\ee{Refl_k}

The defining property of translate \& reflect groups
(`dihedral' groups, `flip systems'\rf{KiLePa03}) is that
any reflection reverses the direction of the translation
\beq
\Refl_{k}\shift = \shift^{-1}\Refl_{k}
\,.
\ee{shiftRefl}
The group multiplication (or `Cayley') table for successive group actions
$\LieEl_i\LieEl_j$ follows:
\beq
\begin{tabular}{c|cc}
            &$\shift_j$        &$\Refl_j$\\\hline
$\shift_i$  &$\shift_{i+j}$     &$\Refl_{j-i}$\\
$\Refl_i$   &$\Refl_{i+j}$     &$\shift_{j-i}$
\end{tabular}
\,.
\ee{eq:DinftyMultTab}
Multiplication either adds up translations,
or shifts and then reverses their direction.
The order in which the elements $\LieEl_i\LieEl_j$ act is right to left,
\ie, a group element acts on the expression to its right.

A crystallographer organizes the \emph{subgroups} of a space group
$\Group$ by means of \emph{Bravais lattices} $\lattice_\mathbf{a}$
\refeq{BravLatt},
sublattices of the lattice $\integers$, each defined here by a 1\dmn\
\emph{Bravais cell} of \emph{period} \cl{}, given by a lattice vector
$\mathbf{a}$ of integer length \cl{},
\beq
\lattice_\mathbf{a} = \{j \mathbf{a} \,|\, j \in \integers\}
\,,
\ee{1DBravLatt}
with the lattice generated by the infinite translation group of all
discrete translations replaced by
\[
  \shift_{j}\to\shift_{j\mathbf{a}}
\]
multiples of $\mathbf{a}$, resulting in
\beq
H_{\mathbf{a}} = \{ \cdots, \shift_{-2 \mathbf{a}}, \shift_{-\mathbf{a}},
1, \shift_{\mathbf{a}}, \shift_{2 \mathbf{a}}, \cdots\}
\,,
\ee{H(n)subgroup}
infinite translation subgroup of $\Cn{\infty}$. You can visualize a
{\lst} invariant under subgroup $H_{\mathbf{a}}$ as a tiling of the
lattice $\integers$ by a generic {\lst} over tile of length \cl{}.

Another family of subgroups of $\Dn{\infty}$ is obtained by substituting
elements of $\Dn{\infty}$ \refeq{D_infty} by
\[
  \shift_{j}\to\shift_{j\mathbf{a}}
\,,\qquad
    \Refl\to \Refl_{k}
  \quad
     0\leq{k}<\cl{}
\,,
\]
resulting in $\cl{}$ infinite dihedral subgroups
\beq
H_{\mathbf{a},k} = \{
\cdots, \shift_{-2 \mathbf{a}}, \Refl_{k}\shift_{-2 \mathbf{a}},
        \shift_{-\mathbf{a}}, \Refl_{k}\shift_{-\mathbf{a}},
        1,                    \Refl_{k},
        \shift_{\mathbf{a}},  \Refl_{k}\shift_{\mathbf{a}},
        \shift_{2\mathbf{a}},\Refl_{k}\shift_{2\mathbf{a}}, \cdots
             \}
\,,
\ee{H(n,k)subgroup}
each given by a {Bravais cell} of period \cl{}, with reflection
across a symmetry point shifted $k$ half-steps, see
\refeq{1dLattRefl1}.

\subsection{Classes}
\label{s:1dLattClass}

    \begin{quote}
Definition:
{\em A class is the set of elements left
invariant by conjugation with all elements $\LieEl$ of the group \Group,
where an element $b$ is \emph{conjugate} to element $a$ {if}

}
\beq
b = \LieEl\,a\,\LieEl^{-1}
\,.
\ee{conjugate}
    \end{quote}

By \refeq{shiftRefl}, a conjugation by any reflection reverses the
direction of translation
\beq
   \Refl_i\shift_j\Refl_{-i} =  \shift_{-j}
\,,
\ee{DinftyInversion}
so every translation pairs up with the equal counter-translation to form
\bea
\mbox{identity class }
    &&\qquad
\{1\}
    \,,\quad\qquad\;
j=0
    \label{DinftyClassId}\\
\mbox{translation classes }
    &&\qquad
\{\shift_j,\shift_{-j}\}
    \,,\quad
j=1,2,3,\cdots
\,.
\label{DinftyClassShift}
\eea
The $\shift_{0}=1$ commutes with all group elements, and is thus always a
class by itself.

From the multiplication table \refeq{eq:DinftyMultTab} it
follows that a conjugate of a reflection
\beq
\shift_i\,\Refl_j\shift_i^{-1}      
= \Refl_{j-2i}
\,, \quad
\Refl_i\Refl_{j}\Refl_i^{-1}  
= \Refl_{2i-j}
\,.
\ee{D_nConj}
is a reflection related to it by a ${2i}$ translation.
Hence the even subscript reflections belong to one class, and the odd
subscript reflections to the other:
\bea
\mbox{even}
    &&\quad
\{\Refl_{2m}\}
\,,\qquad
m\in\integers
    \continue
\mbox{odd}
    &&\quad
\{\Refl_{2m+1}\}
\,.
\label{DinftyClassRefl}
\eea
By \refeq{D_nConj} $r H_{\cl{},k} r^{-1} = H_{\cl{},k-2}$, so for odd
\cl{}, all subgroups $H_{\cl{},k}$ are conjugate subgroups,
and for even \cl{}, $H_{\cl{},k}$ separate into 2 sets of conjugate subgroups,
\bea
\mbox{even}
    &&\quad
\{H_{2m,2j}\}
\,,\qquad
0\leq j<m
    \continue
\mbox{odd}
    &&\quad
\{H_{2m,2j+1}\}
\,,
\label{H(n,k)class}
\eea
each containing $m$ subgroups.

\subsection{Reflections}
\label{s:1dLattRefl}

What's the difference between an `odd' and an `even' reflection?
Every {element} in a class is equivalent to
any other of its {element}s.
So, to understand what everybody in a given class does, it suffices to
work out what a single representative does:
it suffices to analyse the $H_{\cl{},0}$ and $H_{\cl{},1}$
to account for all $H_{\cl{},k}$.

So far,
we have only discussed the abstract structure of the space group
$\Dn{\infty}$  and its subgroups.
    %
\toVideo{youtube.com/embed/2j2K7p-ocu4}
But the difference between an `odd' and
an `even' is easiest to grasp by working out the action of $\Refl_k$ on a
{\lst}.

\begin{figure} \begin{center}
  \begin{minipage}[b]{0.32\textwidth}\begin{center}
\includegraphics[width=\textwidth]{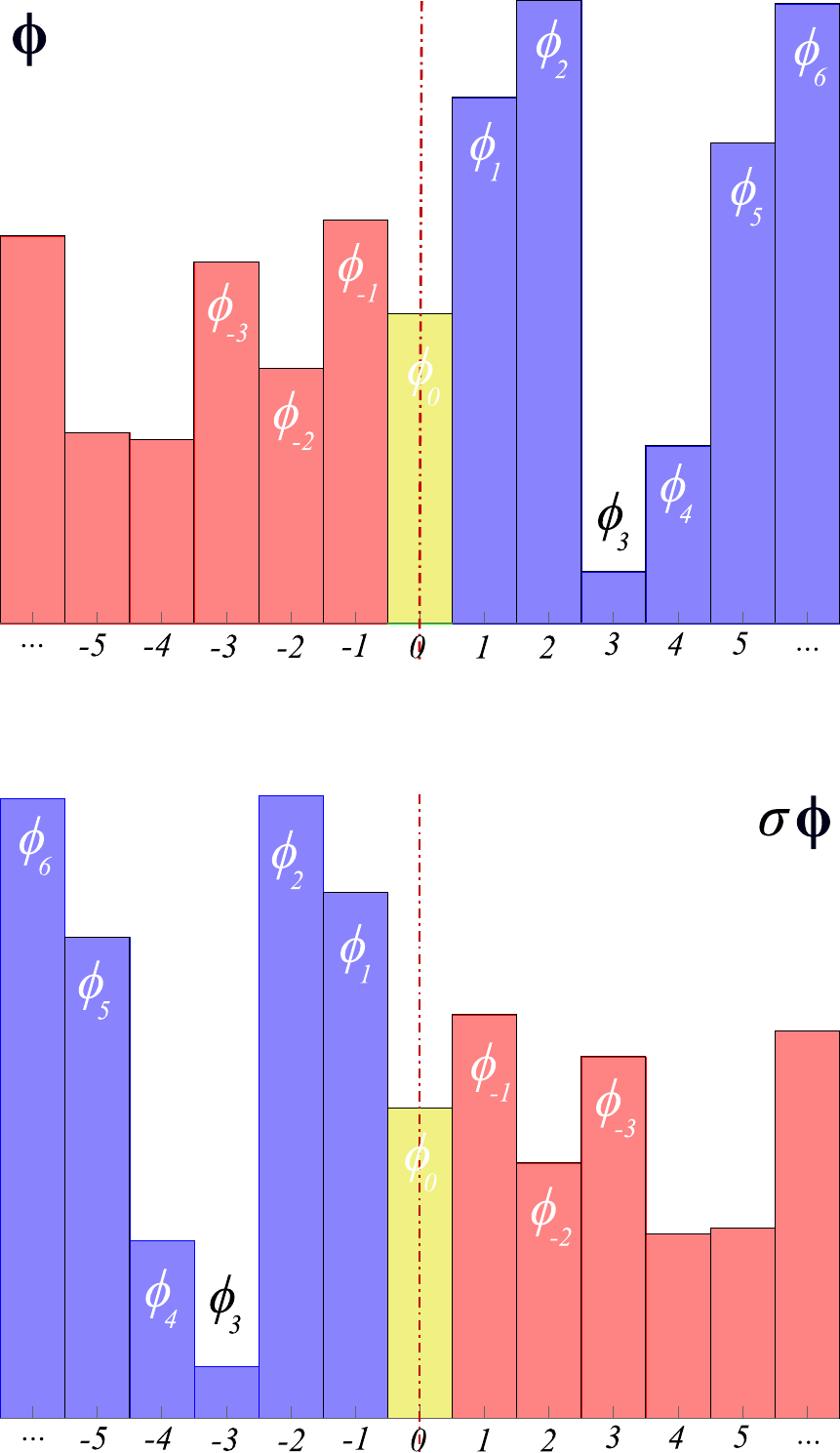}
  \\ (even)
  \end{center}\end{minipage}
\qquad\qquad
  \begin{minipage}[b]{0.32\textwidth}\begin{center}
\includegraphics[width=\textwidth]{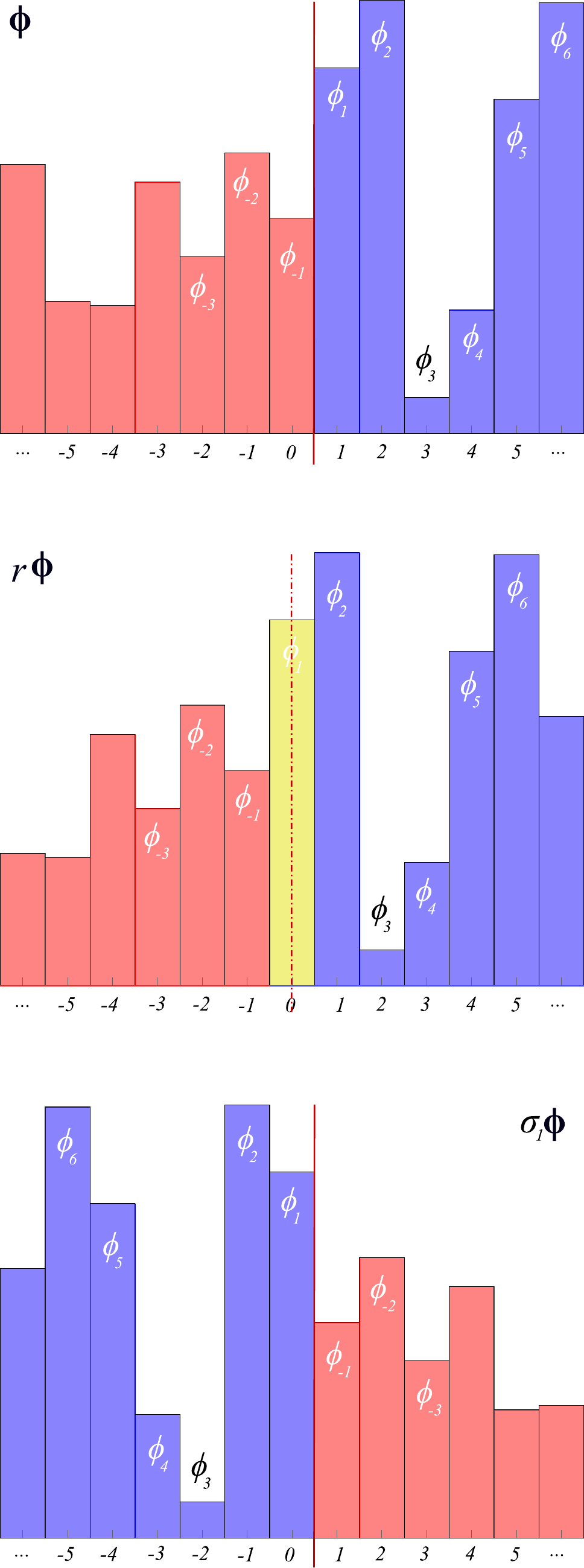}
  \\ (odd)
  \end{center} \end{minipage}
  \end{center}
  \caption{\label{fig:1dLattRefl}
(Color online)~~~
There are two classes 
of {\lst} reflections,
even \refeq{1dLattRefl0} and odd \refeq{1dLattRefl1}.
    (Even)
reflection $\Refl$ exchanges (blue $\ssp_j$) $\leftrightarrow$  (red
$\ssp_{-j}$)
marked by dashed line reflection axis, with lattice site ${0}$ fixed.
    (Odd)
reflection $\Refl_1=\Refl\shift$ swaps the `blues' and the `reds' by a
lattice translation $\Xx\to\shift\Xx$, followed by a reflection $\Refl$.
The result is a reflection across the midpoint of the [01] interval,
marked by full line reflection axis.
See \reffig{FieldConfig}\,(b) for the notation.
Continued in \reffig{fig:1dLatStatC_5}.
}
\end{figure}

\paragraph{Even class.}
Take $\Refl=\Refl_0$ as a representative of all even reflections
$\Refl_{2m}$,  and act on a {\lst} \refeq{1dLattStat}:
\bea
\Xx &=&
\cdots {\ssp}_{-3} {\ssp}_{-2}\,{\ssp}_{-1}\,
       {\ssp}_0\,
      {\ssp}_{1} {\ssp}_{2} {\ssp}_{3} {\ssp}_{4}  \cdots
\continue
\Refl\Xx &=&
\cdots  {\ssp}_{5} {\ssp}_{4} {\ssp}_{3} {\ssp}_{2} {\ssp}_{1}
       \sitebox{{\ssp}_0}\,
      {\ssp}_{-1} {\ssp}_{-2} {\ssp}_{-3}  \cdots
\,,
\label{1dLattRefl0}
\eea
with
\(
\sitebox{{\ssp}_0}
\)
indicating that the field at the lattice site $0$ is
unchanged by the reflection, see \reffig{fig:1dLattRefl}\,(even).

\paragraph{Odd class.}
 Take $\Refl_1$ as a representative of all odd reflections
$\Refl_{2m+1}$.
The result is:
\bea
\Xx &=& ~~~~\,
\cdots {\ssp}_{-3} {\ssp}_{-2} {\ssp}_{-1}
       \,{\underline {\ssp}}{}_0\,\,
      {\ssp}_{1} {\ssp}_{2} {\ssp}_{3} {\ssp}_{4}  \cdots
\continue
\shift\Xx &=& ~\,
\cdots {\ssp}_{-3} {\ssp}_{-2} {\ssp}_{-1}
       {\ssp}_0\,{\underline {\ssp}}{}_{1}\,\,
       {\ssp}_{2} {\ssp}_{3} {\ssp}_{4}  {\ssp}_{5} \cdots
\continue
\Refl_1\Xx =
\Refl\shift\Xx &=& ~~~~
\cdots  {\ssp}_{6} {\ssp}_{5} {\ssp}_{4} {\ssp}_{3} {\ssp}_{2} \,
      {\underline {\ssp}}{}_{1} | \, {\ssp}_0
      {\ssp}_{-1} {\ssp}_{-2} {\ssp}_{-3} \cdots
\,,
\label{1dLattRefl1}
\eea
where ${\underline {\ssp}}{}_{j}$ indicates the field value at the
lattice site $0$, and
\(
|
\)
indicates a reflection across midpoint
between lattice sites $0$ and $1$, see \reffig{fig:1dLattRefl}\,(odd).

More generally, one can say that the subscript $k$ in the
`translate-reflection' \refeq{Refl_k} operation
\(\Refl_{k} =\Refl\shift_k\)
advances the reflection point by $k/2$ steps, and then reflects
across it.

If you do not find the two kinds of reflections intuitive, the
distinction becomes crystal clear once you have a look at the
smallest Bravais lattices,
    %
\toVideo{youtube.com/embed/ue42-tr-tSY}
lattices of periods 3 and 4, \reffig{fig:1dLatStatD3D4}.

\subsection{Symmetries of a system and of its solutions}
\label{s:1dSubLattSymms}

What's the deal about classes?
A `class' is a refinement of our intuitive
notion that ``rotations are rotations, and translations are
translations.''
Translated into a more familiar language,
conjugation \refeq{conjugate} is central
to all of physics: a `law' $F(\Xx)$ is invariant if it
retains its form in all symmetry related coordinate frames,
\beq
F(\Xx)  =  \LieEl^{-1} F(\LieEl\,\Xx)
\,,
\label{dscr:L-inv}
\eeq
where $\LieEl$ is a representation  of group
element $\LieEl\in \Group$.
If this holds, we say that $\Group$ is the \emph{symmetry} of the system.

For example, the `temporal Bernoulli' {\ELe}
\refeq{tempBern}  retains its form under conjugation by any
$\Cn{\infty}$  translation \refeq{C_infty},
\beq
\shift_i({s} 1 - \shift_1)\shift_i^{-1} \Xx
= ({s} 1 - \shift_1) \Xx
\,,
\ee{invBern}
while the Euler–Lagrange second-order difference equations
\refeq{1dTempFT}, `{\templatt}', `{\henlatt}', and `temporal
{$\phi^4$} theory' {\ELe}s \refeq{1dTemplatt},
\refeq{1dHenlatt} and \refeq{1dPhi4} retain their form also
under any  $\Dn{\infty}$ reflection.

\begin{figure} \begin{center}
  \begin{minipage}[b]{0.33\textwidth}\begin{center}
         {(1)}~~\includegraphics[width=\textwidth]{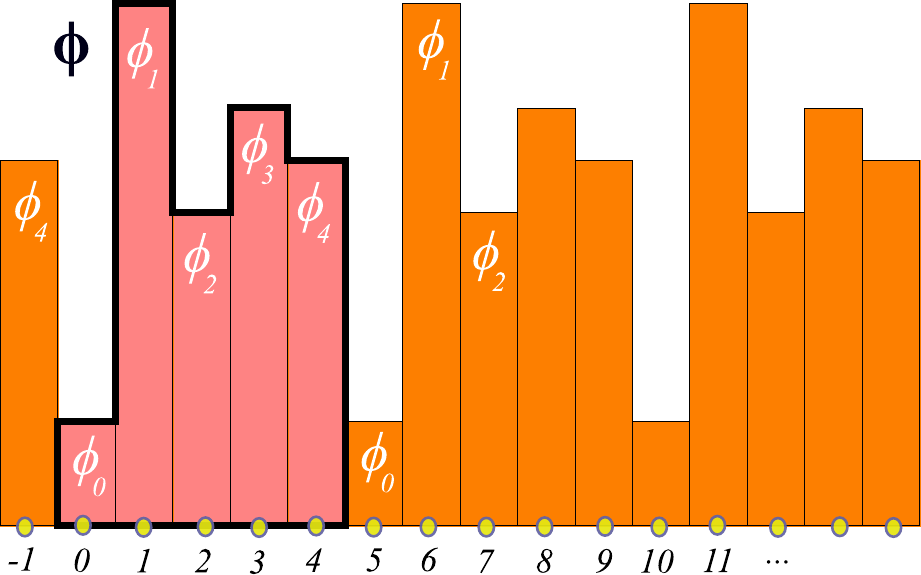}
\\
{($\shift_1$)}~\includegraphics[width=\textwidth]{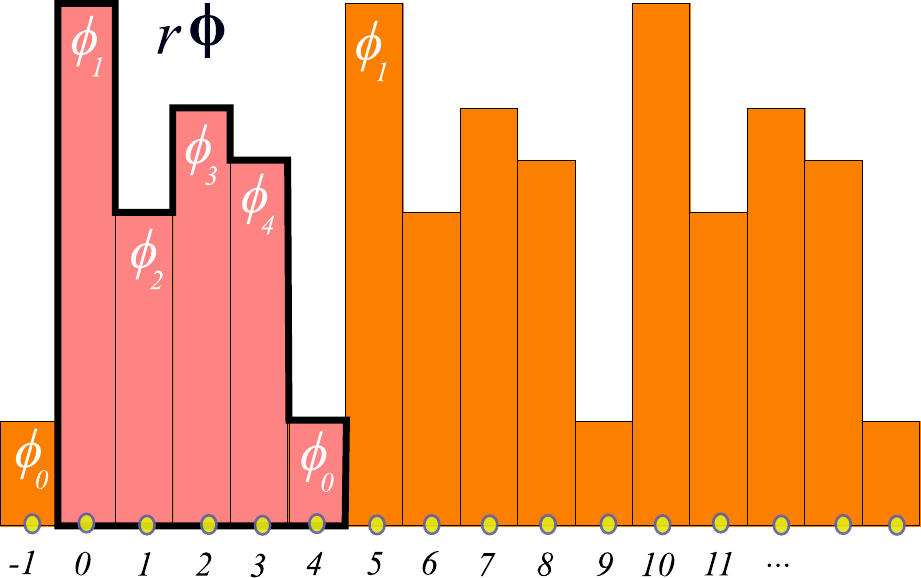}
\\
{($\shift_2$)}~\includegraphics[width=\textwidth]{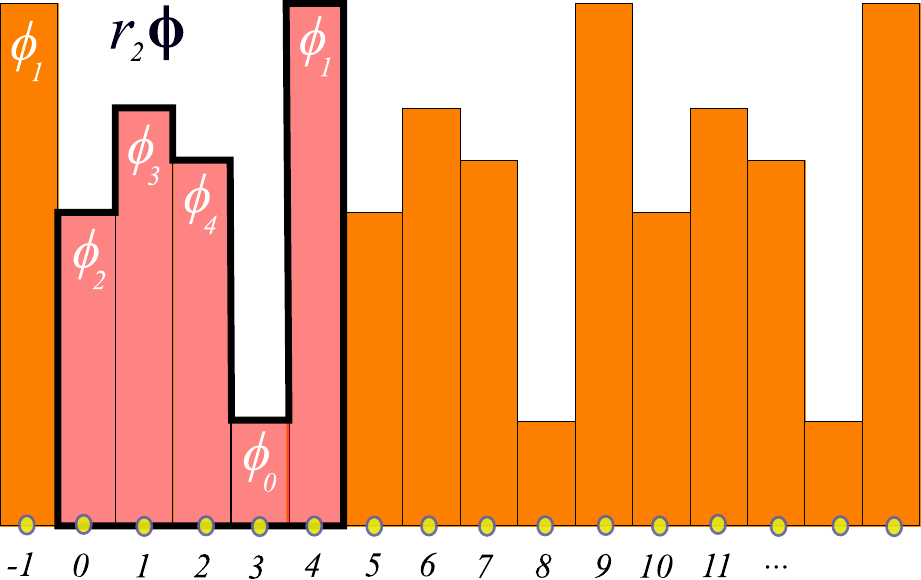}
\\
{($\shift_3$)}~\includegraphics[width=\textwidth]{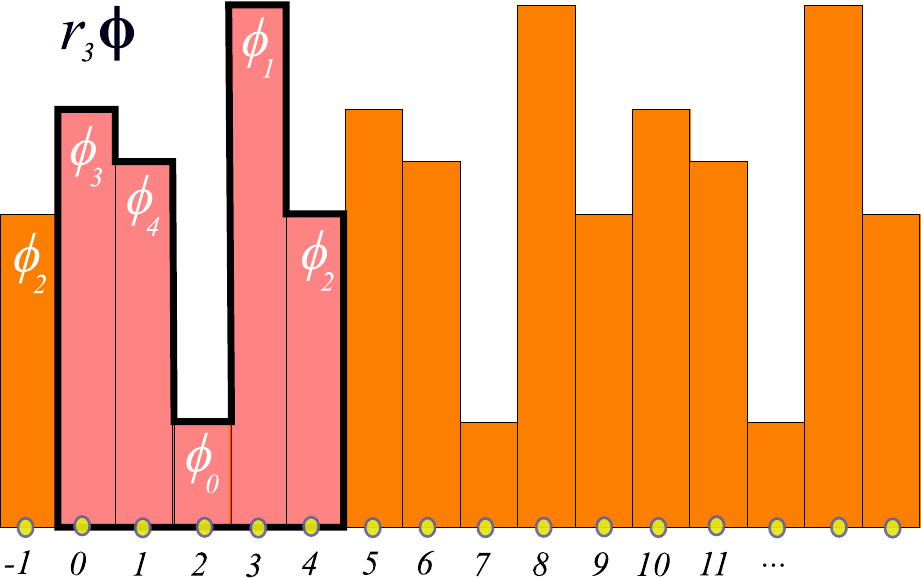}
\\
{($\shift_4$)}~\includegraphics[width=\textwidth]{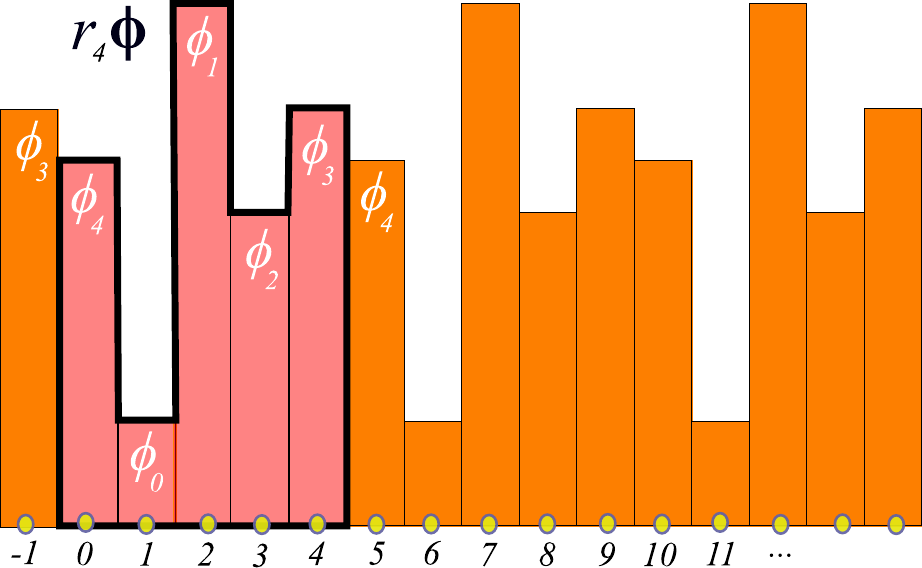}
  \end{center}\end{minipage}
\qquad\quad
  \begin{minipage}[b]{0.33\textwidth}\begin{center}
{($\Refl$)}~~\includegraphics[width=\textwidth]{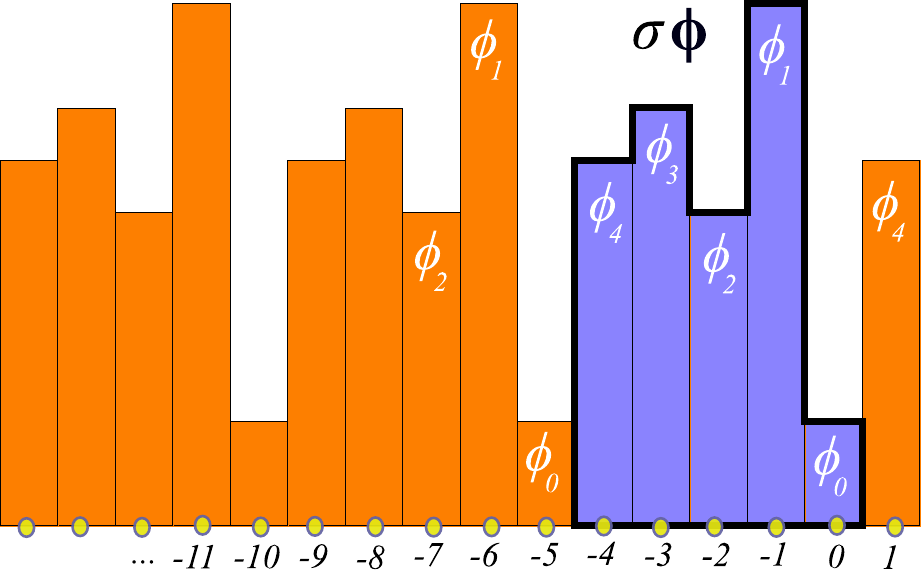}
\\
{($\Refl_1$)}~\includegraphics[width=\textwidth]{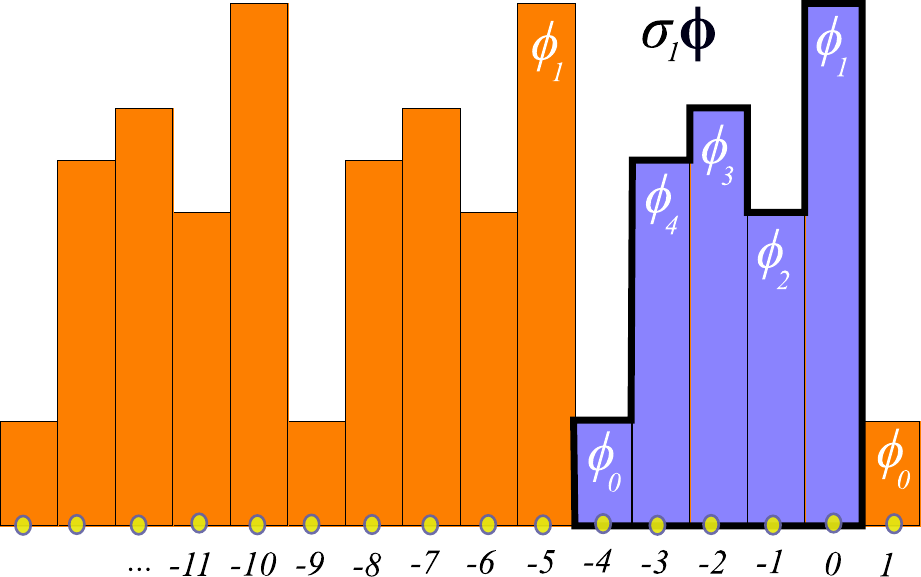}
\\
{($\Refl_2$)}~\includegraphics[width=\textwidth]{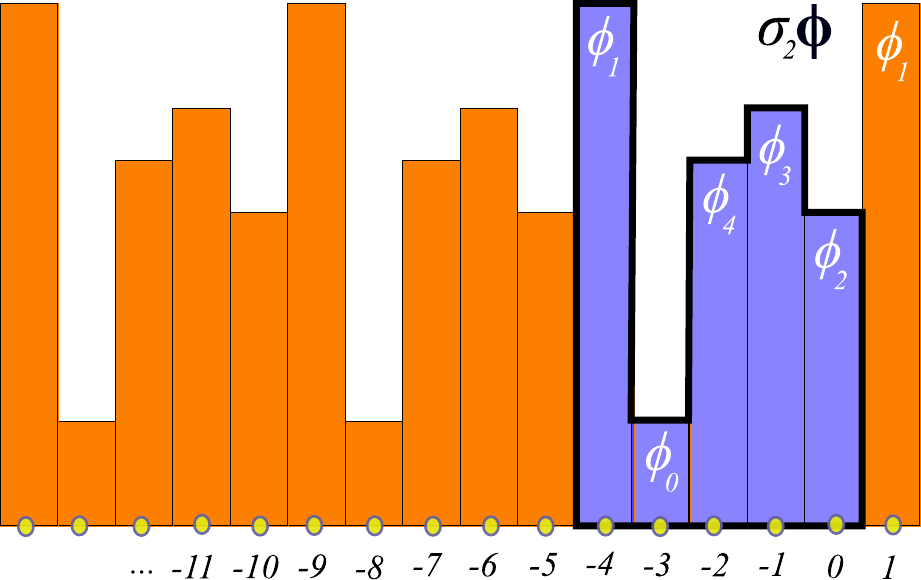}
\\
{($\Refl_3$)}~\includegraphics[width=\textwidth]{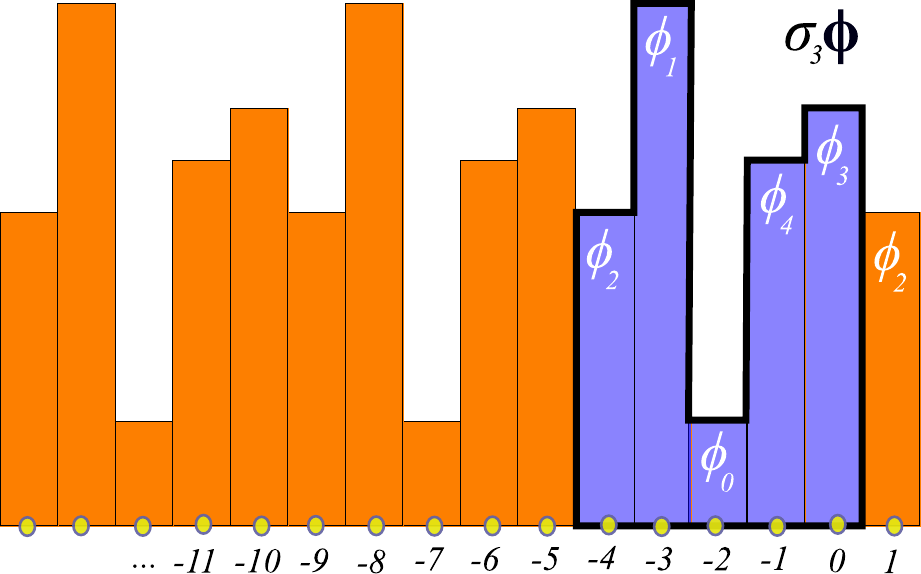}
\\
{($\Refl_4$)}~\includegraphics[width=\textwidth]{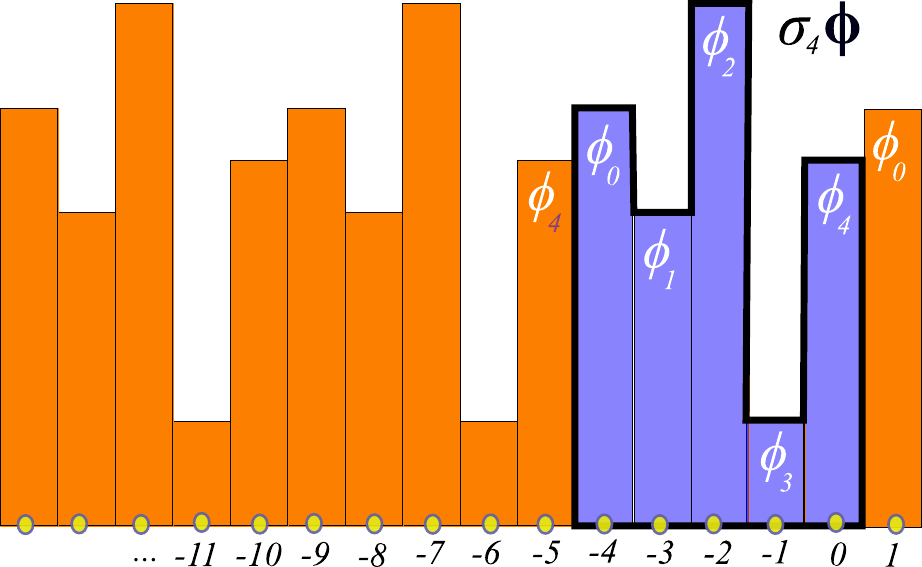}

  \end{center} \end{minipage}
  \end{center}
  \caption{\label{fig:1dLatStatC_5}
(Color online)~~~
(1)
A Bravais cell $\mathbf{a}$ with asymmetric
{\lst}
\(\Xx=\cycle{\ssp_0 \ssp_1 \ssp_2 \ssp_3 \ssp_4}\),
no reflection symmetry, outlined in bold, is invariant under the
translation subgroup $H_{5}$.
Its \Cn{\infty} orbit are the $\cl{}=5$ distinct {\lsts} (1) to
($\shift_4$), obtained by all of the $\Cn{5}$ translations.
Its \Dn{\infty}-orbit are $2\cl{}=10$ distinct {\lsts},
5 translations (1) to ($\shift_4$)
and
5 translate-reflections ($\Refl$) to ($\Refl_4$), obtained by
all of the $\Dn{5}$ actions.
See \reffig{FieldConfig}\,(b) for the notation.
Continued in \reffig{fig:1dLatStatD3D4}.
          }
\end{figure}

Given that $\Group$ is the {symmetry} of the system does not mean that
$\Group$ is also the symmetry of its solutions, or what we here call {\em
\lst}s.
They can satisfy
all of system's symmetries, a subgroup of them, or have no symmetry at
all.
For example, a generic {\lst} \refeq{1dLattStat} sketched in
\reffig{fig:1dLattRefl} has no symmetry beyond the identity, so its
symmetry group is the trivial subgroup $\{e\}$; any translation
$\shift_j$ or reflection $\Refl_k$ maps it into a different, distinct
{\lst}, as shown in \reffig{fig:1dLatStatC_5}.
    %
\toVideo{youtube.com/embed/IgYxosgHXWE}
At the other extreme, the constant {\lst}
$\ssp_j=\ssp$ is invariant under any translation or reflection - its
symmetry group is the full \Group, the symmetry of the system. In
between, there are {\lsts} whose symmetry is a subgroup of
$\Group$.

\subsection{What are `{\lsts}'? Orbits?}
\label{s:LattStates}

For evolution-in-time, every period-\cl{}\ periodic point is a
fixed point of the \cl{}th iterate of the 1 time-step map. In the lattice
formulation, the totality of finite-period {\lsts} is the \emph{set
of fixed points} of all  $H_{\mathbf{a}}$ and  $H_{\mathbf{a},k}$
subgroups of $\Dn{\infty}$.

You can visualize a {\lst} invariant under (`fixed by') subgroup
$H_{\mathbf{a},k}$ as a tiling of the lattice $\integers$ by a
{\lst} tile of length \cl{}, symmetric under reflection $\Refl_k$,
see \reffig{fig:symmLattStates}\,(b-c).

    \begin{quote}
Definition: {\em
{\em Orbit} or \emph{$\Group$-orbit} of a {\lst} $\Xx$ is the
set of all {\lsts}
\beq
    \pS_\Xx = \{\LieEl\,\Xx \mid \LieEl \in {\Group}\}
\ee{GroupOrbDisc}
into which $\Xx$ is mapped under the action of group $\Group$.
We label the orbit $\pS_\Xx$ by any {\lst} $\Xx$ belonging to
it.
            }
    \end{quote}
As an example, the $\Dn{\infty}$ orbit of the period-5 {\lst}
is shown in \reffig{fig:1dLatStatC_5}.

    \begin{quote}
Definition: Symmetry of a solution.
{\em
We shall refer to the maximal subgroup $\Group_\Xx \subseteq \Group$ of
actions which permute {\lsts} within the orbit $\pS_\Xx$, but leave the
orbit invariant, as the \emph{symmetry} $\Group_\Xx$ of the orbit
$\pS_\Xx$,
\beq
\Group_\Xx =
   \{ \LieEl \in \Group_\Xx \mid \LieEl \pS_\Xx = \pS_\Xx \}
\,.
\ee{stabilSet}
}
    \end{quote}
An orbit $\pS_\Xx$ is $\Group_\Xx$-{\em symmetric}
({\em symmetric}, {\em set-wise symmetric}, {\em self-dual})
if the action of elements of $\Group_\Xx$ on
the set of {\lsts} $\pS_\Xx$ reproduces the orbit.
    \begin{quote}
Definition: Index
{\em
of orbit $\pS_\Xx$ is given by
}
\beq
m_\Xx=|\Group|/|\Group_{\Xx}|
\,.
\ee{GroupOrbMult}
    \end{quote}
(See Wikipedia\rf{wikiIndex}
and Dummit and Foote\rf{DumFoot03}.)

\bigskip

And now, a pleasant surprise, obvious upon an inspection of
\reffigs{fig:1dLatStatC_5}{fig:symmLattStates}: what happens
in the Bravais cell, stays in the Bravais cell.
Even though
the lattices \lattice, $\lattice_{\mathbf{a}}$ are infinite,
and their symmetries
$\Dn{\infty}$, $H_{\mathbf{a}}$, $H_{\mathbf{a},k}$ are
\emph{infinite} groups, the Bravais {\lsts}'
\emph{orbits} are \emph{finite}, described by the finite group
permutations of the infinite lattice curled up into a Bravais cell periodic
$\cl{}$-site ring.


\begin{figure}
\begin{center}
  \begin{minipage}[b]{0.30\textwidth}\begin{center}
  \setlength{\unitlength}{1.00\textwidth}
  \begin{picture}(1,3.01432686)%
    \setlength\tabcolsep{0pt}%
    \put(0,0){\includegraphics[width=\unitlength,page=1]{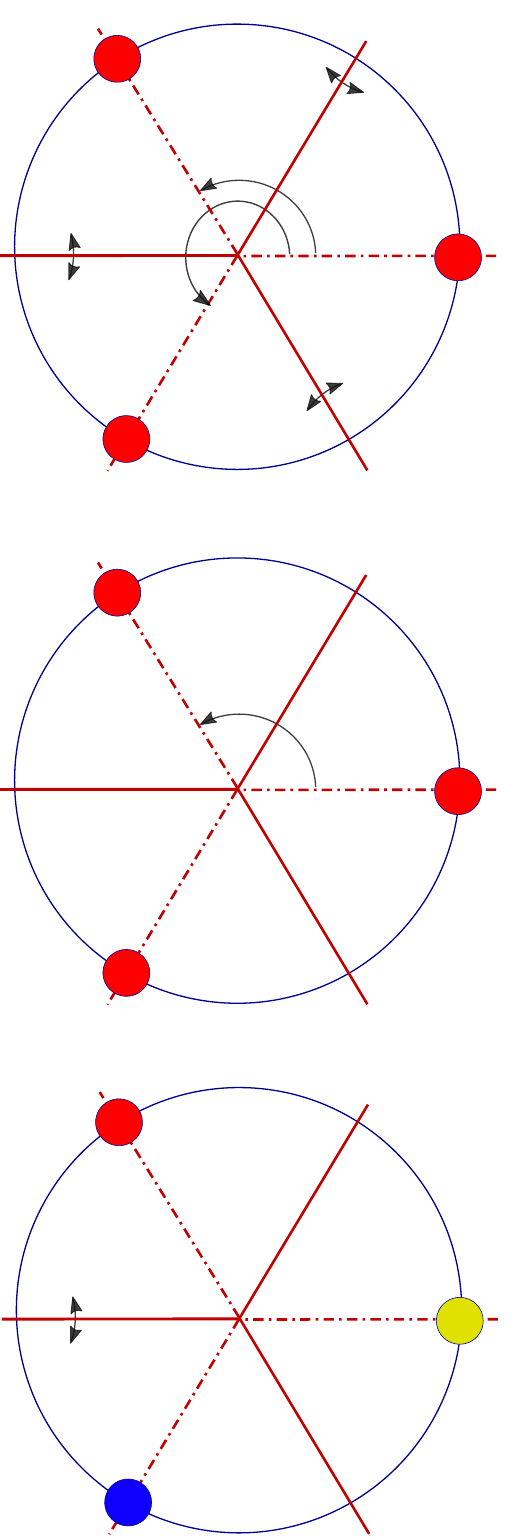}}%
    \put(0.31337215,2.59963103){\color[rgb]{0.1372549,0.12156863,0.1254902}\makebox(0,0)[lt]{\smash{$\shift$}}}%
    \put(0.41392193,2.35766425){\color[rgb]{0.1372549,0.12156863,0.1254902}\makebox(0,0)[lt]{\smash{$\shift_2$}}}%
    \put(0.116656,2.59005801){\color[rgb]{0.1372549,0.12156863,0.1254902}\makebox(0,0)[lt]{\smash{$\Refl$}}}%
    \put(0.69928915,2.77077246){\color[rgb]{0.1372549,0.12156863,0.1254902}\makebox(0,0)[lt]{\smash{$\Refl_1$}}}%
    \put(0.51597418,2.16979492){\color[rgb]{0.1372549,0.12156863,0.1254902}\makebox(0,0)[lt]{\smash{$\Refl_2$}}}%
    \put(0.94010872,2.57439466){\color[rgb]{0.1372549,0.12156863,0.1254902}\makebox(0,0)[lt]{\smash{$\ssp_0$}}}%
    \put(0.25021686,2.96750521){\color[rgb]{0.1372549,0.12156863,0.1254902}\makebox(0,0)[lt]{\smash{$\ssp_1$}}}%
    \put(0.32548233,2.18423413){\color[rgb]{0.1372549,0.12156863,0.1254902}\makebox(0,0)[lt]{\smash{$\ssp_2$}}}%
    \put(0.95966259,2.44353222){\color[rgb]{0.1372549,0.12156863,0.1254902}\makebox(0,0)[lt]{\smash{$0$}}}%
    \put(0.12026137,2.91808733){\color[rgb]{0.1372549,0.12156863,0.1254902}\makebox(0,0)[lt]{\smash{$1$}}}%
    \put(0.13556959,2.08378882){\color[rgb]{0.1372549,0.12156863,0.1254902}\makebox(0,0)[lt]{\smash{$2$}}}%
    \put(0.31337215,1.55356868){\color[rgb]{0.1372549,0.12156863,0.1254902}\makebox(0,0)[lt]{\smash{$\shift$}}}%
    \put(0.94010872,1.52833231){\color[rgb]{0.1372549,0.12156863,0.1254902}\makebox(0,0)[lt]{\smash{$\ssp_1$}}}%
    \put(0.25021686,1.92144286){\color[rgb]{0.1372549,0.12156863,0.1254902}\makebox(0,0)[lt]{\smash{$\ssp_2$}}}%
    \put(0.32548233,1.13817178){\color[rgb]{0.1372549,0.12156863,0.1254902}\makebox(0,0)[lt]{\smash{$\ssp_0$}}}%
    \put(0.95966259,1.39746987){\color[rgb]{0.1372549,0.12156863,0.1254902}\makebox(0,0)[lt]{\smash{$0$}}}%
    \put(0.12026137,1.87202498){\color[rgb]{0.1372549,0.12156863,0.1254902}\makebox(0,0)[lt]{\smash{$1$}}}%
    \put(0.13556959,1.03772648){\color[rgb]{0.1372549,0.12156863,0.1254902}\makebox(0,0)[lt]{\smash{$2$}}}%
    \put(0.12012811,0.50626931){\color[rgb]{0.1372549,0.12156863,0.1254902}\makebox(0,0)[lt]{\smash{$\Refl$}}}%
    \put(0.94358083,0.49060596){\color[rgb]{0.1372549,0.12156863,0.1254902}\makebox(0,0)[lt]{\smash{$\ssp_0$}}}%
    \put(0.25368897,0.8837165){\color[rgb]{0.1372549,0.12156863,0.1254902}\makebox(0,0)[lt]{\smash{$\ssp_2$}}}%
    \put(0.32895444,0.10044543){\color[rgb]{0.1372549,0.12156863,0.1254902}\makebox(0,0)[lt]{\smash{$\ssp_1$}}}%
    \put(0.9631347,0.35974352){\color[rgb]{0.1372549,0.12156863,0.1254902}\makebox(0,0)[lt]{\smash{$0$}}}%
    \put(0.12373348,0.83429863){\color[rgb]{0.1372549,0.12156863,0.1254902}\makebox(0,0)[lt]{\smash{$1$}}}%
    \put(0.1390417,0.00000012){\color[rgb]{0.1372549,0.12156863,0.1254902}\makebox(0,0)[lt]{\smash{$2$}}}%
  \end{picture}  \\ $(\Dn{3})$
  \end{center}\end{minipage}
\qquad\qquad
  \begin{minipage}[b]{0.305\textwidth}\begin{center}
  \setlength{\unitlength}{1.00\textwidth}
  \begin{picture}(1,3.03643837)%
    \setlength\tabcolsep{0pt}%
    \put(0.54129223,2.67451538){\color[rgb]{0.1372549,0.12156863,0.1254902}\makebox(0,0)[lt]{\smash{$\shift$}}}%
    \put(0,0){\includegraphics[width=\unitlength,page=1]{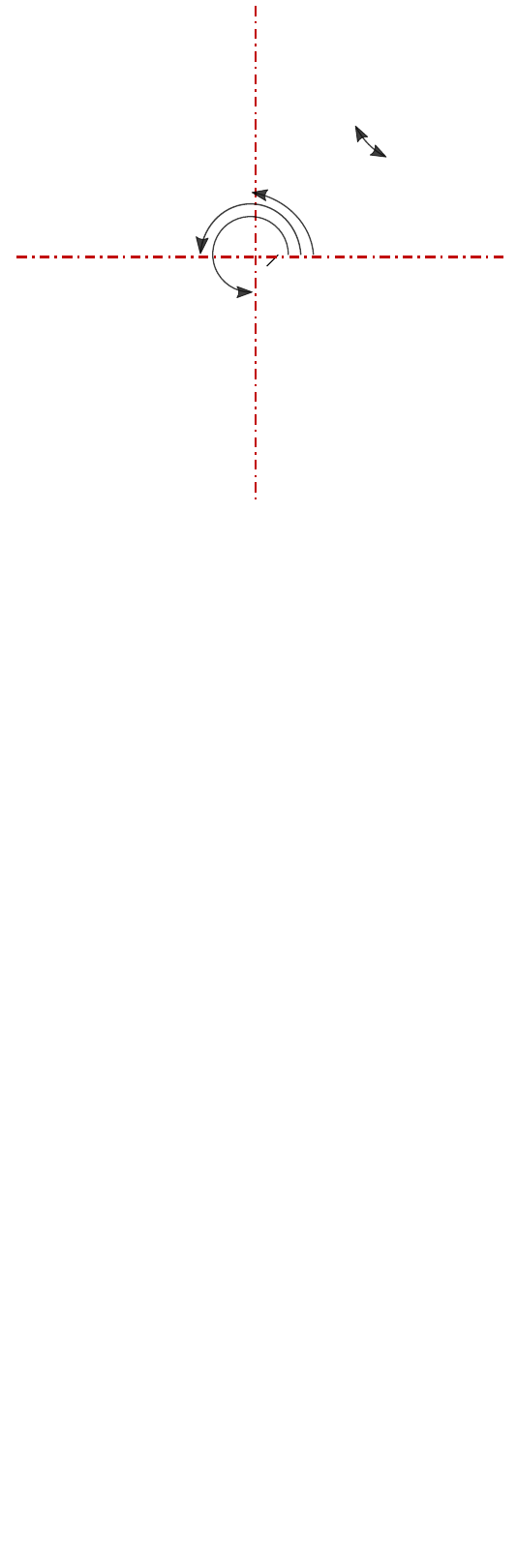}}%
    \put(0.31525021,2.59209199){\color[rgb]{0.1372549,0.12156863,0.1254902}\makebox(0,0)[lt]{\smash{$\shift_2$}}}%
    \put(0.40560025,2.39674053){\color[rgb]{0.1372549,0.12156863,0.1254902}\makebox(0,0)[lt]{\smash{$\shift_3$}}}%
    \put(0.82752268,2.45361935){\color[rgb]{0.1372549,0.12156863,0.1254902}\makebox(0,0)[lt]{\smash{$\Refl$}}}%
    \put(0.76172139,2.67388076){\color[rgb]{0.1372549,0.12156863,0.1254902}\makebox(0,0)[lt]{\smash{$\Refl_1$}}}%
    \put(0.5582712,2.85268367){\color[rgb]{0.1372549,0.12156863,0.1254902}\makebox(0,0)[lt]{\smash{$\Refl_2$}}}%
    \put(0.29865836,2.83491117){\color[rgb]{0.1372549,0.12156863,0.1254902}\makebox(0,0)[lt]{\smash{$\Refl_3$}}}%
    \put(0,0){\includegraphics[width=\unitlength,page=2]{1dLatStatD4.pdf}}%
    \put(0.93632269,2.44407772){\color[rgb]{0.1372549,0.12156863,0.1254902}\makebox(0,0)[lt]{\smash{$0$}}}%
    \put(0.39331567,2.97321897){\color[rgb]{0.1372549,0.12156863,0.1254902}\makebox(0,0)[lt]{\smash{$1$}}}%
    \put(0.38295558,2.0753459){\color[rgb]{0.1372549,0.12156863,0.1254902}\makebox(0,0)[lt]{\smash{$3$}}}%
    \put(-0.00069027,2.42499839){\color[rgb]{0.1372549,0.12156863,0.1254902}\makebox(0,0)[lt]{\smash{$2$}}}%
    \put(0.94600174,2.58758294){\color[rgb]{0.1372549,0.12156863,0.1254902}\makebox(0,0)[lt]{\smash{$\ssp_0$}}}%
    \put(0.12237582,2.59448962){\color[rgb]{0.1372549,0.12156863,0.1254902}\makebox(0,0)[lt]{\smash{$\ssp_2$}}}%
    \put(0.5410955,2.1913101){\color[rgb]{0.1372549,0.12156863,0.1254902}\makebox(0,0)[lt]{\smash{$\ssp_{3}$}}}%
    \put(0.54972883,2.99162582){\color[rgb]{0.1372549,0.12156863,0.1254902}\makebox(0,0)[lt]{\smash{$\ssp_1$}}}%
    \put(0,0){\includegraphics[width=\unitlength,page=3]{1dLatStatD4.pdf}}%
    \put(0.82383473,1.43453815){\color[rgb]{0.1372549,0.12156863,0.1254902}\makebox(0,0)[lt]{\smash{$\Refl$}}}%
    \put(0,0){\includegraphics[width=\unitlength,page=4]{1dLatStatD4.pdf}}%
    \put(0.93263474,1.42499652){\color[rgb]{0.1372549,0.12156863,0.1254902}\makebox(0,0)[lt]{\smash{$0$}}}%
    \put(0.38962772,1.95413783){\color[rgb]{0.1372549,0.12156863,0.1254902}\makebox(0,0)[lt]{\smash{$1$}}}%
    \put(0.37926763,1.05626474){\color[rgb]{0.1372549,0.12156863,0.1254902}\makebox(0,0)[lt]{\smash{$3$}}}%
    \put(-0.00437824,1.40591715){\color[rgb]{0.1372549,0.12156863,0.1254902}\makebox(0,0)[lt]{\smash{$2$}}}%
    \put(0.94231378,1.5685017){\color[rgb]{0.1372549,0.12156863,0.1254902}\makebox(0,0)[lt]{\smash{$\ssp_0$}}}%
    \put(0.11868786,1.57540842){\color[rgb]{0.1372549,0.12156863,0.1254902}\makebox(0,0)[lt]{\smash{$\ssp_2$}}}%
    \put(0.53740751,1.17222886){\color[rgb]{0.1372549,0.12156863,0.1254902}\makebox(0,0)[lt]{\smash{$\ssp_1$}}}%
    \put(0.54604091,1.97254465){\color[rgb]{0.1372549,0.12156863,0.1254902}\makebox(0,0)[lt]{\smash{$\ssp_{3}$}}}%
    \put(0,0){\includegraphics[width=\unitlength,page=5]{1dLatStatD4.pdf}}%
    \put(0.75803344,0.60478566){\color[rgb]{0.1372549,0.12156863,0.1254902}\makebox(0,0)[lt]{\smash{$\Refl_1$}}}%
    \put(0,0){\includegraphics[width=\unitlength,page=6]{1dLatStatD4.pdf}}%
    \put(0.93263474,0.37498258){\color[rgb]{0.1372549,0.12156863,0.1254902}\makebox(0,0)[lt]{\smash{$0$}}}%
    \put(0.38962772,0.90412381){\color[rgb]{0.1372549,0.12156863,0.1254902}\makebox(0,0)[lt]{\smash{$1$}}}%
    \put(0.37926763,0.00625079){\color[rgb]{0.1372549,0.12156863,0.1254902}\makebox(0,0)[lt]{\smash{$3$}}}%
    \put(-0.00437822,0.35590321){\color[rgb]{0.1372549,0.12156863,0.1254902}\makebox(0,0)[lt]{\smash{$2$}}}%
    \put(0.94231378,0.5184879){\color[rgb]{0.1372549,0.12156863,0.1254902}\makebox(0,0)[lt]{\smash{$\ssp_1$}}}%
    \put(0.11868787,0.52539448){\color[rgb]{0.1372549,0.12156863,0.1254902}\makebox(0,0)[lt]{\smash{$\ssp_{3}$}}}%
    \put(0.53740751,0.12221507){\color[rgb]{0.1372549,0.12156863,0.1254902}\makebox(0,0)[lt]{\smash{$\ssp_2$}}}%
    \put(0.54604091,0.92253071){\color[rgb]{0.1372549,0.12156863,0.1254902}\makebox(0,0)[lt]{\smash{$\ssp_0$}}}%
  \end{picture}  \\$(\Dn{4})$
  \end{center} \end{minipage}
  \end{center}
  \caption{\label{fig:1dLatStatD3D4}
(Color online)~~~
Consider a period-$\cl{}$ Bravais cell tiling of a 1\dmn\ lattice
{\lattice}. With {\lattice} curled into a ring of $\cl{}$ lattice sites,
actions of the {infinite dihedral group} $\Dn{\infty}$ reduce to
translational and reflection symmetries of
    $(\Dn{3})$
an equilateral triangle,  $\cl{}=3$ lattice sites;
    $(\Dn{4})$
a square,                  $\cl{}=4$ lattice sites;
all group operations that
overlie an $\cl{}$-sided regular polygon onto itself.
The $\cl{}$ translations $\shift_j$ permute the sites cyclically.
The $\cl{}$ dihedral group \Dn{\cl{}} translate-reflect $\Refl_k$
elements \refeq{DnElements} reflect the sites across reflection axes,
exchanging red and blue sites.
For even $\cl{}$, an even reflection (dashed line reflection axis), here
$\Refl$, leaves
a pair of opposite sites fixed (marked yellow),
while an odd reflection axis (full line), here
$\Refl_1$, bisects the opposite edges, and
flips all sites.
For odd $\cl{}$, every reflection half-axis leaves a site fixed (dashed
line), and bisects the opposite edge (full line).
This periodic ring visualization makes it obvious that any symmetric
{\lst} is reflection invariant across two points on the lattice,
see \reffig{fig:symmLattStates}.
}
\end{figure}

Indeed, to grasp everything one needs to know about translations
$\shift_j$ (for regular polygons, `rotations'),
and
reflections $\Refl_k$,
it suffices to understand the symmetries of
an equilateral triangle (dihedral group \Dn{3})
and
a square (dihedral group \Dn{4}), depicted in \reffig{fig:1dLatStatD3D4}.
It is clear by inspection that an $\cl{}$-sided regular polygon has
$\cl{}$-fold translational symmetry and $\cl{}$ reflection symmetry axes.
The group of such symmetries is the finite dihedral group
\beq
\Dn{\cl{}} = \{1,\Refl,\shift,\Refl_{1},\shift_{2},\Refl_{2},\
           \cdots,
           \shift_{\cl{}-1},\Refl_{\cl{}-1}\}
\ee{DnElements}
of order $2\cl{}$.
A half of its elements are the $\cl{}$ cyclic group \Cn{n} translations
$\shift_{j}$.
The other half are the $\cl{}$ reflections $\Refl_k$, one for the
reflection across each symmetry axis.
The group multiplication table is the same as the $\Dn{\infty}$
\refeq{eq:DinftyMultTab}, but with all subscripts mod $\cl{}$.
As in \refeq{DinftyInversion},
conjugation by any reflection reverses the direction of translation
\beq
   \Refl_i\shift_j\Refl_{-i} =  \shift_{\cl{}-j}
\,,\qquad 0<j<\cl{}
\,,
\ee{D_nInversion} 
so every translation pairs up with the equal counter-translation to form
a 2-element class \refeq{DinftyClassShift}. 

The distinction between the classes of even and odd reflections
\refeq{DinftyClassRefl} is visually self-evident by inspection of
\reffig{fig:1dLatStatD3D4}:
the symmetry axes either connect opposite lattice sites, or bisect the
edges, or both, if $\cl{}$ is odd (a triangle, for example).
One can say that $k$ in the
`translate-reflection' \refeq{Refl_k} operation
\(
\Refl_{k} 
\) advances the reflection point by $k$ 1/2 steps, and then reflects
across it.

For a polygon with an \emph{odd} number of
lattice sites (a triangle, for example), we see by contemplating the
triangle of \reffig{fig:1dLatStatD3D4},  as well as by taking  mod $\cl{}$ of the
conjugation relation \refeq{D_nConj}, that
all reflections are in the same conjugacy class $\{\Refl_{j}\}$:
 there is no splitting into odd and even cases, in
contrast to the infinite lattice case \refeq{DinftyClassRefl}.

For a polygon with an \emph{even} number of lattice sites (a square, for
example), one must distinguish
the `long' axes that connect lattice sites (we label them by even numbers
$0,2,\cdots$)
from
the `short' symmetry axes that bisect opposite edges (labelled by odd
numbers $1,3,\cdots$).
The corresponding reflections belong to
different \Dn{\cl{}} (subclasses of
\refeq{DinftyClassRefl}),
\bea
\mbox{even reflections}
    &&\quad
\{\Refl,\Refl_{2},\Refl_{4},\cdots,\Refl_{\cl{}/2}\}
    \continue
\mbox{odd reflections}
    &&\quad
\{\Refl_{1},\Refl_{3},\cdots,\Refl_{\cl{}/2+1}\}
\,.
\label{DnClassRefl}
\eea

\subsection{Symmetries of {\lsts}}
\label{s:LattStateSyms}

\begin{figure}
  \centering
{$(a)$}\;
\includegraphics[width=0.40\textwidth]{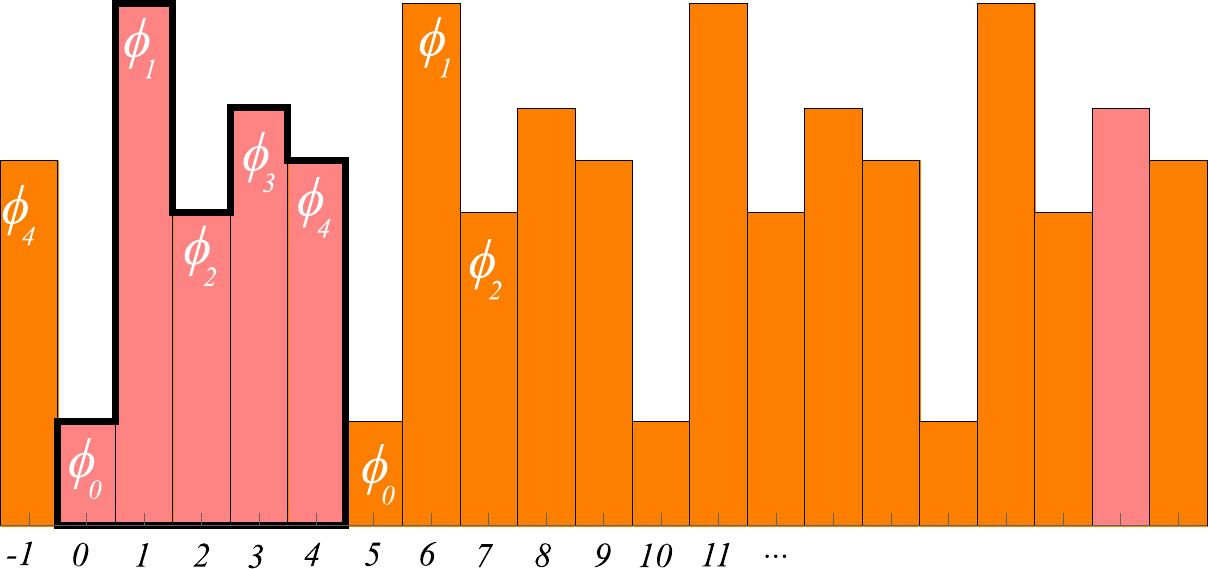}\quad~
{$(o)$}
\includegraphics[width=0.40\textwidth]{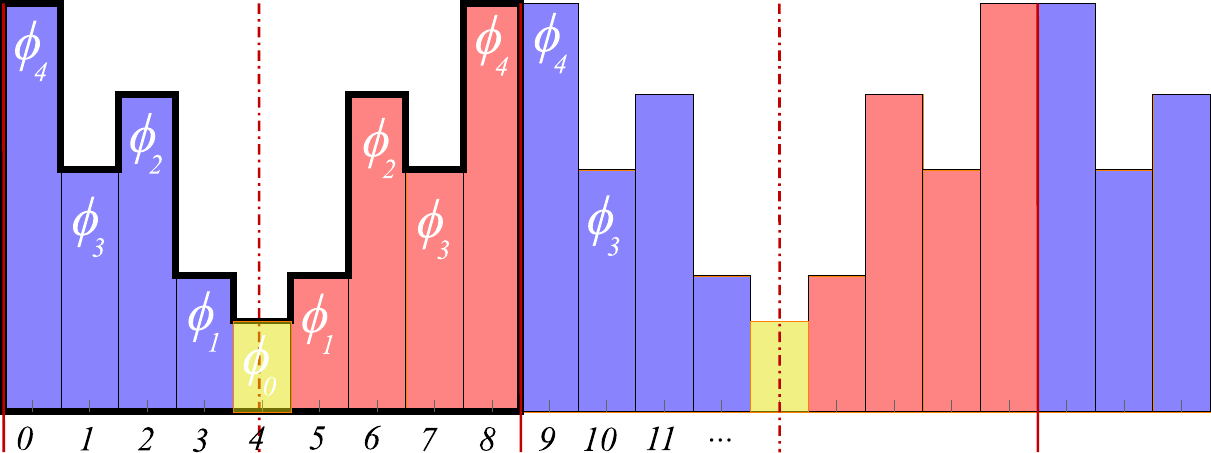} 
\\ 
{$(ee)$}
\includegraphics[width=0.40\textwidth]{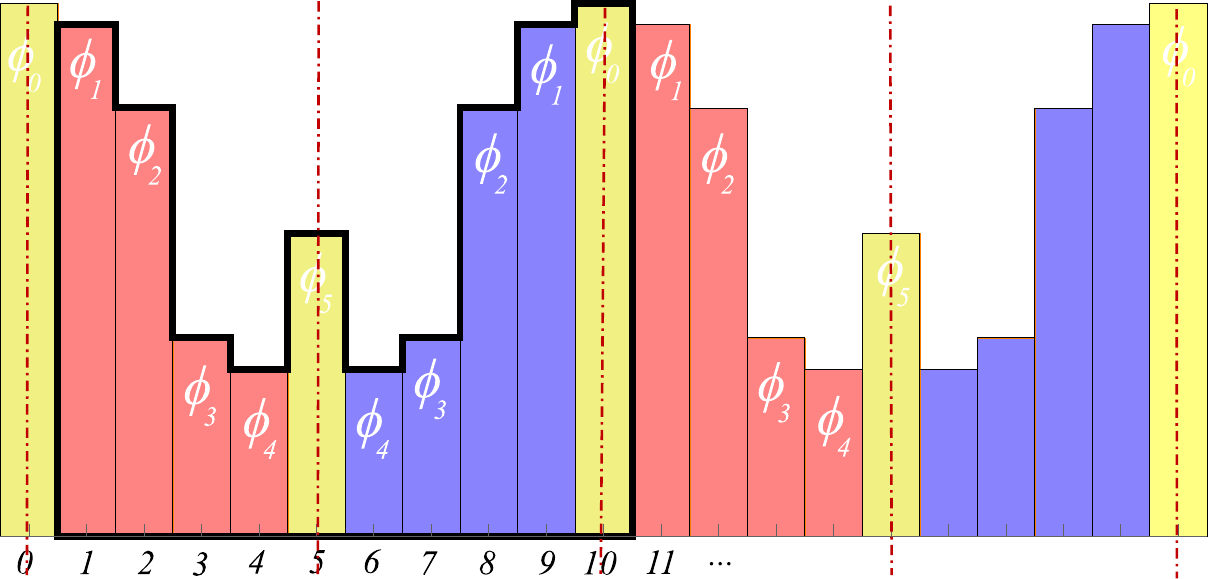}\quad
{$(eo)$}
\includegraphics[width=0.40\textwidth]{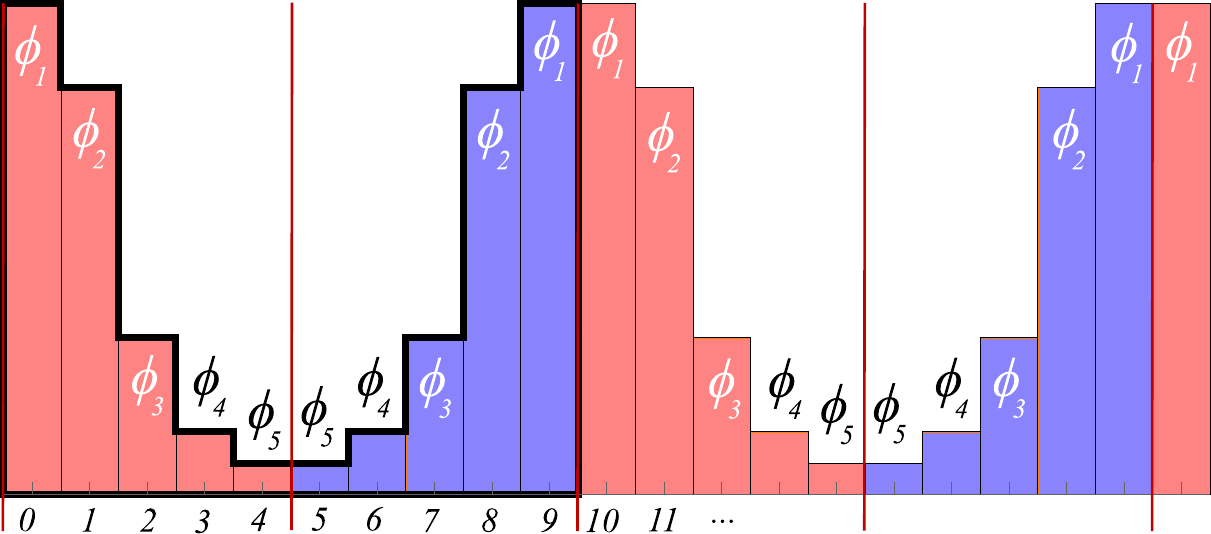} 

  \caption{\label{fig:symmLattStates}
(Color online)~~~
A Bravais {\lst} \Xx\ has one of the
4 possible symmetries, illustrated by:
$(a)$ {\em No reflection symmetry:}
    an $H_{5}$ invariant period-5 {\lst} \refeq{reflSymNo}. For its
    \Group-orbit, see \reffig{fig:1dLatStatC_5}.
$(o)$ {\em Odd period, reflection-symmetric:}
    an $H_{9,8}$ invariant period-9 {\lst} \refeq{reflSymOdd},
    reflection symmetric over the lattice sites interval [8-9]  midpoint
    and over the lattice site~4.
$(ee)$ {\em Even period, even reflection-symmetric:}
    an $H_{10,0}$  invariant period-10 {\lst} \refeq{reflSymEvens0},
    reflection symmetric over lattice sites 0 and 5.
$(eo)$ {\em Even period, odd reflection-symmetric:}
    an $H_{10,9}$  invariant period-10 {\lst} \refeq{reflSymEvens1},
    reflection symmetric over the [4-5] and [9-10] interval midpoints.
Horizontal: lattice sites labelled by $\zeit\in\integers$.
Vertical: value of field $\ssp_\zeit$, plotted as a bar centred at
lattice site $\zeit$. Time reversed \brick s indicated in blue, boundary
sites in yellow.
Even reflection axes dashed, odd reflections full line.
          }
\end{figure}

A Bravais {\lst} \Xx\ has one of the four symmetries:
    %
\toVideo{youtube.com/embed/mvOLj5s69RY}

\bea
    && \mbox{ asymmetric, no reflection symmetry }
    \continue
(a) \quad &&
\cycle{\ssp_0 \ssp_1 \ssp_2 \ssp_3 \cdots \ssp_{\cl{}-1}}
\label{reflSymNo} \\ 
    &&
\mbox{ index } m_\Xx=2\cl{}
    \nnu 
\eea
{\lst} invariant under the translation group
$H_{\cl{}}$.
Its \Group-orbit, generated by all actions of \Dn{\infty},
results  in  $2\cl{}$ distinct,  \Dn{\cl{}} related {\lsts}.
This is illustrated by the $H_{5}$-invariant {\lst} \Xx\ of
\reffig{fig:1dLatStatC_5}. 
Its \Dn{5} orbit are $2\cl{}=10$ {\lsts}, 5  translations
and 5 translate-reflections.

Next, the reflection-symmetric {\lsts}.
As illustrated in \reffigs{fig:1dLattRefl}{fig:1dLatStatD3D4}, there are
two classes \refeq{DinftyClassRefl} of {\lst} reflections: even, across a
lattice site, and odd, across the mid-point between a pair of adjacent
lattice sites.
However, as is evident by inspection of \reffig{fig:1dLatStatD3D4},
curling up the lattice {\lattice} into a Bravais cell periodic
$\cl{}$-site ring implies that an axis cuts the ring twice, and
constrains the possible reflection points to three configurations:

\bea
    && \mbox{ odd period } \cl{}=2m+1
    \continue
(o) \quad &&
\cycle{\sitebox{\ssp_0} \ssp_1 \ssp_2 \cdots \ssp_{m}|\ssp_{m}\cdots \ssp_2 \ssp_1}
\label{reflSymOdd} \\
    &&
\mbox{ index } m_\Xx=\cl{}
    \nnu 
\eea
{\lst} invariant under the dihedral group $H_{\cl{},k}$,
illustrated by the $H_{9,8}$ invariant {\lst} \Xx\
of \reffig{fig:symmLattStates}\,$(o)$.
\bea
    && \mbox{ even period } \cl{}=2m+2\,, \mbox{ even reflection } k
    \continue
(ee) \quad &&
\cycle{\sitebox{\ssp_0} \ssp_1 \ssp_2 \cdots \ssp_{m}
        \sitebox{\ssp_{m+1}} \ssp_{m} \cdots \ssp_2 \ssp_1}
\label{reflSymEvens0} \\
    &&
\mbox{ index } m_\Xx=\cl{}
    \nnu 
\eea
{\lst} invariant under the dihedral group $H_{\cl{},k}$,
$k$ even,
illustrated by the $H_{10,0}$ invariant {\lst} \Xx\
of \reffig{fig:symmLattStates}\,$(ee)$.
\bea
    && \mbox{ even period } \cl{}=2m\,, \mbox{ odd reflection } k
    \continue
(eo) \quad &&
\cycle{\ssp_1 \ssp_2 \ssp_3 \cdots \ssp_{m}| \ssp_{m}\cdots \ssp_2 \ssp_1|}
\label{reflSymEvens1} \\
    &&
\mbox{ index } m_\Xx=\cl{}
    \nnu 
\eea
{\lst} invariant under the dihedral group $H_{\cl{},k}$,
$k$ odd,
illustrated by the $H_{10,9}$ invariant {\lst} \Xx\
of \reffig{fig:symmLattStates}\,$(eo)$.

The {\lst} {symmetry} $\Group_\Xx$ \refeq{stabilSet} of the
above $(o)$--$(eo)$ reflection-symmetric {\lsts} \Xx\ is the reflection
group $\Dn{1}=\{1,\Refl_k\}$. This symmetry means
two things:
\begin{itemize}
  \item[(1)]
The \Dn{\infty} orbits of reflection-symmetric {\lsts} contain only
translations, as any reflection amounts to a cyclic group
\Cn{\cl{}} translation.
(Reflect a {\lst} in \reffig{fig:symmLattStates}\,(b-d)
over any lattice site or mid-interval: the result is its translation.)
  \item[(2)]
The prime {\lst} is a `half' of the Bravais cell,
the length ${m}$ orbit,
\beq
\tilde{\Xx}      = (\ssp_1 \ssp_2 \ssp_3 \cdots \ssp_{m})
\,,
\ee{primeLattStat}
give or take some boundary sites.
\end{itemize}
To develop intuition about how one reconstructs the period-$\cl{}$ orbit from
this length-$m$ \brick\ it is helpful to have a look at explicit matrix
representation of the dihedral group $\Dn{\cl{}}$ actions.

\subsection{Permutation representation}
\label{sect:permReps}

A {\lst} {\Xx} over a Bravais cell $\mathbf{a}$ can be assembled
into an $\cl{}$\dmn\ vector whose components are lattice site fields
\beq
\transp{\Xx} = (\ssp_0,\ssp_1,\ssp_2,\ssp_3,\cdots,\ssp_{\cl{}-1})
\,.
\ee{1dLattStatVec}
Matrices that reshuffle the
components of such vectors form the {\em permutation representation} of a
finite group \Group. They give us a different
perspective on the above three kinds of symmetric solutions.

The permutation representation of 1-step lattice translation $\shift$
acts on a Bravais {\lst} by the off-diagonal
$[\cl{}\!\times\!\cl{}]$ matrix \refeq{hopMatrix}. This is a cyclic
\Cn{\cl{}} permutation that translates the {\lst} $\Xx$
"for\-ward-in-time" by one site,
\[
\transp{(\shift\Xx)}=(\ssp_1,\ssp_2,\cdots,\ssp_{\cl{}-1},\ssp_0)
\,.
\]
A permutation representation of a \Dn{\cl{}} translate-reflect operation
is essentially an anti-diagonal matrix that reverses the order of site
fields, up to a cyclic permutation
\[
\transp{(\Refl_k\Xx)}=(\ssp_{\cl{}-1},\cdots,\ssp_2,\ssp_1,\ssp_0)
\,.
\]

\emph{Even periods}:
The shortest even period symmetric  {\lst} is the period-2 {\lst}
\(
\transp{\Xx_p}=(\ssp_0,\ssp_{1})
\)
such as the \templatt\ \refeq{catFundPar2}.
For example, for \henlatt\ \refeq{Hen3term} there is only one period-2
prime orbit, consisting of {\lst}
\beq
\Xx_p
= \frac{1}{a}
\left(\begin{array}{c}
 -1-\sqrt{a-3} \cr
 -1+\sqrt{a-3}
\end{array}\right)
\,,
\ee{henLst2}
and its translation $\shift\Xx_p$. Its symmetry,
\(
\transp{\Xx_p}=(\sitebox{\ssp_0}\,\sitebox{\ssp_1})
\)
is of $(ee)$ type \refeq{reflSymEvens0}, indicated as yellow lattice
sites fields in \reffig{fig:symmLattStates}\,$(ee)$.
Its {\jacobianOrb} is of the nonlinear field theory form \refeq{jMorb1dFT}
\beq
\jMorb =
 \left(\begin{array}{cc}
  s_0 & -2 \\
 -2 &  s_1
 \end{array} \right)
\,.
\ee{henFundPar2}
This {\lst} is `all boundary
points', too short to illustrate a symmetry reduction to a $D_{\cl{}}$
prime orbit. Still, as we show in \reffig{fig:recipCatCn}\,(b), its relation
to the block circulant structure of repeated-tile {\jacobianOrb}
\refeq{orbJprimeRpt} and its contribution to $\phi^3$ field theory
spectrum is instructive.

\emph{Odd periods}:
In odd dimensions, the $\cl{}$ translate-reflect matrices of \Dn{\cl{}} are
related by translations \refeq{D_nConj}.
For example, for a period-3 {\lst}
without symmetry
\(
\transp{\Xx} = (\ssp_0,\ssp_1,\ssp_2)
\,,
\)
they are
\[
\Refl=
\left(
\begin{array}{ccc}
 1 & 0 & 0 \\
 0 & 0 & 1 \\
 0 & 1 & 0
\end{array}
\right)
    \,,\quad
{\Refl_1}=
\left(
\begin{array}{ccc}
 0 & 1 & 0 \\
 1 & 0 & 0 \\
 0 & 0 & 1
\end{array}
\right)
    \,,\quad
\Refl_2=
\left(
\begin{array}{ccc}
 0 & 0 & 1 \\
 0 & 1 & 0 \\
 1 & 0 & 0
\end{array}
\right)
\,.
\]
In agreement with \refeq{reflSymOdd}, \reffig{fig:1dLatStatD3D4} and
\reffig{fig:symmLattStates}\,$(o)$, these reflections keep one lattice site fixed
(for each permutation matrix $\Refl_k$ there is only one `1' on the
diagonal), swap the rest.

To get some insight into the length-${m}$ prime {\lst}
$\tilde{\Xx}$ \refeq{primeLattStat},
consider next a period-5 reflection symmetric {\lsts} that tile the infinite lattice
\lattice\ with a reflection-fixed $\sitebox{\ssp_0}$, and a length-2
{\brick} $\tilde{\Xx}=(\ssp_1,\ssp_2)$,
\beq
\transp{\Xx} =
    (\sitebox{\ssp_0} \ssp_1 \ssp_2 | \ssp_2 \ssp_1)
\,.
\ee{symmCycD5}
Actions of \Dn{5} permutation representation illustrates that
the fixed {\lsts} {\Xx} of $\Refl_{k}$ are related by cyclic translations:
\bea
\Refl\Xx &=&
\left(
\begin{array}{ccccc}
 1 & 0 & 0 & 0 & 0 \\
 0 & 0 & 0 & 0 & 1 \\
 0 & 0 & 0 & 1 & 0 \\
 0 & 0 & 1 & 0 & 0 \\
 0 & 1 & 0 & 0 & 0
\end{array}
\right)
\left(\begin{array}{c}
 \sitebox{\ssp_0}\cr
 \ssp_1\cr
 \ssp_2\cr
 \ssp_2\cr
 \ssp_1\cr
\end{array}\right)
=
\left(\begin{array}{c}
 \sitebox{\ssp_0}\cr
 \ssp_1\cr
 \ssp_2\cr
 \ssp_2\cr
 \ssp_1\cr
\end{array}\right)
        \continue
\Refl_{4}(\shift_{-2}\Xx )
     &=&
\left(
\begin{array}{ccccc}
 0 & 0 & 0 & 0 & 1 \\
 0 & 0 & 0 & 1 & 0 \\
 0 & 0 & 1 & 0 & 0 \\
 0 & 1 & 0 & 0 & 0\\
 1 & 0 & 0 & 0 & 0
\end{array}
\right)
\left(\begin{array}{c}
 \ssp_2\cr
 \ssp_1\cr
 \sitebox{\ssp_0}\cr
 \ssp_1\cr
 \ssp_2\cr
\end{array}\right)
=
\left(\begin{array}{c}
 \ssp_2\cr
 \ssp_1\cr
 \sitebox{\ssp_0}\cr
 \ssp_1\cr
 \ssp_2\cr
\end{array}\right)
\,.
\label{symmCycD5Refl}
\eea

What is the orbit stability of such \lst?
The symmetry conditions are the Bravais \lst\ 5-periodicity
mod 5, and the even reflection across
$\sitebox{\ssp_0}$:
\beq
\ssp_{i} = \ssp_{i+5}
    \,, \quad
\ssp_{-i} = \ssp_{i}
\,.
\ee{symmCycD5bcs}
A {\lst} satisfies the {\ELe} \refeq{1dTempFT} 
\beq
- \ssp_{\zeit-1} + 2\ssp_{\zeit} - \ssp_{\zeit+1}
+ V'(\ssp_{\zeit}) 
    =
0
\,,
\ee{1dTempFTa} 
on the period-5 Bravais cell,
\bea
- \ssp_{1} + 2\ssp_{0} - \ssp_{1} + V'(\ssp_{0}) &=& 0 \continue
- \ssp_{0} + 2\ssp_{1} - \ssp_{2} + V'(\ssp_{1}) &=& 0 \continue
- \ssp_{1} + 2\ssp_{2} - \ssp_{2} + V'(\ssp_{2}) &=& 0 \continue
- \ssp_{1} + 2\ssp_{2} - \ssp_{2} + V'(\ssp_{2}) &=& 0 \continue
- \ssp_{0} + 2\ssp_{1} - \ssp_{2} + V'(\ssp_{1}) &=& 0
\,,
\label{symmCycD5eqs5} 
\eea
where we have used \refeq{symmCycD5bcs}.
The result are symmetry reduced equations,
 modified by the two reflection \bcs,
\bea
- 2\ssp_{1} + 2\ssp_{0}           + V'(\ssp_{0}) &=& 0 \continue
- \ssp_{0} + 2\ssp_{1} - \ssp_{2} + V'(\ssp_{1}) &=& 0 \continue
- \ssp_{1} + \ssp_{2}             + V'(\ssp_{2}) &=& 0
\,,
\label{symmCycD5eqs} 
\eea
with an asymmetric 3\dmn\ {\jacobianOrb} \refeq{jMorb1dFT}
\bea
\jMorb_o &=&
\left(\begin{array}{ccc}
{s}_0 & -2 & 0 \\
 -1 & {s}_1 & -1 \\
 0 & -1 & {s}_2-1
\end{array}\right)
\,.
\label{jacobianOrbD5} 
\eea
So, for a reflection-symmetric \lst\ of odd period, one has to
impose the even $\sitebox{\ssp_0}$ and odd $|$ reflection \bcs\
in order to define the {\jacobianOrb} for the {prime} orbit $\tilde{\Xx}$
\refeq{primeLattStat}.

The form of {\jacobianOrbs} for all \bcs\ of \refsect{s:LattStateSyms},
specialized to \templatt\ but easily generalized to general nonlinear
field theories, is given in \refappe{sect:SymmReducedJacobian}.

\paragraph{Why bother?}
\wwwcb{}\rf{ChaosBook} bemoans more than 20 times that in it the
time-reversal symmetric orbits are not accounted for (here that is
accomplished in \refsect{s:Lind1d}).
But for long periods, almost all {\lsts} are of asymmetric type
\refeq{reflSymNo}. Why do we
obsess about symmetric {\lsts} so much? How important are they?

The reason is that {\po} expansions are dominated by short orbits, with
the longer ones only providing exponentially small corrections. But
almost all short-period {\lsts} are symmetric; for example, for the
$\phi^3$ field theory, the first two asymmetric {\lsts} of are of period
6.

\section{Reciprocal lattice}
\label{s:recip1d} 

If the {\jacobianOrb} is invariant under time
translation, its eigenvalue spectrum and {\HillDet} can be efficiently
computed using tools of crystallography,
such as the discrete Fourier
transform, a discretization approach that goes all the way back to
Hill's 1886 paper\rf{Hill86}.

Think of a solution of a discrete time dynamical system as a 1\dmn\
{\lst} with the field on each site labeled by integer
time.
A time period-$\cl{}$ {\lst} lives on a discrete 1-torus (a ring or chain or
necklace) of period-$\cl{}$, and if system's law is time-invariant, its orbit,
the set of {\lsts} related to it by cyclic translations, are physically
equivalent (\reffig{fig:1dLatStatC_5}).
The symmetry is the cyclic group \Cn{n}, and one only needs to count and
distinguish \Cn{n} \emph{orbits}, compute only one {\lst} per each
orbit.

The smart way to do this is by a discrete Fourier transform.
Were the lattice $d$\dmn\ \refeq{BravLatt}, defined by Bravais cell
vectors $\{\mathbf{a}\}$, a crystallographer would immediately move to
the \emph{reciprocal} lattice,
\( 
\tilde{\lattice}_{\mathbf{b}} = \{k \mathbf{b}\,|\, k \in \mathbb{Z}\}
\,,
\) 
with {reciprocal}
lattice basis vectors $\{\mathbf{b}\}$ satisfing
\( 
\mathbf{b}\!\cdot\!\mathbf{a} = 2 \pi
\,.
\) 
On the {reciprocal} lattice translations are
quotiented out, and calculations are restricted to a finite
{Brilluoin zone} (Bloch's theorem of
solid state physics). Here we work on a 1\dmn\ lattice with unit
lattice spacing 1, so the reciprocal lattice spacing is $2\pi/1=2\pi$, with
the (first) Brillouin zone from $k=-\pi$ to $k=\pi$
(we give an example in \reffig{fig:recipCatCn}).

The  cyclic group \Cn{n} elements
are generated by
the $[\cl{}\!\times\!\cl{}]$ shift matrix
\refeq{hopMatrix} which
 translates a {\lst} \refeq{1dLattStatC_n}  for\-ward-in-time by one site,
$\transp{(\shift \Xx)}=(\ssp_1,\ssp_2,\cdots,\ssp_{\cl{}-1},\ssp_0)$.
After $\cl{}$ shifts, the {\lst} returns to the initial
state, yielding the characteristic equation for the matrix $\shift$
\beq
\shift^\cl{}-\id=0
\,,
\ee{shift2n}
whose eigenvalues are $\cl{}$th roots of unity, with the $\cl{}$ complex
eigenvectors also built from roots of unity
\bea
\{\lambda_k\} &=& \{1, \omega, \omega^2,\cdots, \omega^{\cl{}-1}\}
                \,,\qquad\qquad\quad
                  \omega=\e^{2\pi\mathrm{i}/\cl{}}
                \continue
\tilde{e}_k   &=&
    \frac{1}{\sqrt{\cl{}}}
    (1, \omega^k, \omega^{2k}, \ldots, \omega^{k(\cl{}-1)})
    \,,\qquad k=0, 1,\ldots, \cl{}-1
\,.
\label{FourierModes}
\eea

\subsection{Reciprocal {\lsts}}
\label{sect:ReciprLst}

In the $\{\tilde{e}_k\}$ Fourier basis, a real $\cl{}$\dmn\
{\lst} vector $\Xx$ is mapped onto a $\cl{}$\dmn\ complex
{reciprocal} lattice  vector
\beq
\tilde{\Xx} = (\cssp_{0},\cssp_{1},\cssp_{2},\dots,\cssp_{\cl{}-1})
\,,
\eeq
with the $k$th Fourier mode of magnitude $|\cssp_k|$ and phase $e^{i\theta_k}$.

On the reciprocal lattice, the shift matrix is diagonal,
$\shift_{jk}=\omega^k\,\delta_{jk}$, and the `time' dynamics is
breathtakingly simple: no matter what the dynamical system is, in
one time step $\Xx \to \shift \Xx$, the $k$th Fourier
mode phase is incremented by a fraction of the circle,
\bea
(\cssp_{0},\cssp_{1},\cssp_{2},\dots,\cssp_{\cl{}-1}) &\to&
    (\cssp_{0},\omega\cssp_{1},\omega^2\cssp_{2},\dots,
                          \omega^{\cl{}-1}\cssp_{\cl{}-1})
\continue
e^{\mathrm{i}\theta_k} &\to& e^{\mathrm{i}(\theta_k+2\pi{k}/\cl{})}
\,,
\label{recipCircl}
\eea
so reciprocal {\lsts} literally run in circles; for non-zero $k$ and
$|\cssp_k|$, all reciprocal {\lsts} lie on vertices of regular complex
plane $\cl{}$-gons, inscribed in circles of radius $|\cssp_k|$, one
circle for each orbit.

\begin{figure}
  \centering
            \begin{minipage}[c]{0.3\textwidth}\begin{center}
\includegraphics[width=1.0\textwidth]{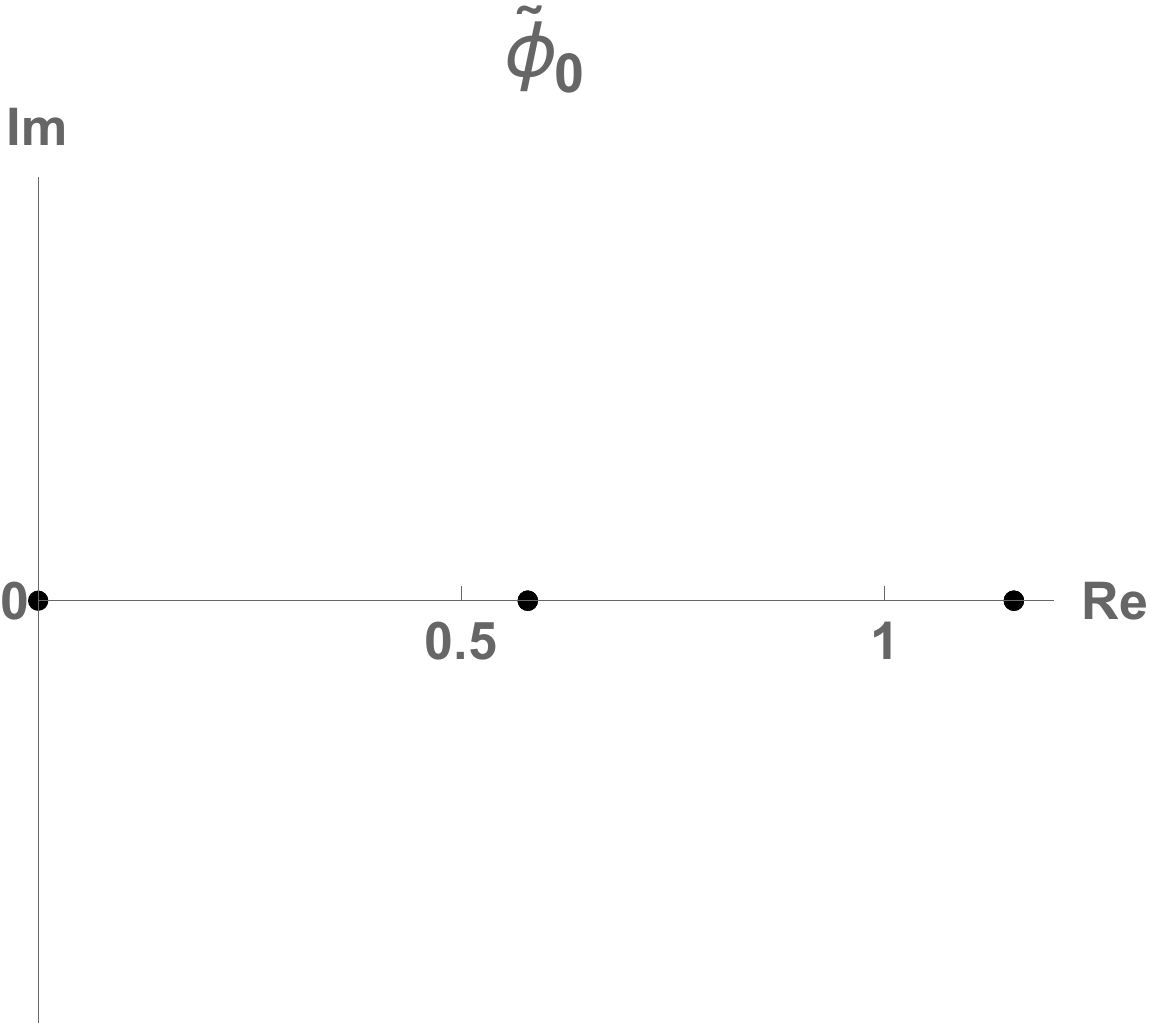}\\(a)   
            \end{center}\end{minipage}
            \begin{minipage}[c]{0.3\textwidth}\begin{center}
\includegraphics[width=1.0\textwidth]{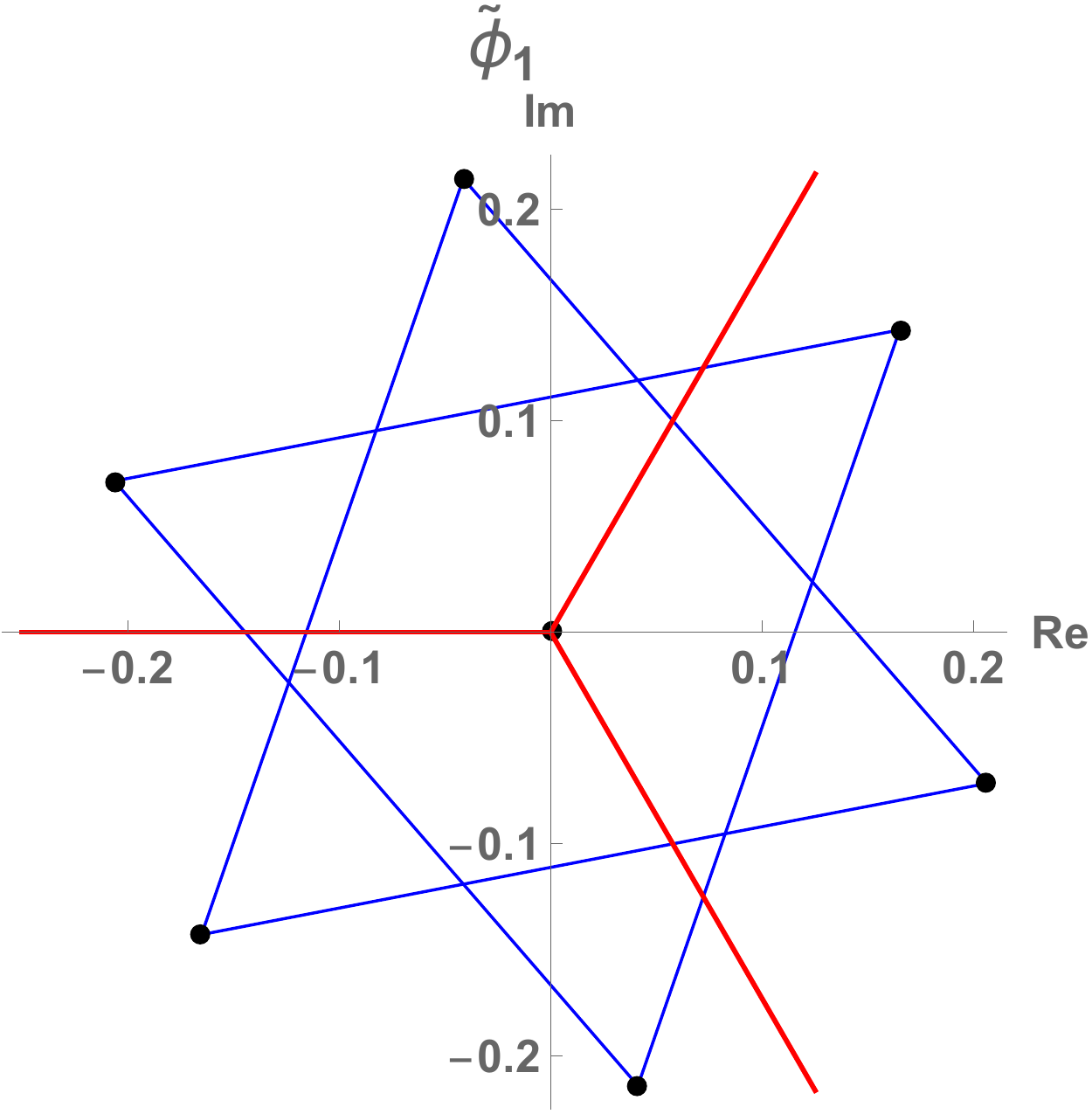}\\(b)   
            \end{center}\end{minipage}
            \begin{minipage}[c]{0.3\textwidth}\begin{center}
\includegraphics[width=1.0\textwidth]{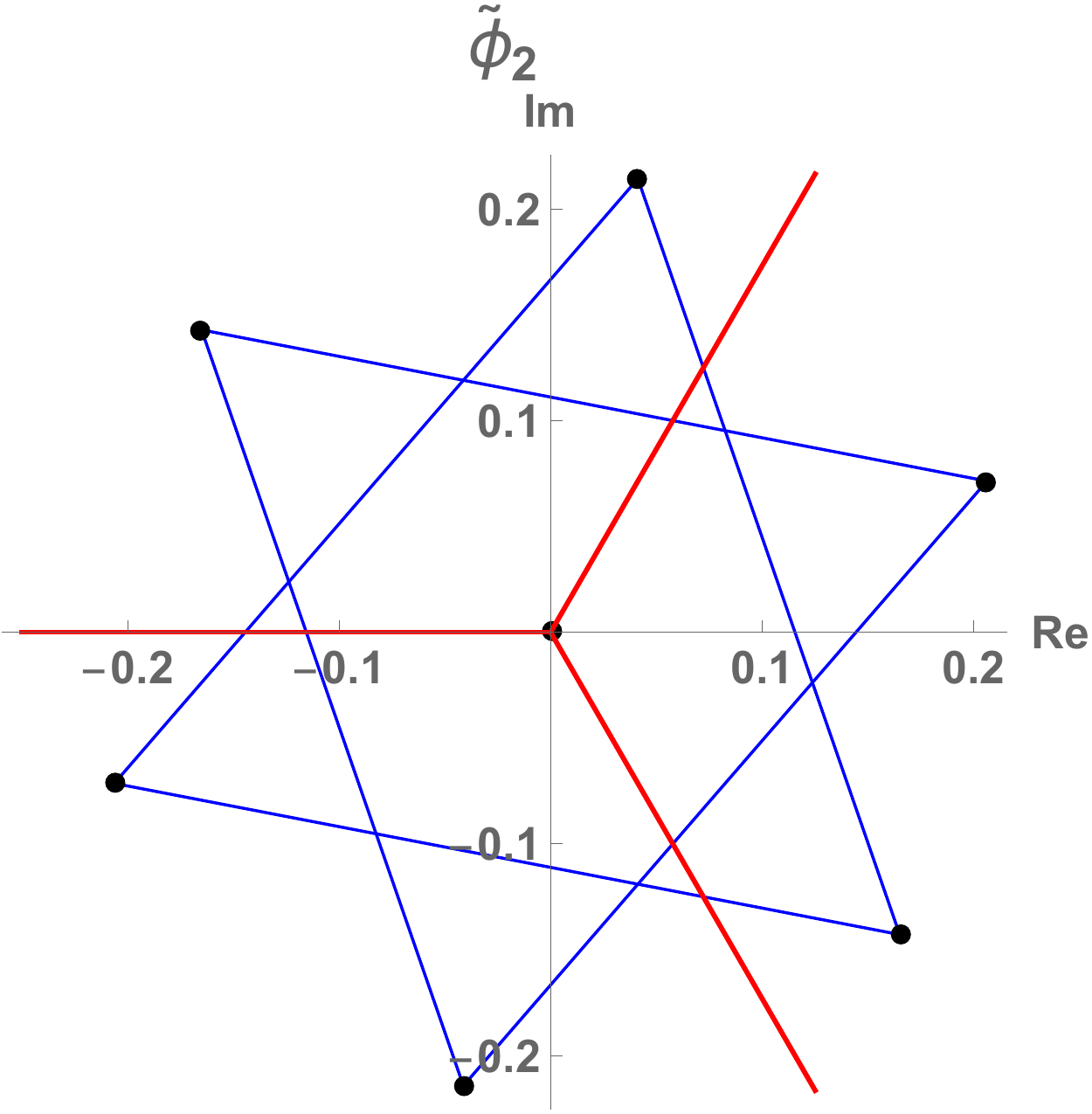}\\(c)   
            \end{center}\end{minipage}
  \caption{\label{fig:BernC3}
(Color online)~~~
The reciprocal lattice $(\cssp_{0},\cssp_{1},\cssp_{2})$ Fourier
components of the 7 $\Cn{3}$-equivariant period-3
{\lsts}, $s=2$ {temporal Bernoulli system} \refeq{1stepDiffEq}. In
$\cssp_1$ and $\cssp_2$ complex planes, reciprocal {\lsts} lie on
vertices of the 2 equilateral triangles, one for each
\Cn{3} orbit, while the component at the origin is the fixed point
$\Xx=(0,0,0)$.
The \Cn{3} fundamental domain indicated by red border lines
contains non-zero reciprocal
{\lsts} whose phases lie in the $[-2\pi/6,2\pi/6)$ wedge, one reciprocal
{\lst} for each distinct \Cn{3} orbit.
}
\end{figure}

As a concrete example, consider the period-3 {\lsts} of
the temporal Bernoulli \refeq{1stepDiffEq} for stretching parameter $s=2$.
It is a linear problem and all {\lsts} are easily computed
by hand, one for each symbol \brick\ $\Mm$. There is always the
fixed point {\lst} $(0,0,0)$ at the origin, and the remaining {\lsts}
belong to $M_3=2$ period-3 orbits, where $M_\cl{}$ is the number of orbits of
period $\cl{}$.
Discrete Fourier transform
maps these 2 orbits into reciprocal lattice $(\cssp_{0},\cssp_{1},\cssp_{2})$
triangles, see \reffig{fig:BernC3}. The time-step $\shift$ acts on the $\cssp_1$, $\cssp_2$
components by complex 1/3-circle phase rotations $\exp(2 \pi
\mathrm{i}/3)$ and $\exp(4 \pi \mathrm{i}/3)$, respectively: reciprocal
{\lsts} connected by blue lines in \reffig{fig:BernC3}
lie on a circle and belong to the same orbit.
In this example the two orbits happen to lie on the
same circle, as they are related by the {\em internal} $\Dn{1}: \ssp_i
\to 1-\ssp_i$ symmetry of the Bernoulli system, see
\refsect{s:InternalSymm}.

\begin{figure}
  \centering
            \begin{minipage}[c]{0.25\textwidth}\begin{center}
\includegraphics[width=1.0\textwidth]{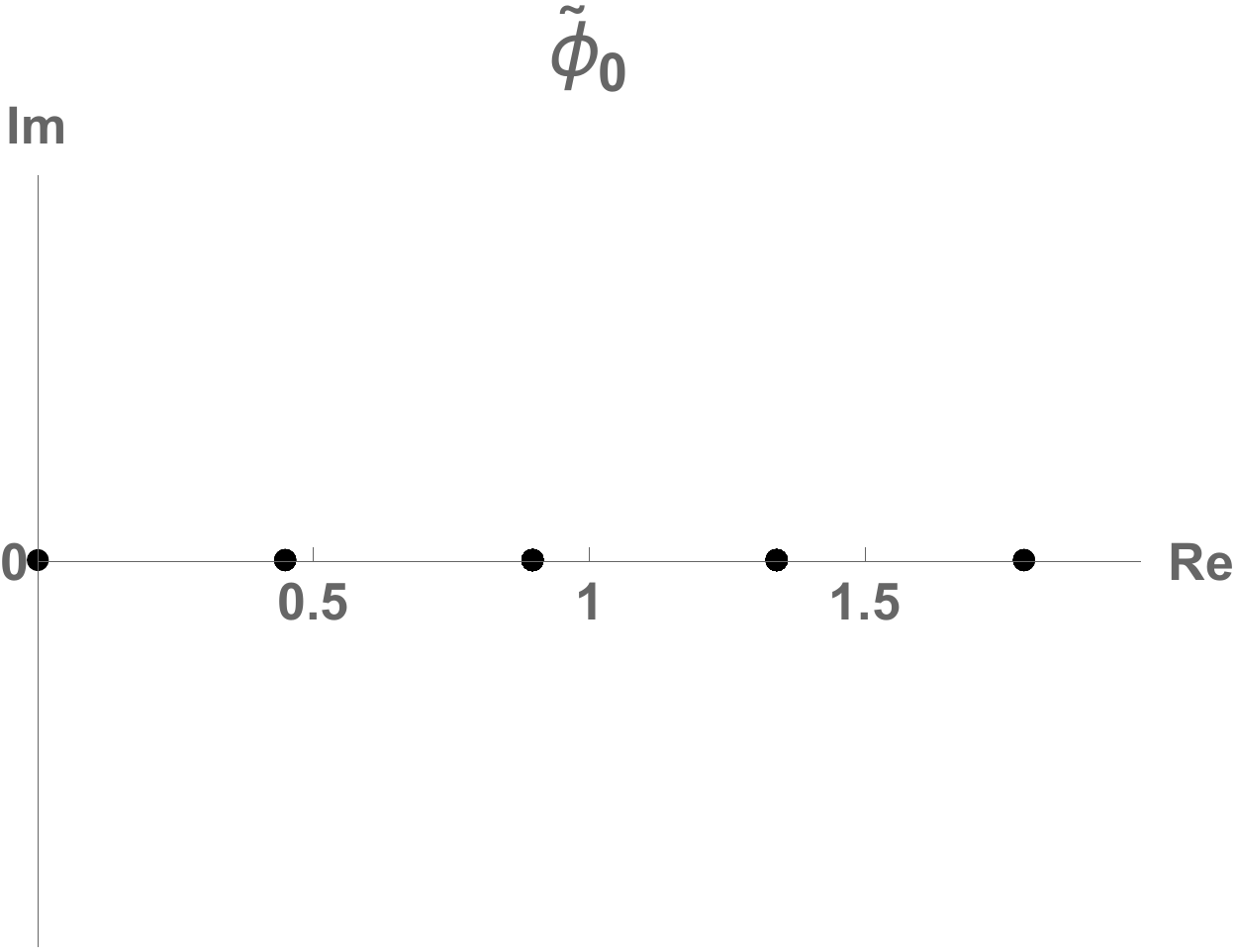}\\(a)
            \end{center}\end{minipage}  
            \begin{minipage}[c]{0.25\textwidth}\begin{center}
\includegraphics[width=1.0\textwidth]{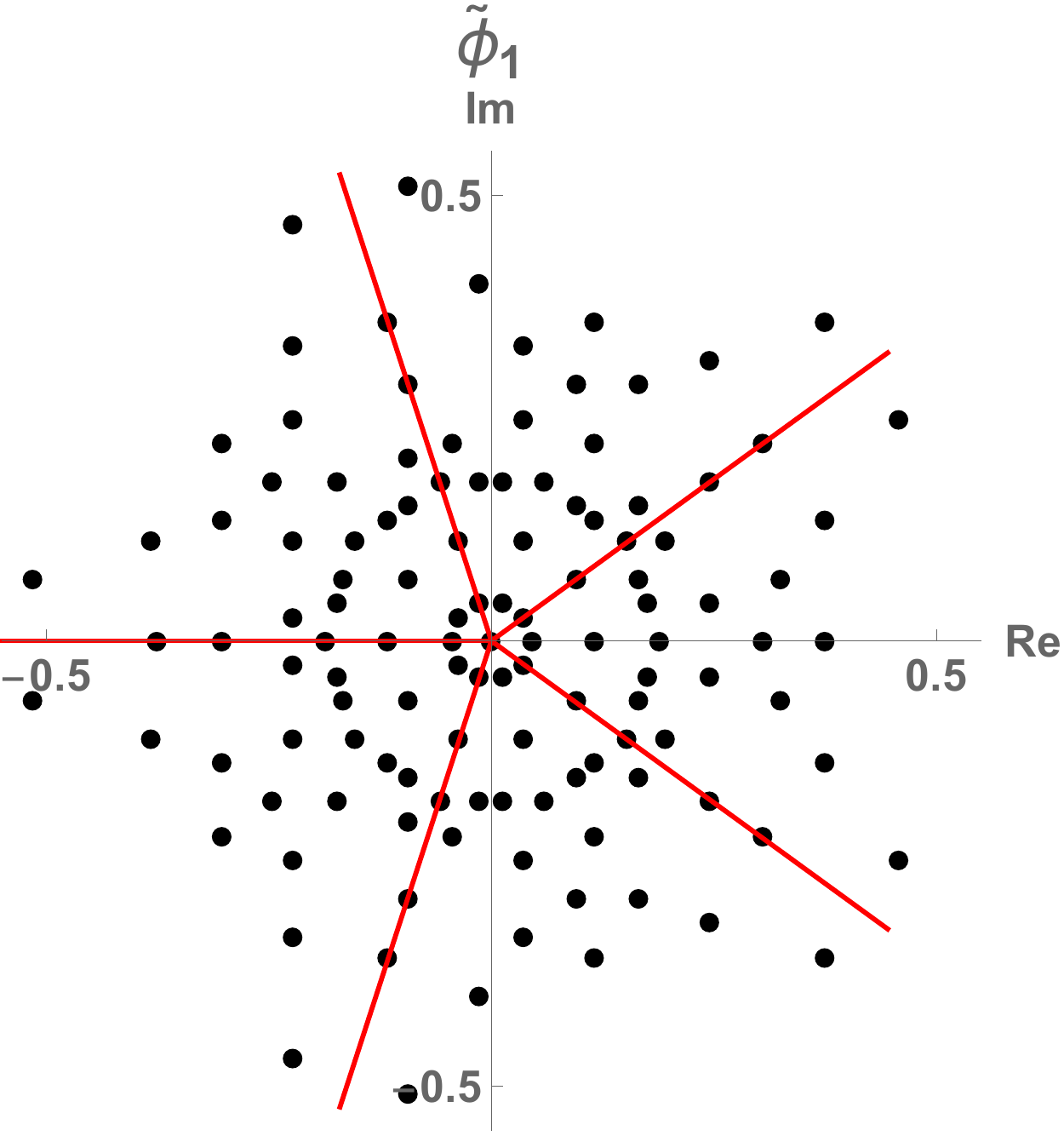}\\(b)
            \end{center}\end{minipage}  
            \begin{minipage}[c]{0.25\textwidth}\begin{center}
\includegraphics[width=1.0\textwidth]{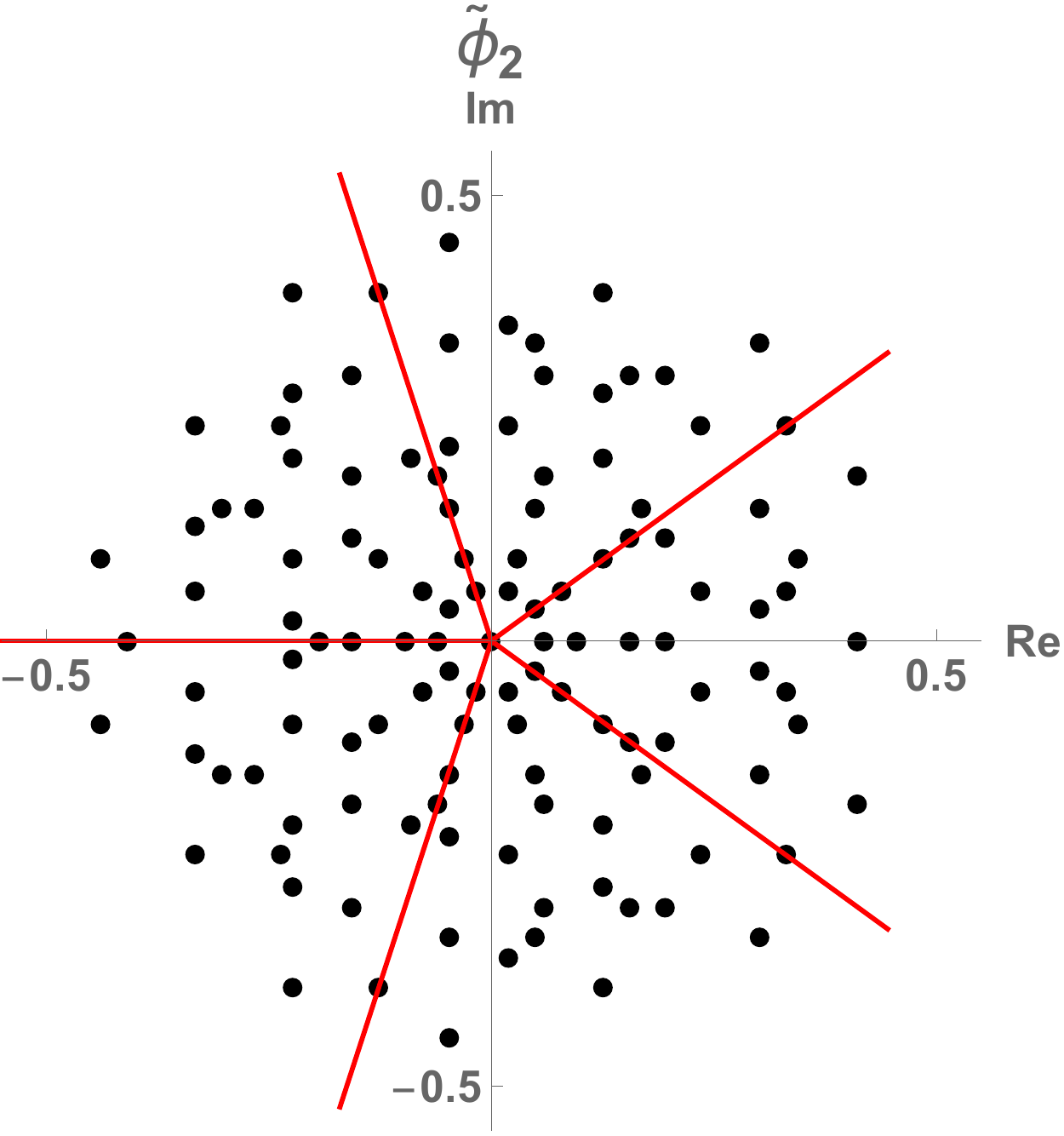}\\(c)
            \end{center}\end{minipage}  
~~~
\\
            \begin{minipage}[c]{0.25\textwidth}\begin{center}
\includegraphics[width=1.0\textwidth]{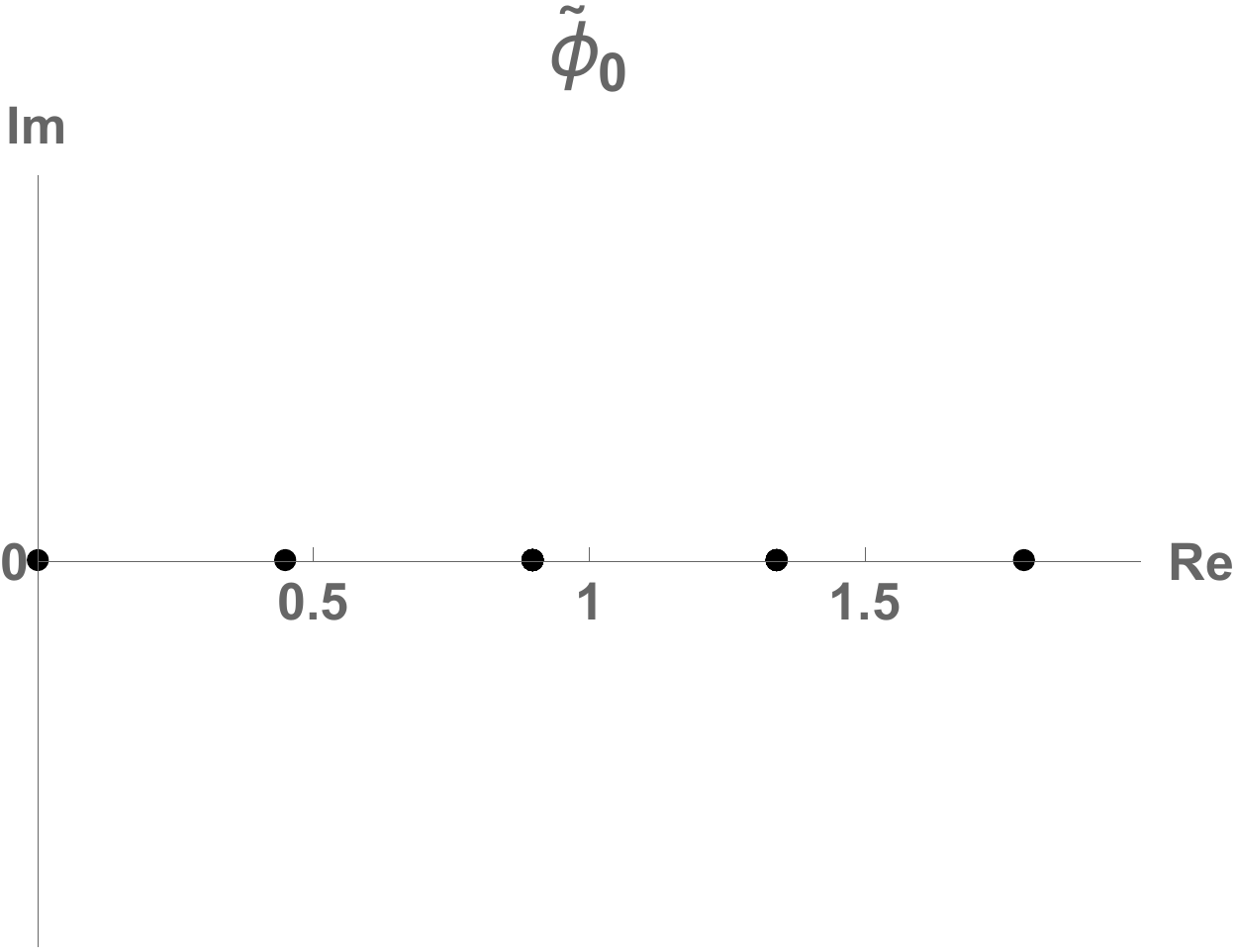}\\(d)
            \end{center}\end{minipage}  
            \begin{minipage}[c]{0.25\textwidth}\begin{center}
\includegraphics[width=1.0\textwidth]{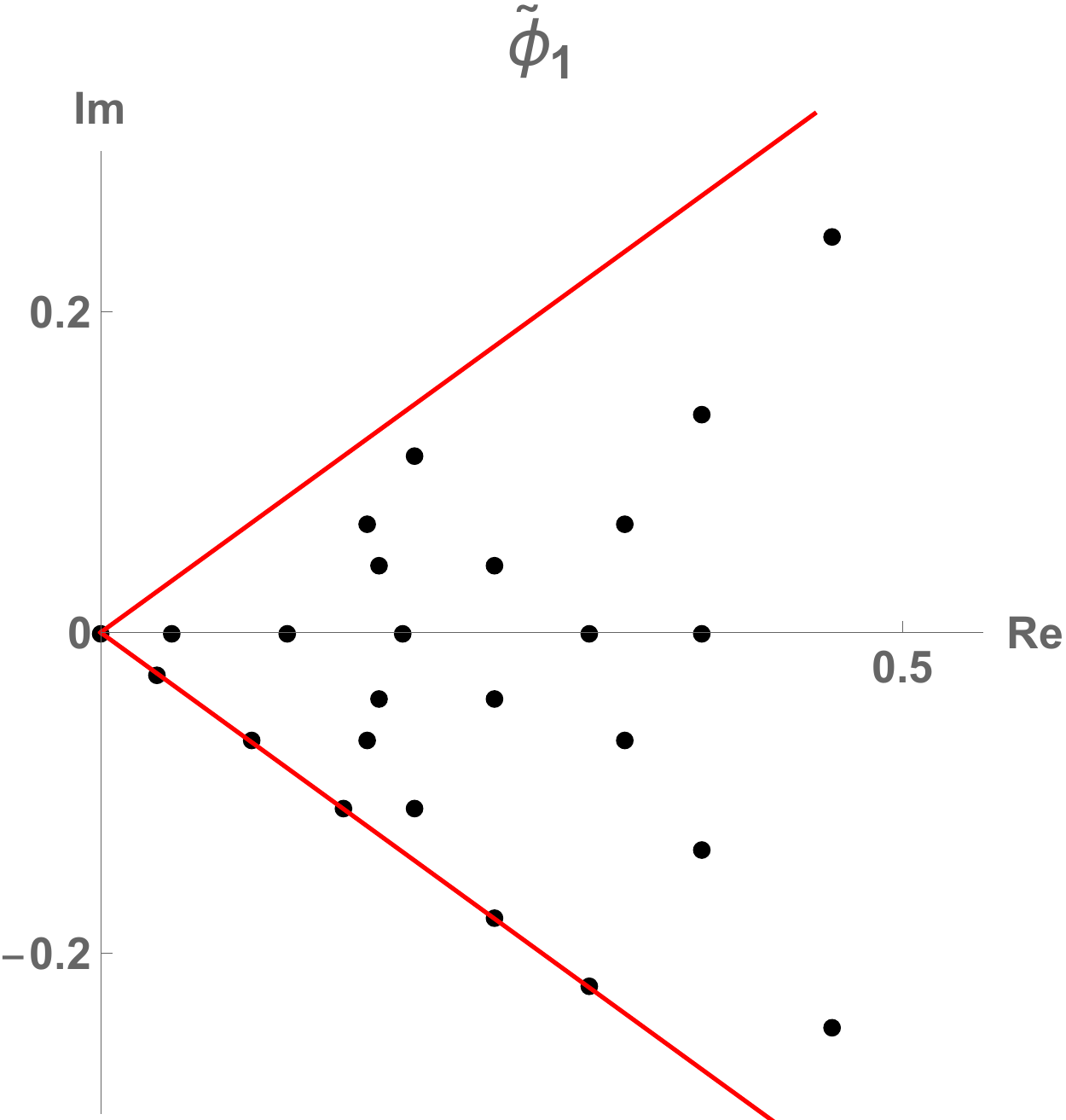}\\(e)
            \end{center}\end{minipage}  
            \begin{minipage}[c]{0.25\textwidth}\begin{center}
\includegraphics[width=1.0\textwidth]{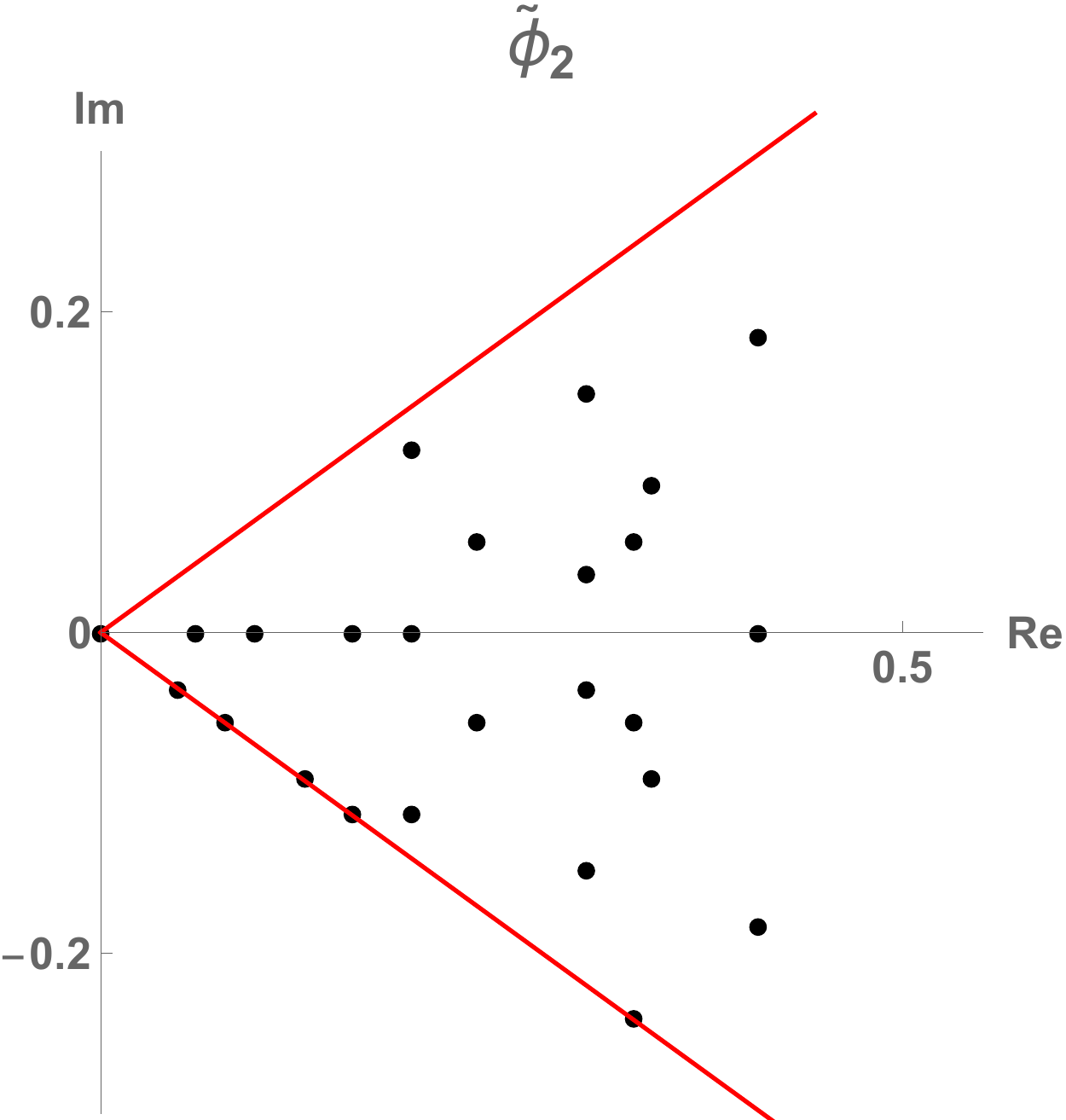}\\(f)
            \end{center}\end{minipage}  
\\ 
  \caption{\label{fig:catC5} 
(Color online)~~~
The 121 period-5 reciprocal {\lsts}
of the $s=3$ \templatt\ \refeq{catMapNewt}.
    (a,b,c)
The reciprocal lattice $\cssp_{0},\cssp_{1},\cssp_{2}$ complex planes.
The state at the origin is the fixed point $(0,0,0,0,0)$.
As in \reffig{fig:BernC3}, all non-zero reciprocal {\lsts} lie on
vertices of regular pentagons (not drawn here) that form orbits under \Cn{5}
cyclic permutations.
    (d,e,f)
The \Cn{5} fundamental domain contains $M_5=24$ non-zero  reciprocal
{\lsts} whose phases lie in the $[-2\pi/10,2\pi/10)$ wedge, one
reciprocal {\lst} for each distinct \Cn{5} orbit.
          }
\end{figure}

\subsubsection{\Cn{\cl{}} fundamental domain.}
\label{sect:CnReciprLattFD}

Divide each $k>0$ complex $\cssp_k$ plane of a period-$\cl{}$ reciprocal
\lst\ into $\cl{}$ equal wedges, and call one of them the `fundamental
domain', for example the wedge bordered by $[-\pi/\cl{},\pi/\cl{})$.
Under $\cl{}$ discrete rotations the fundamental domain completely tiles
the complex plane.
We exclude the fixed point {\lst} $\cssp_k=0$ at the origin from the
domain, as it belongs to a time invariant subspace.

As is clear by inspection of \reffig{fig:BernC3}, every $k$th complex
plane regular $\cl{}$-gon has precisely one vertex, \ie, a single
reciprocal \lst\ per each orbit, in the interior of the fundamental
domain, or on its border. For the period-3 reciprocal {\lsts} shown in
\reffig{fig:BernC3}, there are 2 points in both $k=2$ and $k=3$ complex
plane fundamental domain, one \lst\ for each period-3 orbit.

Consider next the 121
period-5 reciprocal {\lsts} of the $s=3$ \templatt\ \refeq{catMapNewt},
\reffig{fig:catC5}\,(a,b,c). Excluding the fixed point $\cssp_k=0$
\lst\ at the origin, there are $M_5=24$ reciprocal {\lsts} in the
\reffig{fig:catC5}\,(d,e,f) fundamental domain, each representing
$\cl{}=5$ {\lsts} in its orbit, so the total number of {\lsts} is
$N_5=1+5\,M_5 = 1+5\times24=121$, in agreement with
\reftab{tab:lattstateCountCat}.

The set of period-5 reciprocal {\lsts} of \refFig{fig:catC5} clearly
exhibits symmetries beyond the cyclic \Cn{5}, in particular under reflections
across the axes drawn in red. It also
turns out that for the \templatt\ the $\cssp_{1}$ complex plane looks the same as the
$\cssp_{4}$ complex plane, and $\cssp_{2}$ the same as $\cssp_{3}$, so we
do not plot them here. Furthermore, for period $\cl{}$ not a prime number, some
$|\cssp_k|$ might vanish. We shall return to these symmetries in
\refsect{s:DnRecipLatt}.

\subsection{Spectra of \jacobianOrbs}
\label{sect:Hillrecip1d}

As the period of a {\lst} gets longer, the {\jacobianOrb}
becomes larger and the {\HillDet} becomes harder to compute.
For a period-\cl{} \lst, $\jMorb$ is a matrix with \cl{} eigenvalues and
eigenvectors.
    %
\toVideo{youtube.com/embed/w67SSTkEXQw}
    %
What are they? What are the magnitudes of these eigenvalues?

\subsubsection{Spectra of translationally invariant {\jacobianOrbs}.}
\label{s:jacOrbSpectr}

{\JacobianOrbs} of the {temporal Bernoulli} \refeq{bernFixPoint} and
{\templatt} \refeq{tempCatFix} consist of only identity matrix and cyclic
shift matrix, whose eigenvectors are discrete Fourier basis
\refeq{FourierModes}, so they are diagonalized by discrete Fourier transform.
In the space of reciprocal {\lsts}, {\jacobianOrbs} \refeq{bernFixPoint}
and \refeq{tempCatFix} are diagonal, each diagonal element an eigenvalues of
{\jacobianOrbs}, for the {temporal Bernoulli}
\bea
({s}\id - {\shift})\,\tilde{e}_k
= ({s} - \omega^{k})\,\tilde{e}_k
\,,
\label{BernEigen}
\eea
and for the \templatt
\bea
(-\shift+{s}\,\id-\shift^{-1})\,\tilde{e}_k
=
\left[{s} - 2\cos\left(\frac{2\pi k}{\cl{}}\right)\right]\,\tilde{e}_k \, .
\label{CatEigen}
\eea
Determinants are products of eigenvalues, so the {temporal Bernoulli}
{\HillDet} for any period-\cl{} {\lst} is
\bea
\Det({s}\id - {\shift})
=
\prod_{k=0}^{\cl{}-1} ({s} - \omega^{k})
=
s^{\cl{}} - 1
\,,
\eea
in agreement with the time-evolution count \refeq{noPerPtsBm} and Hill's
formula calculation \refeq{detBern}.
The {\templatt} {\HillDet} is
\bea
\Det(-\shift+{s}\,\id-\shift^{-1})
    =
\prod_{k=0}^{\cl{}-1} \left[{s} - 2\cos\left(\frac{2\pi k}{\cl{}}\right)\right]
    =
2\,T_{\cl{}} \left({s}/{2}\right) - 2
\,,
\eea
confirming the $2$nd-order inhomogeneous difference equation calculation
\refeq{POsChebyshev}.

\begin{figure}\begin{center}
            \begin{minipage}[c]{0.45\textwidth}\begin{center}
\includegraphics[width=1.0\textwidth]{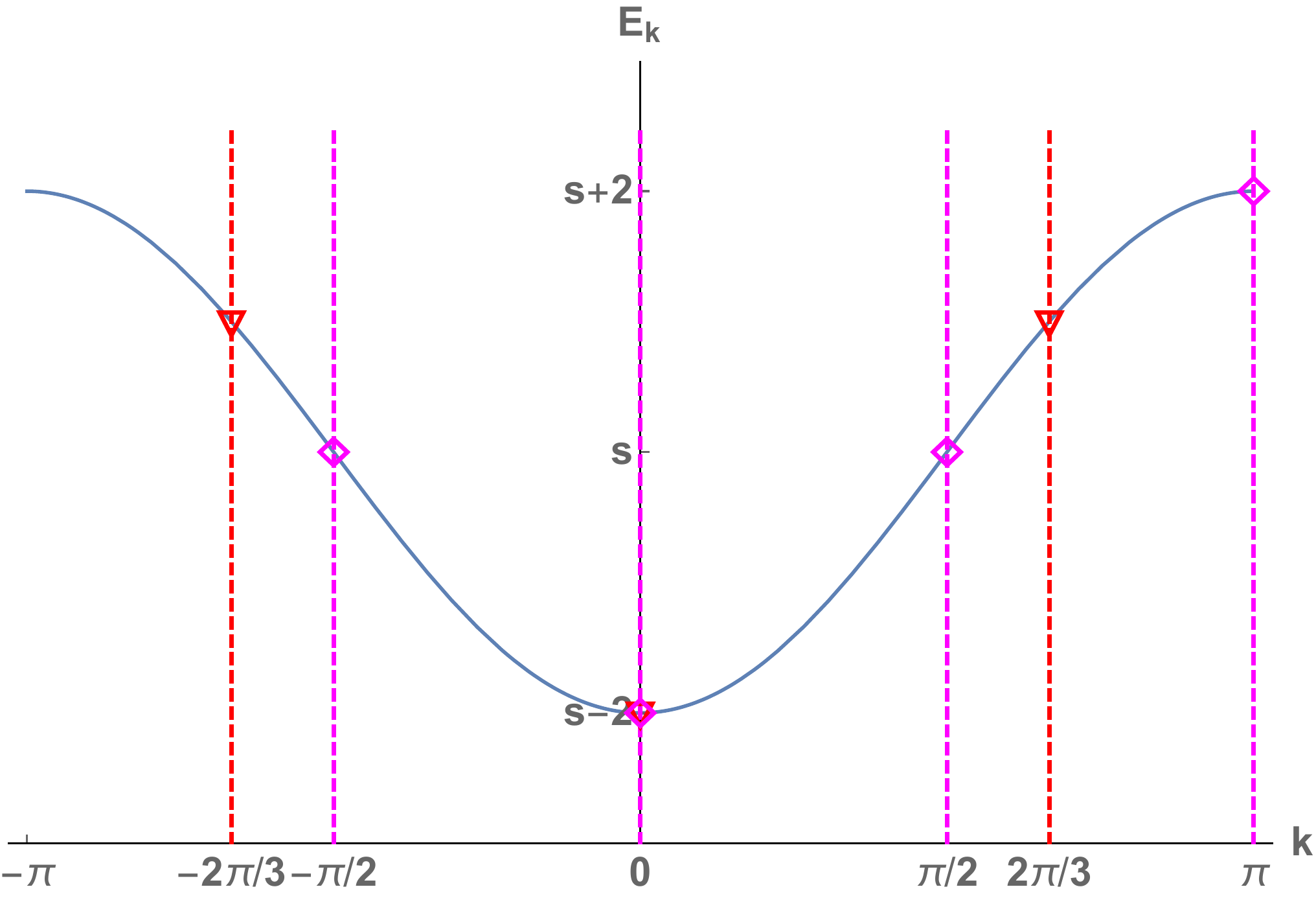}\\(a)
            \end{center}\end{minipage} 
\qquad
            \begin{minipage}[c]{0.45\textwidth}\begin{center}
\includegraphics[width=1.0\textwidth]{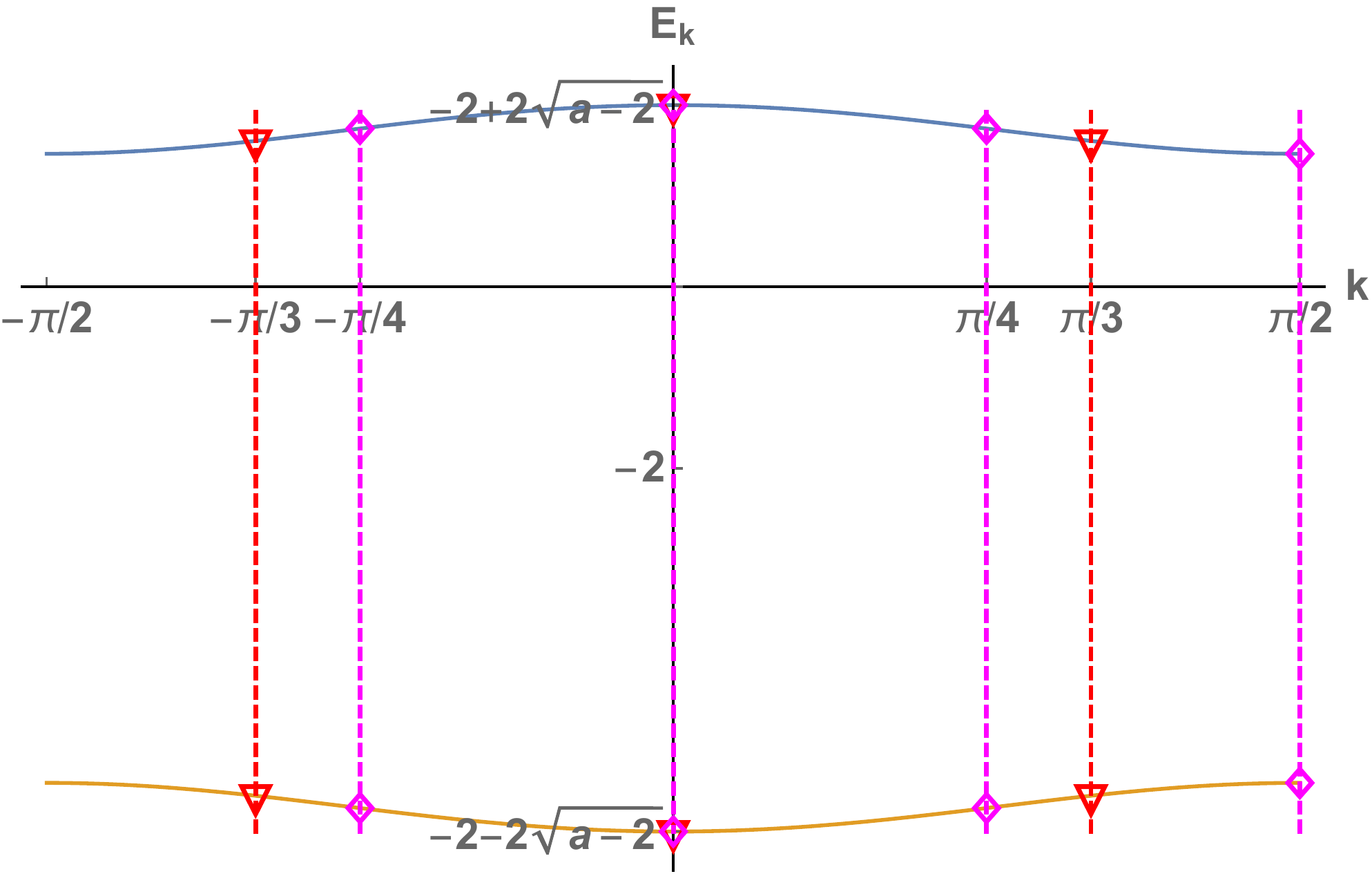}\\(b)
            \end{center}\end{minipage} 
\end{center}
  \caption{\label{fig:recipCatCn} 
(Color online)~~~
Infinite lattice \jacobianOrb\ first Brillouin zone spectra, as functions
of the reciprocal lattice wavenumber $k$. For time-reversal invariant
systems the spectra are $k\to -k$ symmetric.
(a)
The \templatt\ $E(k)$ spectrum \refeq{templattSpect}.
Any period-\cl{} \lst\ spectrum consists of \cl{} discrete points embedded into
$E(k)$, for example
period-3 (red triangles) and
period-4 (magenta diamonds) {\lsts} eigenvalues \refeq{catSpectPer-n}.
There are only 4 reciprocal {\lsts}, as $k=\pi$ and $k=-\pi$ differ by a
reciprocal lattice translation, and are counted only once.
(b)
The \henlatt\ $E(k)_{\pm}$ spectrum \refeq{HenSpectPer2}
of the infinite lattice tiled by athe period-2 {\lst}, together with
the eigenvalues of repeats \refeq{HenSpectPer2rep} for 1st repeat,
3rd repeat (red triangles) and
4th repeat (magenta diamonds).
          }
\end{figure}

The {\jacobianOrb} \refeq{jacobianOrb} of a period-\cl{} {\lst}
is an $[\cl{}\!\times\!\cl{}]$ matrix.
For an infinite {\lst}, the {\jacobianOrb} is an infinite\dmn\ linear operator.
For example, the {\templatt} infinite\dmn\ {\jacobianOrb} has the form
\refeq{tempCatFix}, where the time translation operator $\shift$ implements the
translation on the infinite lattice. The {\jacobianOrb}
commutes with the time translation operator, so its symmetry is the
{infinite cyclic group}
of integer lattice translations \refeq{C_infty}. By Bloch's theorem
an
eigenstate of the linear operator $\jMorb$ is of form
\beq
\psi_{k}(\zeit) = e^{\mathrm{i}k\zeit} u_{k}(\zeit)
\,,
\ee{BlochTheorem}
where $u_k(\zeit)$ is a period-1 periodic function.
The {\jacobianOrb} only acts on the field over the integer lattice, on which the periodic
function $u_k(\zeit)$ is a constant. The eigenstates
\beq
(\psi_{k})_\zeit = \psi_{k}(\zeit) = e^{\mathrm{i}k\zeit}
    \,,\quad \zeit \in \integers
 \,,
\ee{BlochLatt}
are plane waves on the lattice.
The wavenumber $k$ of the eigenstate $\psi_k$ can always be restricted to
the first Brillouin zone, $k\in[-\pi,\pi)$. Action of the {\jacobianOrb}
\refeq{tempCatFix} on the wavenumber $k$ eigenstate \refeq{BlochLatt} yields
eigenvalue
\beq
E(k) \psi_k= \jMorb \psi_k = (s - 2\cos k)\psi_k
\,,
\ee{templattSpect}
plotted in the first Brillouin zone in \reffig{fig:recipCatCn}\,(a).
The infinite lattice spectrum contains all
eigenvalues of
{\jacobianOrbs} for any period-\cl{} {\lst}. The period-\cl{} boundary
condition is
\beq
\psi_k(\zeit + \cl{}) = \psi_k(\zeit) \,,
\ee{PeriodicEigenstate}
hence the wavenumber
$k$ is restricted to the \cl{} first Brillouin zone  values
 $k = 2 \pi l/\cl{}$, where
$l$ is an integer, with eigenvalues
\beq
E_k \psi_k= \jMorb \psi_k = (s - 2\cos k)\psi_k \,,
\qquad k = \frac{2 \pi l}{\cl{}}\,, \quad l \in \integers \,,
\ee{FiniteStateEigenvalue}
in agreement with \refeq{CatEigen}.
For example, for the period-1, period-3 and the period-4 {\lsts} the
wavenumbers and corresponding eigenvalues are, respectively
\bea
(k_0)   &=& (0)\,,\quad
(\lambda_{0})=(s-2)
        \continue
(k_{-1},k_0,k_1) &=& (-\frac{2\pi}{3},0,\frac{2\pi}{3})
        \continue
(\lambda_{-1},\lambda_{0},\lambda_{1})
                     &=&
        (s+1,s-2,s+1)
        \continue
(k_{-1},k_0,k_1,k_2) &=& (-\frac{\pi}{4},0,\frac{\pi}{4},\frac{\pi}{2})        \continue
(\lambda_{-1},\lambda_{0},\lambda_{1},\lambda_{2})
                         &=&
   (s,s-2,s,s+2)
\,,
\label{catSpectPer-n}
\eea
plotted in \reffig{fig:recipCatCn}\,(a).

In summary:
for a field theory with translationally invariant, uniform stretching
parameter $s$ {\jacobianOrbs}, eigenvalues of any \lst\ are embedded in a
single first Brillouin zone spectrum $E(k)$, the blue curve in
\reffig{fig:recipCatCn}\,(a). As long as the Klein-Gordon mass $\mu^2=s-2>0$
is positive, all {\lsts} are unstable, and the field theory is chaotic.

\subsubsection{Spectra of nonlinear field theories.}
\label{s:jacOrbSpectrNL}

For a nonlinear field theory, the {\jacobianOrb} \refeq{jMorb1dFT} of a
general prime
{\lst} is not translationally invariant. The {\jacobianOrb} of a repeat of a
period-\cl{} prime {\lst}, however, is a tri-diagonal block circulant matrix
\refeq{orbJprimeRpt}, which commutes with the translation operator
$\shift^{\cl{}}$.
For an infinite {\lst} tiled by repeats of a period-\cl{} prime {\lst} $\Xx_p$
(see \reffig{fig:1dLatStatC_5}\,(1) for a sketch),
the infinite\dmn\ {\jacobianOrb} is invariant under
$\shift^\cl{}$
translation subgroup \refeq{H(n)subgroup}. Now we can
apply Bloch's theorem to Bravais lattice $\cl{}\integers$, with an eigenstate of the
{\jacobianOrb} a plane wave times a periodic function \refeq{BlochTheorem},
where $u_k(t)$ is a periodic function with period \cl{}. The wavenumber $k$ is
restricted in the first Brillouin zone of the Bravais lattice $\cl{}\integers$,
$k \in [-\pi/\cl{},\pi/\cl{})$. Using the Bloch's theorem
\refeq{BlochTheorem} we can find the eigenvalue
spectrum of the {\jacobianOrb} of a prime {\lst} $\Xx_p$'s repeats.

This infinite lattice spectrum contains eigenvalues of {\jacobianOrb} of the infinite
repeat of the prime {\lst} $\Xx_p$. To find eigenvalues of the {\jacobianOrb} of the
${m}$th repeat of $\Xx_p$ \refeq{orbJprimeRpt}, we need to only use the period-$({m}\cl{})$ eigenstates,
which satisfy:
\beq
\psi_k(\zeit + {m}\cl{}) = \psi_k(\zeit) \,.
\ee{PeriodmnEigenstate}
The wavenumber $k$ must exist on the reciprocal lattice of the Bravais lattice spanned
by $({m}\cl{})$, \ie, $k = 2\pi l/({m}\cl{})$, where $l$ is an integer.

As an example, consider the {\henlatt} period-2 \lst\ $\Xx_p$
\refeq{henLst2}, with a 2\dmn\ repeating block \refeq{henFundPar2}
{\jacobianOrb}.
The {\jacobianOrb} of an infinite lattice state tiled by repeats of
$\Xx_p$ is the infinite\dmn\ linear operator
\beq
\jMorb =
\left(
\begin{array}{ccccc}
 \ddots & \ddots &  &  &  \\
 \ddots & s_1 & -1 &  &  \\
  & -1 & s_0 & -1 &  \\
  &  & -1 & s_1 & \ddots \\
  &  &  & \ddots & \ddots \\
\end{array}
\right)
        \,,\quad
\left(\begin{array}{c}
 s_0 \cr
 s_1
\end{array}\right)
=
\left(\begin{array}{c}
 -2-2\sqrt{a-3} \cr
 -2+2\sqrt{a-3}
\end{array}\right)
\label{orbJprimeInfRpt}
\eeq
whose eigenstates are plane waves of form \refeq{BlochTheorem}, where
$u_{k}(\zeit)$ is periodic with period 2. Now there
are two families of $\jMorb$ eigenvalues:
\beq
E(k)^{\pm} \psi_k= \jMorb \psi_k = -2 (1 \pm \sqrt{a - 3 + \cos^2 k})\,\psi_k
\,,
\ee{HenSpectPer2}
plotted in the first Brillouin zone $k\in[-\pi/2,\pi/2)$ in \reffig{fig:recipCatCn}\,(b).

The eigenvalues of the {\jacobianOrbs} of the $m$th repeat of $\Xx_p$ are
$2m$ points in the spectrum \refeq{HenSpectPer2}, at $k=2\pi l/(2m)$,
where $l$ is an integer.
For example, for the single repeat (see {\jacobianOrb} \refeq{henFundPar2}),
the 3rd repeat and the 4th repeat the wavenumbers
and corresponding eigenvalues are, respectively
\bea
(k_0)   &=& (0)\,,\quad
(\lambda^\pm_{0})=(-2\pm 2\sqrt{a-2})
        \continue
(k_{-1},k_0,k_1) &=& (-\frac{\pi}{3},0,\frac{\pi}{3})
        \continue
(\lambda^\pm_{-1},\lambda^\pm_{0},\lambda^\pm_{1})
                     &=&
        (-2 \pm \sqrt{4a-11}, -2 \pm 2\sqrt{a-2}, -2 \pm \sqrt{4a-11})
        \continue
(k_{-1},k_0,k_1,k_2) &=& (-\frac{\pi}{4},0,\frac{\pi}{4},\pi/2)        \continue
(\lambda^\pm_{-1},\lambda^\pm_{0},\lambda^\pm_{1},\lambda^\pm_{2})
                         &=&
   (-2 \pm \sqrt{4a-10}, -2 \pm 2\sqrt{a-2},
                        \ceq
   -2 \pm \sqrt{4a-10}, -2 \pm 2\sqrt{a-3})
\,,
\label{HenSpectPer2rep}
\eea
plotted in \reffig{fig:recipCatCn}\,(b).

In summary:
for a nonlinear field theory, each period-\cl{} prime \lst\ $\Xx_p$ has
up to \cl{} distinct  eigenvalues $\{\lambda^{(j)}\}$, each with its own infinite
lattice spectrum family $E(k)^{(j)}$ into which eigenvalues of all repeats of $\Xx_p$
are embedded.

\subsection{Reciprocal lattice visualization of system's symmetries}
\label{s:DnRecipLatt}

\begin{figure}
  \centering
              \begin{minipage}[c]{0.4\textwidth}\begin{center}
\includegraphics[width=1.0\textwidth]{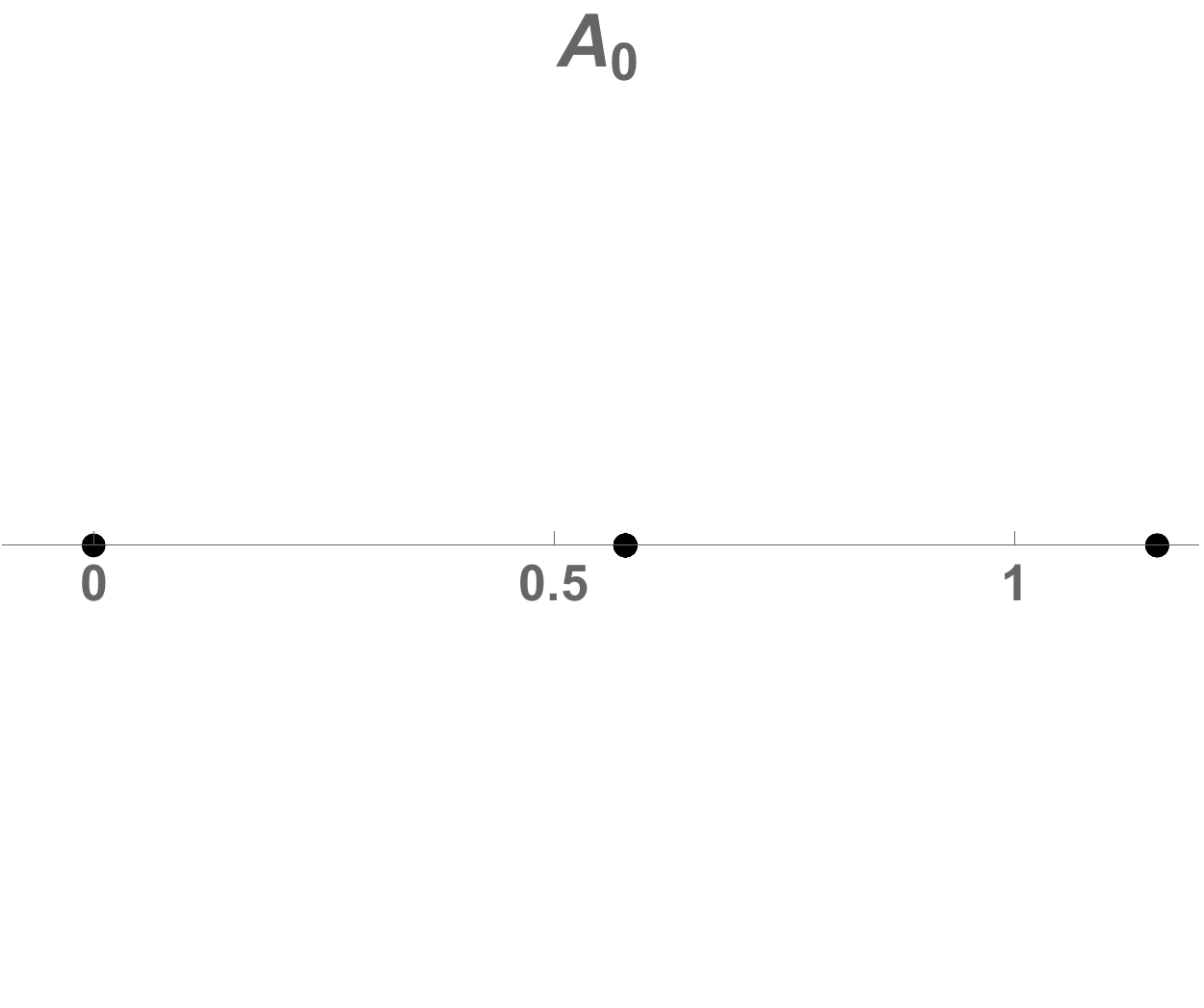}\\(a) 
            \end{center}\end{minipage}
            \begin{minipage}[c]{0.4\textwidth}\begin{center}
\includegraphics[width=1.0\textwidth]{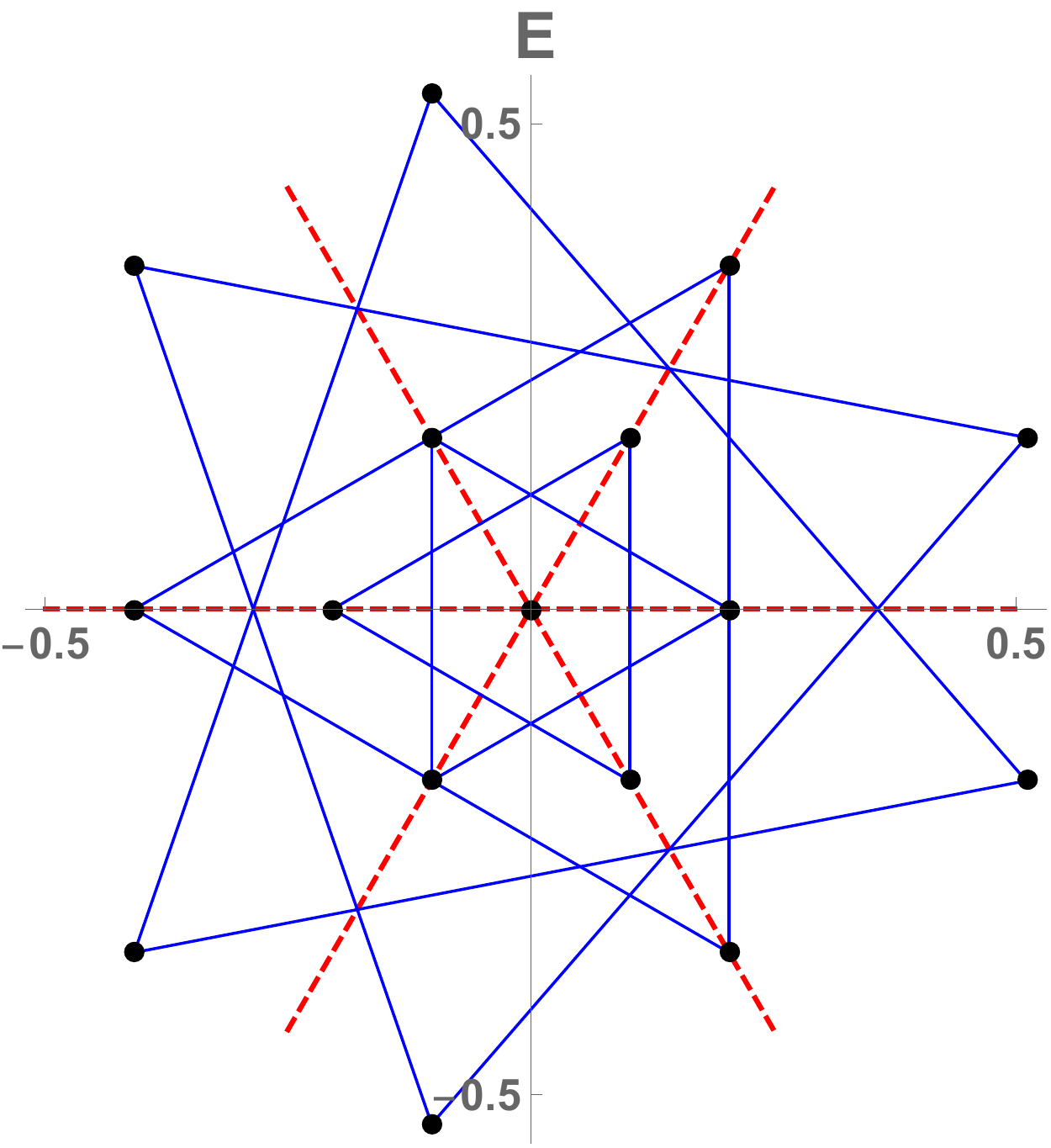}\\(b) 
            \end{center}\end{minipage}
  \caption{\label{fig:catD3}
(Color online)~~~
Period-3 reciprocal {\lsts} of the $s=3$ \templatt, plotted in the $\Dn{3}$
permutation representation irreducible subspaces $A_0+E$.
(a)
$A_0$ representation is the one-dimensional, time invariant
 `center of mass' of the orbit.
(b) In contrast to the
\Cn{\cl{}} complex irreps of
\reffig{fig:BernC3}, the 2\dmn\ irreps $E_k$ of $\Dn{\cl{}}$ are real.
In the $E_k$ planes
reciprocal {\lsts} related by cyclic permutations and reflections lie
on vertices of $\cl{}$-gons or pairs of $\cl{}$-gons. For $\Dn{3}$ they
lie on triangles,
with cyclic sets connected by blue lines.
The 2 big triangles are a single \Dn{3} 6-{\lsts} orbit, what for
\Cn{3} is a pair of 3-{\lsts} orbits, with time reversal symmetry ignored.
The remaining 3 smaller triangles are 3 time-reversal symmetric orbits.
The small pair in the center would be two distinct radius $\Dn{3}$
orbits, where they not related by the internal symmetry $\Dn{1}: \ssp_i
\to 1-\ssp_i$ specific to the \catlatt\ (see \refsect{s:InternalSymm}).
The state in the center is the fixed point $\ssp_0=0$.
Red dashed lines are the reflection axes of the $\Dn{3}$ group.
Note that there are 4 reciprocal \lst s on each reflection axis.
}
\end{figure}

When we block-diagonalize the permutation
representation of \refsect{sect:permReps} into $\Dn{\cl{}}$ irreps, the {\lsts} are projected into the
irrep subspaces. An example are the period-3 reciprocal {\lsts} of $s=3$ \templatt\
shown in \reffig{fig:catD3}. The permutation representation
is block diagonalized by basis vectors: $e_0=1/\sqrt{3}(1,1,1)$,
$e_1=\sqrt{2/3}(1,\cos(2\pi/3),\cos(4\pi/3))$ and
$e_2=\sqrt{2/3}(0,\sin(2\pi/3),\sin(4\pi/3))$. The basis vector $e_0$
spans the subspace of the  symmetric 1\dmn\ irrep $A_0$, invariant under $\Dn{3}$ actions.
In the 2\dmn\ irrep $E$ subspace spanned
by basis vectors $(e_1,e_2)$ the translation operator $\shift$ rotates the
reciprocal {\lsts} clockwise by $2\pi/3$, while the reflection operator $\Refl$ reflects the {\lsts}
across the horizontal axis. In \reffig{fig:catD3}
{\lsts} that are related by translations are connected by blue lines.
The red dashed lines are reflection operators' reflection axes.
The 2 big triangles are
related by reflection, and thus form a 6 reciprocal \lst s $\Dn{3}$-orbit.
The remaining 3 triangles
are 3 orbits of symmetric {\lsts}, self dual under reflection.

\subsubsection{\Dn{\cl{}} fundamental domain.}

If the space of the field configuration has \Dn{\cl{}} symmetry,
the subspace of the 2\dmn\ irrep $E_1$ can be divided
into $2\cl{}$ copies by the irrep. One can choose the fundamental domain to be the region
with polar angle between 0 and $\pi/\cl{}$, assuming that the horizontal axis is one of the
reflection axis of the irrep $E_1$. Each orbit only appears once in the fundamental domain,
as shown in \reffig{fig:catD3}. Note that the two translational orbits related by
the time reflection are a single orbit of the dihedral group.

What happens when {\lsts} appear on the boundary of the fundamental domain?
There are two possible situations. The first situation is that the {\lst} belongs to an
orbit with index less than the order of the symmetry group. For example,
in \reffig{fig:catD3} subspace of $E$, there are 3 points in the fundamental domain
with polar angle equal to 0 or $\pi/3$. These 3 points are representative {\lsts} of orbits
with time reflection symmetry. The indices of these orbits are 3 instead of 6.

The second situation is that the index of
the orbit of the {\lst} is equal to the order of the symmetry group
but the component in the subspace is 0. For example,
\[
\Xx = \frac{1}{104} (17, 51, 49, 43, 25, 75)
\]
is a period-6 {\lst} of the $s=3$ temporal Bernoulli
\refeq{1dBernLatt}. Using the discrete Fourier transform
this {\lst} becomes:
\[
\cssp =
\left(\frac{5}{2 \sqrt{6}},0,\frac{-5-3 i \sqrt{3}}{13
   \sqrt{6}},-\frac{\sqrt{3}}{4\sqrt{2}},\frac{-5+3 i \sqrt{3}}{13 \sqrt{6}},0\right) \,.
\]
This is a period-6 {\lst}. It belongs to an orbit that
contains 6 different {\lsts}. The $k=1$ component of this {\lst} is 0, which is on the boundary of the fundamental domain. To put this kind of
{\lsts} into the fundamental domain one needs to divide other subspaces.
For this {\lst} the $k=2$ and $k=3$ components are not 0. The irreps divide
the $k=2$ subspace into 3 copies and the $k=3$ subspace into 2 copies. One way to
choose the fundamental domain in these subspaces is: the argument of the component
in the $k=1$ subspace is $0\leq\arg(\cssp_{1})<\pi/3$; if the $k=1$ component
is 0, the arguments of the components
in the $k=2$ and $k=3$ subspaces are $0\leq\arg(\cssp_{2})<2\pi/3{}$ and
$0\leq\arg(\cssp_{3})<\pi$. Each orbit is guaranteed to visit this fundamental
domain exactly once.

Time-reversal invariant field configurations are confined to subspaces of
the system's $\infty$\dmn\ \statesp, as only `half' of the Bravais cell
field values are independent, the other half being a repeat in the
reverse order. The {\jacobianOrb} of the corresponding  prime {\lst} is
evaluated in a time-reversal invariant subspace. An example of such prime
{\lst} {\jacobianOrb}  was given in \refeq{jacobianOrbD5}.
\refAppe{sect:SymmReducedJacobian} lists all {\jacobianOrbs} of the
{\templatt} prime {\lst} with time-reversal symmetry. The corresponding
{\HillDet}s, together with the fundamental fact of
\refsect{sect:fundFact}, are then used to count the time-reversal
symmetric {\lsts}.

\section{Lind zeta function}
\label{s:Lind1d}

For a discrete time dynamical system
\(
\ssp_{\zeit+1}=\flow{}{\ssp_{\zeit}}
\,,\)
period-$\cl{}$ solutions (in present context, the period-$\cl{}$ Bravais
cell {\lsts}) are \emph{fixed point}s of the $\cl{}$th iterate map
$\map^\cl{}$.
While the symmetry of a time-invariant dynamical `law' $\map$ is
the {infinite cyclic group} \Cn{\infty},
the group of all temporal lattice translations, for a period-$\cl{}$
solution the symmetry is $\cl{}$-steps translation subgroup $H_{\cl{}}$.

This observation motivates definition of a general group-theoretic
\emph{fixed-points counting} generating  function that relates the number
of {\lsts} to the number of prime orbits for any \statesp\ map
\(
\map : \pS \to \pS
\)
with a symmetry group $\Group$,
    %
\toVideo{youtube.com/embed/QAus4P4p7lA} 
a counting function that applies equally
well to multi\dmn\ lattice field theories\rf{CL18} as to the 1\dmn\
theories considered here.

Let $\Group$ be a group whose action
$\alpha: \Group \times \pS \to \pS$
permutes elements of a set $\pS$.
The Lind zeta function\rf{Lind96} is then defined as
\beq
\zeta_{Lind}(t) =
\exp \Big( \sum_{H} \;
            \frac{N_{H}}{|\Group/{H}|}t^{|\Group/H|}
      \Big)
\,,
\ee{LindZeta}
where the sum is over all subgroups $H$ of $\Group$
of index $|G/H| < \infty$, and $N_{H}$ is the number of states
in $\pS$ invariant under action of (\ie, fixed points of) subgroup $H$,
\beq
N_{H} =
   |\{\Xx \in \pS : \mbox{ all } h \in H \quad \alpha(h,\Xx) = \Xx\}|
\,.
\ee{LindN_H} 
The index $|\Group/{H}|$ is best explained by working out a
few examples.

In the lattice field theory setting, $\pS$ is the set of all
{\lsts}.
For 1\dmn\ lattice field theories, the group $\Group$ is either the
{infinite cyclic group}  \Cn{\infty} of all lattice translations \refeq{C_infty},
or
the {infinite dihedral group} $\Dn{\infty}$  of all translations and
reflections \refeq{D_infty}.
Their finite-index subgroups $H$ are,
respectively, infinite translation subgroups $H_{\mathbf{a}}$
\refeq{H(n)subgroup}, or $H_{\mathbf{a}}$ and infinite dihedral subgroups
$H_{\mathbf{a},k}$ \refeq{H(n,k)subgroup}.

\subsection{\Cn{\infty} or Artin-Mazur zeta function}
\label{sect:ArtinMazur}

Assume that the symmetry group of a given discrete time dynamical system
is the group of temporal lattice translations \Cn{\infty}, \ie, that the
system's defining law is time invariant. The infinite translation
subgroup $H_{\mathbf{a}}$ \refeq{H(n)subgroup} leaves {Bravais cell} of
{period} \cl{} \refeq{1DBravLatt} invariant, so the sum over $H$ in
\refeq{LindZeta} can be replaced by the sum over periods $\cl{}$, with
$N_\cl{}$ the number of {\lsts} of period $\cl{}$. For \Cn{\infty}
 subgroups $H_{\cl{}}$ (contemplate the \Cn{\infty} column of
\reffig{fig:1dLatStatC_5}) 
the index is
\beq
|\Cn{\infty}/H_\cl{}|=\cl{}
\,.
\ee{CinfHnMult}
The Lind zeta function
\refeq{LindZeta} now takes form of the Artin-Mazur zeta
func\-tion\rf{ArtMaz65,CBcount}
\beq
\zeta_{\mbox{\footnotesize AM}}(z) =
     \exp\Big(\sum_{\cl{}=1}^\infty
\frac{N_\cl{}}{\cl{}} z^\cl{}
         \Big)
\,,
\ee{AMzeta}
which, from symmetry perspective, is the statement that the law
governing the dynamical system is time invariant.

The number of \emph{prime} orbits of period $\cl{}$ can be computed
recursively by subtracting repeats of shorter prime orbits\rf{CBcount},
\beq
M_\cl{}\,=\,\frac{1}{\cl{}}
            \Big( N_\cl{} - \sum _{d|\cl{}}^{d<\cl{}}\,d M_d \Big)
\,,
\ee{primeCount}
where $d$'s are all divisors of $\cl{}$.

In what sense is $\zeta_{\mbox{\footnotesize AM}}(z)$ a {\lst}
counts generating  function? Given the $\zeta_{\mbox{\footnotesize
AM}}(z)$, the number of periodic points of period $\cl{}$ is given by its
logarithmic derivative (see
\HREF{http://chaosbook.org/chapters/ChaosBook.pdf\#section.18.7}
{ChaosBook}\rf{CBcount})
\beq
\sum_{\cl{}=1}N_\cl{} z^\cl{}
    = \frac{1}{\zeta_{\mbox{\footnotesize AM}}}
            \,z\frac{d}{dz} \zeta_{\mbox{\footnotesize AM}}
\,.
\ee{zetatop-N}
Examples of $\zeta_{\mbox{\footnotesize AM}}(z)$ {\lst} counts for scalar
lattice field theories studied here are given in
\refappe{sect:AMZetaCounting}.

\subsection{\Dn{\infty} or Kim-Lee-Park zeta function}
\label{sect:KiLePa}

If the assumed symmetry $\Group$ is not the maximal symmetry group, let's
say we assume only $\Group=\Cn{\infty}$ whereas the full symmetry is
$\Dn{\infty}$, Lind zeta function \refeq{LindZeta} reduces to the
Artin-Mazur zeta \refeq{Isola90-13} which counts reflection
symmetry-related {\lsts} as belonging to separate `prime orbits', a
problem that repeatedly bedevils the \wwwcb{} exposition of \po\ theory.
    %
\toVideo{youtube.com/embed/Y1j1mNq-RxE} 

So our next task is to evaluate Lind zeta function when the symmetry
group $\Group$  of temporal lattice of a given dynamical system is the
{infinite dihedral group} $\Dn{\infty}$, the group of all translations
and reflections, \ie, system's defining law is time and
time-reversal invariant.
For the infinite translation $H_{\mathbf{a}}$ subgroup
\refeq{H(n)subgroup}
the index is (as illustrated by \reffig{fig:1dLatStatC_5})
\beq
|\Dn{\infty}/H_{\cl{}}| =  2\cl{}
\,.
\ee{HnMult}
As explained in \refsect{s:LattStateSyms}, the \Dn{\infty} orbits of
reflection-symmetric {\lsts} contain only translations, so the
index of each infinite dihedral subgroup $H_{\mathbf{a},k}$
\refeq{H(n,k)subgroup} is
\beq
|\Dn{\infty}/H_{\cl{},k}| = \cl{}
\,.
\ee{HnkMult}

The Lind zeta function \refeq{LindZeta} now has contributions
from periodic {\lsts}, whose index is \refeq{HnMult},
and symmetric periodic {\lsts}, index \refeq{HnkMult}:
\beq
\zeta_{\mbox{\footnotesize \Dn{\infty}}}(t) =
    \exp\Big(
\sum_{\cl{}=1}^{\infty}\frac{N_{\cl{}}}{2}\,\frac{t^{2\cl{}}}{\cl{}}
          +
\sum_{\cl{}=1}^{\infty}\sum_{k=1}^{n-1}
N_{\cl{},k}\frac{t^{\cl{}}}{\cl{}}
      \Big)
\,.
\ee{DinftyZeta}
${N_{\cl{}}}$ counts the number of {\lsts} invariant under the translation group
$H_{n}$,  while ${N_{\cl{},k}}$ counts the the number of {\lsts}
invariant under the dihedral group $H_{n,k}$.
So the first sum yields the square root of the {Artin-Mazur} zeta
func\-tion $\sqrt{\zeta_{\mbox{\footnotesize AM}}(t^2)}$. The
zeta function is factorized into:
\beq
\zeta_{\mbox{\footnotesize \Dn{\infty}}}(t) =
    \sqrt{\zeta_{\mbox{\footnotesize AM}}(t^2)}\,
    \zeta_{\mbox{\footnotesize s}}(t)
\,.
\ee{DinftyZetaAM}
The second factor $\zeta_{\mbox{\footnotesize s}}(t)$ is the contribution from
the time-reversal symmetric {\lsts}.
\beq
\zeta_{\mbox{\footnotesize s}}(t) =
\exp \Big(\sum_{\cl{}=1}^{\infty} \, \sum_{k=0}^{\cl{}-1}\,
                     \frac{N_{\cl{},k}}{\cl{}}t^{\cl{}} \Big)
\,.
\ee{KLPzeta}

The number of reflection-symmetric {\lsts} does not depend on the
location of the reflection point $k$, only on the type of symmetry (see
the class counts \refeq{H(n,k)class} and \refeq{DnClassRefl}), so
\beq
N_{\cl{},k} =
          \left\{
            \begin{array}{ll}
\, N_{\cl{}, 0} \qquad \mbox{ if } \cl{} \mbox{ odd} \\ 
\, N_{\cl{}, 0} \qquad \mbox{ if } \cl{} \mbox{ and } k \mbox{ are even} \\ 
\, N_{\cl{}, 1} \qquad \mbox{ if } \cl{} \mbox{ even and } k \mbox{ is odd}
\,,
        \end{array}
           \right.
\ee{N_nkCount}
and the Lind zeta function takes the form
that we refer to as the Kim-Lee-Park\rf{KiLePa03} zeta function
\beq
\zeta_{\mbox{\footnotesize \Dn{\infty}}}(t)
    = \sqrt{\zeta_{\mbox{\footnotesize AM}}(t^2)} \; e^{h(t)},
\ee{KLPzetaFact}
where
\beq
h(t) = \sum_{m=1}^{\infty} \left\{
       N^s_{2m-1, 0}\,t^{2m-1}
       + \left(N^s_{2m,0}+N^s_{2m,1}\right)\,\frac{ t^{2m}}{2}
                               \right\}
\,.
\ee{KLPzetaExp}

\subsubsection{Euler product form of Kim-Lee-Park zeta function.}
\label{sect:KiLePaEuler}

The \statesp\
\pS\ of a $\Dn{\infty}$ invariant dynamical system is the union
\beq
\pS = \pS_{a} \cup \pS_{s}
\,,
\ee{pSunion2}
where $\pS_{a}$ is the set of pairs of asymmetric orbits \refeq{reflSymNo},
each element of the set a forward-in-time orbit and the time-reversed orbit,
and $\pS_{s}$ is the set of time reversal symmetric orbits, invariant under
reflections \refeqs{reflSymOdd}{reflSymEvens1}.
Kim \etal\rf{KiLePa03}
show that the contribution of a single prime orbit $p$  to
the Kim-Lee-Park zeta function is:
\bea
1/\zeta_{\mbox{\footnotesize \Dn{\infty}}}(t)|_{p}=
          \left\{
            \begin{array}{ll}
\, 1 - t^{\cl{p}} \qquad\qquad\qquad\qquad\quad
                    \mbox{ if } p \in \pS_{a} \,, \\[1ex]
\, \sqrt{1 - t^{2\cl{p}}} \,\exp\left(-\frac{t^{\cl{p}}}{1-t^{\cl{p}}}
\right)     \qquad \mbox{ if } p \in \pS_{s}
\,,
        \end{array}
           \right.
\label{KLPzetap}
\eea
with the zeta function written as a product over prime orbits:
\beq
1/\zeta_{\mbox{\footnotesize \Dn{\infty}}}(t)=
\prod_{p_a\in\pS_{a}} (1 - t^{\cl{p_a}})
\prod_{p_s\in\pS_{s}} \sqrt{1 - t^{2\cl{p_s}}}
\exp\left(-\frac{t^{\cl{p_s}}}{1-t^{\cl{p_s}}}\right)
\,,
\ee{KLPzetaProd}
to be expanded as a power series in $t$.

The Euler product form of {\tzeta s} makes it explicit that they count
{\em prime orbits}, \ie, {sets} of equivalent {\lsts} related by
symmetries of the problem.
\refAppe{sect:LC21catCounts} verifies by explicit \templatt\ calculations
that the Kim-Lee-Park zeta function indeed counts
{infinite dihedral group} $\Dn{\infty}$ orbits and the corresponding {\lsts}.

\section{Summary}
\label{s:summary}

How to think about matters {\spt}? As our intuition about motions of
fluids is so much better than about turbulent quantum field theories,
here we briefly describe the recent lack of progress in turbulence that
underpins ideas developed in this paper.

While dynamics of a turbulent system might appear so complex to defy any
precise description, the laws of motion drive a spatially extended system
through a repertoire of recognizable unstable patterns (clouds, say),
each defined over a finite  {\spt} region\rf{GuBuCv17,GudorfThesis}.
The local dynamics, observed through such finite spatiotemporal windows,
can be thought of as a visitation sequence of a finite repertoire of
finite patterns.
To make predictions,  one needs to know how often a
given pattern  occurs, and that is a purview of \po\ theory\rf{focusPOT}.
The early 2000's rapid progress in the description of turbulence in terms of such
`{\ecs}s'\rf{KawKida01,science04,GHCW07} has since slowed down to a crawl
due to our inability to extend the theory and the computations to
spatially large or infinite computational domains\rf{WFSBC15}.

In dynamics, we have no fear of the infinite extent in time. That is \po\
theory's\rf{ChaosBook} raison d'\^{e}tre; the dynamics itself describes the
infinite time invariant sets by a hierarchical succession of \po s, of
longer and longer finite periods (with no artificial external
periodicity imposed along the time axis). And, since 1990's we know how
to deal with spatially extended, temporally infinite regions by tiling
them with {\spt}ly periodic tiles\rf{Christiansen97,FE03,GHCW07,CL18,GuBuCv17}.
A time {\po} is computed in a finite time, with period \period{}, but its
repeats ``tile'' the time axis for all times. Similarly, a {\spt}ly
periodic ``tile'' or ``\po'' is computed on a finite spatial region
\speriod{}, for a finite period \period{}, but its repeats in both time and
space directions tile the infinite spacetime.

These ideas open a path to determining exact solutions on \emph{spatially
infinite} regions, and many physical turbulent flows come equipped
with $D$ spatial translational symmetries.
For example, in a pipe flow at
transitional Reynolds number, the azimuthal and radial directions
(measured in viscosity length units) are compact, while the pipe length
is infinite. If the theory is recast as a $d$\dmn\ space-time theory,
\(d= D +1\,,\)
{\spt}ly translational invariant recurrent solutions are $d$-tori
(and \emph{not} the $1$\dmn\ temporal \po s of the traditional {\po} theory),
and the symbolic dynamics is likewise $d$\dmn\
(rather than a 1\dmn\ temporal string of symbols).

This changes everything. Instead of studying time evolution of a chaotic
system, one now studies the repertoire of {\spt} patterns compatible with
a given PDE, or, in the discretized spacetime, a given partial difference
equation.
    %
\toVideo{youtube.com/embed/gDNjOGGBJZY} 
There is no more \emph{time} evolution in this vision of nonlinear
\emph{dynamics}! Instead, there is a \statesp\ of all {\spt} patterns,
and the likelihood that a given finite {\spt}ly pattern can appear, like
the mischievous grin of Cheshire cat, anywhere, anytime in the turbulent
evolution of a flow. A bold vision, but how does it work?
\\

It is in this context of working out the geometry of Hopf's\rf{hopf48}
infinite-dimensional {\statesp}s of turbulent fields (\emph{not} 3\dmn\
visualizations of fluids!), that we find the lessons learned from
discretized field theories very helpful.
Already 1\dmn\ lattice discretization teaches us so much that it merits
this paper by itself, with the intricacies of higher dimensional
Bravais lattices reserved for the sequel\rf{CL18}.

We have learned that, in order to describe turbulence, one has to think
globally but act locally. Turbulence is described by a catalogue of
\spt\ patterns ({`{\lsts}'} in the present, discretized field
theory context; `tiles' in the PDE settings\rf{CL18,GuBuCv17}), each a
numerically exact \emph{global} solution satisfying
\emph{local} deterministic {\ELe}s lattice site by site.
Stripped to its essentials, the problem is to systematically enumerate
them, compute them, and determine their relative importance:

\begin{enumerate}
              \item
\emph{{\Lst}s.} Each solution $\Xx_c$ is a zero of
a global {\em fixed point} condition
\[
F[\Xx_c] = 0
\,.
\]
Together, these solutions form the deterministic scaffold, the
$\infty$\dmn\ \statesp\ of  \spt\ `patterns' explored by deterministic
(or semiclassical or stochastic) turbulence.
              \item
{\em Global stability} is given by the {\jacobianOrb}
\[
\jMorb_{ij} =\frac{\delta F[\Xx_c]_i}{\delta \ssp_j}
\,.
\]
In the field-theoretical formulation there is no evolution in time;
Hill's formulas relate the two notions of stability.
              \item
{\em {\HillDet}s} \[
\Det\jMorb
\]
determine the numbers of \spt\ orbits and the weight of each. In the
discrete spacetime,  the time-{\po}s of dynamical systems theory are
replaced by periodic $d$\dmn\ {Bravais cell} tilings ($d$-tori) of
spacetime, each weighted by the inverse of its instability, its Hill
determinant.
The weighted sum over \spt\ patterns leads to chaotic field theory
partition sums \refeq{ClassPartitF} much like those of solid state, field
theory, and statistical mechanics.

              \item
{\em Symmetries.} The reciprocal {\lsts} are organized by rules
of crystallography. In particular, from a \spt\ field theory
perspective, `time'-reversal is a purely crystallographic notion, a
reflection point group, leading to a dihedral symmetry quotienting of
time-reversible theories and to the associated

              \item
 topological {\em zeta functions} of, to us, surprising form.
In principle, zeta functions encapsulate all the predictions of the theory. In
contrast to conventional solid state calculations, the hyperbolic shadowing
of large Bravais cells by smaller ones\rf{GutOsi15,GHJSC16} ensures that
the predictions of the theory are dominated by the smallest cells.
            \end{enumerate}

What lies ahead? We know how to enumerate 2\dmn\ spacetime Bravais
lattices, count and compute {\lsts} and their Hill
determinants\rf{CL18}, but at present we have no {\dzeta}
analogous to the Lind zeta counting function for the
1- and higher\dmn\ deterministic lattice field theories.

\ack
We are grateful to Karol {\.{Z}}yczkowski, Sven Gnutzmann, Henning
Schomerus and Thomas Guhr for including us in this remembrance and
celebration of Fritz Haake.

Fritz was not the only doubter. During the 2016 Kadanoff Symposium Steve
Shenker asked ``What is chaos in quantum field theory?'', but as P.~C.
was getting ready to formulate an answer, Pavel Wiegmann butted in and
settled the matter from the Soviet angle, with details elaborated by
Boris Gutkin during 2015-2016 visit to Georgia Tech. But when it comes to
a spacetime field theory, Yasha Sinai's forward-in-time partitions only
get in the way. The \spt\ chaotic field theory sketched here is our
answer to Shenker's question, the first seed of which was planted decades
earlier by Wiegmann's throwaway remark that one can have systems with
several ``time'' dimensions, \ie, treat all translational symmetries on
equal footing. Along the way we have received and misunderstood much
in-person sage tutoring from Rafael De La Llave. Matt Gudorf has furthered
the project more than anybody, and we have enjoyed interactions with
Sidney Williams.

Further across the spacetime,
the first seed of doubt was planted by Mitchell Feigenbaum, who
admired Leibnitz and  in 1975 asked P.~C.: ``Do you
really believe that clouds are supercomputers in the sky?''.
We have since been inspired and helped along by
Robert MacKay's 1982 work on time-reversal,
MacKay and James Meiss 1983 Hill's formula,
Alessandro Torcini and Antonio Politi path breaking
1990's chronotopic theory of spatiotemporal chaos,
Serge Aubry's 1990 anti-integrability framework,
Douglas Lind's 1996 multi\dmn\ zeta functions,
David Sterling and
Meiss' 1999 multi\dmn\ \spt\ \Henon\ field theory,
Divakar Viswanath' 2002 Lindstedt-Poincar{\'e} theory,
Carl Dettmann, Ronnie Mainieri and G\'abor Vattay 1998 and
Domenico Lippolis and Jeffrey Heninger
2010's stochastic field theory,
Young-One Kim, Jungseob Lee and Kyewon K. Park's 2003 flip zeta function,
Jason Gallas' 2005 \HenonMap\ orbit polynomials,
Fabian Waleffe, Genta Kawahara,
Bruno Eckhardt, John Gibson, Laurette Tuckerman, Rich Kerswell, Marc Avila,
Ashley Willis,  and Jonathan Halcrow 2000's,
and Mike Schatz and Roman Grigoriev 2020's `exact coherent structures',
Evangelos Siminos and Burak Budanur's 2010's symmetry reductions,
Michael Aizenman's 2016 Ihara zeta functions,
and
Sara Solla, Juan Gallego, Matt Perich and Lee Miller 2018 neural
manifolds.
This paper thus sets up the necessary underpinnings for the quantum field
theory of the \catlatt, the details of which we leave to our friends Jon
Keating and Marcos Saraceno.

This research was initiated during the KITP UC Santa Barbara 2017 {\em
Recurrent Flows: The Clockwork Behind Turbulence} program, supported in
part by the National Science Foundation under Grant No. NSF PHY-1748958.
The work of H.~L. was fully supported by the family of late G. Robinson, Jr..
No actual cats, graduate or undergraduate 
were harmed during this research.




\appendix

\section{Historical context}
\label{s:HillHistory}

                                                        \toCB
Anyone who had ever thought of an integer, or looked at a crystal, or
discretized a PDE, or constructed a secure cryptographic key, or
truncated a Fourier series, or solved turbulence, eventually writes a
paper much like this one. Our reading list is currently at 3,000
references, and there is not much here what no eye has seen,
what no ear has heard, and what no human mind has conceived.

Still, we did not know that the dynamicist's Arnold cat
(see \toChaosBook{exmple.14.12}{example~14.12}) is but the Klein-Gordon field
theory \refeq{freeAction} in disguise, and that the lattice form
\refeq{catMapNewt} of the theory is so much more elegant than the cat map
\refeq{catMap} or the {\HenonMap} \refeq{HenMap}.
A cat is Hooke's wild, `anti-harmonic' sister.
For $s<2$ Hooke rules, with restoring oscillations around the resting
state.
For $s>2$ cats rule, with exponential runaway, wrapped globally around a
\statesp\ torus. {Cat} is to {chaos} what {harmonic oscillator} is to
{order}. There is no more fundamental example of chaos in mechanics.

For small stretching parameter values, $s<2$, discretized
Euler–\-Lagrange equation \refeq{1dTempFT} describes a set of coupled
penduli, with oscillatory solutions,
\toVideo{youtube.com/embed/qzsBtAm5FHc}  
known as the discrete Helmholtz
equation in applied math\rf{DiHaHu01,Lick89,FetWal03}, as  the
tight-binding model, the discrete Schr{\"o}dinger equation,
the Harper's or Azbel-Hofstadter model in solid
state physics\rf{Peierls33,MiDuWh92,Cserti00,Economou06,CsSzDa11},  and
the critical almost Mathieu operator in mathematical physics\rf{Simon82},
with quadratic action \refeq{freeAction} written as Hamiltonian
\[ 
H=\sum_\ell\ket{\ell}\epsilon_0\bra{\ell}
  + \sum_{\ell m}\ket{\ell}U_{\ell m}\bra{m}
\,,\quad
   U_{\ell m} = \left\{
     \begin{array}{ll}
         U & \ell,m \mbox{ nearest neighbors}\\
         0 & \mbox{otherwise}
     \end{array}
             \right.
\] 
with the stretching factor ${s}=-\epsilon_0/U$ in
\refeq{PerViv2.2}.
%
Equilibria or steady solutions of the $d$-dimensional Frenkel-Kontorova
Hamiltonian lattice\rf{AuAb90,MraRin12},
and
discrete breather solutions\rf{BouSko12} of the discrete nonlinear
Klein-Gordon system
that
describe the motion of particles under the competing influence of an
onsite potential field and the nearest neighbor attraction
\beq
\frac{d^2 \ssp_z}{dt^2} - \Box\,\ssp_z + V'(\ssp_z)
    = 0 \,,\quad z\in\mathbb{Z}^d
\,,
\ee{FKHam}
are also examples of what we here call a `nonlinear lattice field theory'.
In contrast to the above mostly oscillatory, often weakly coupled mechanical
systems, the $d$\dmn\ spatiotemporal, everywhere hyperbolic discretized
strongly coupled field
theory developed here\rf{GHJSC16,CL18} is a descendant of Gutkin's
many-particle quantum chaos\rf{GutOsi15}.

Much of the work on 1\dmn\ $\ssp^3$ (\henlatt) and $\ssp^4$ field theories
has already been carried out in 1990's, within the {\em anti-integrability}
framework\rf{AuAb90,aub95ant,StMeiss98}. Our work is a continuation of
Torcini, Politi and collaborators%
\rf{PolTor92,PolTor92b,LePoTo96,LePoTo97,PoToLe98,GiLePo95,JiPoTo13}
1990's \emph{chronotopic approach} to spatiotemporal chaos, and the multi\dmn\
\spt\ $\ssp^3$ field theory pioneered by
Sterling\rf{SterlingThesis99} in his 1999
\HREF{https://www.proquest.com/docview/304508605} {PhD thesis}. Sterling
studies {\HenonMap} lattices in both Hamiltonian and Lagrangian
formulations, and introduces
the multidimensional
`symbol tensor'
(Bunimovich and Sinai\rf{BunSin88} `symbolic representation',
Coutinho and Fernandez\rf{CouFer97} `\spt\ code',
Just\rf{Just01} `symbol lattice',
MacKay\rf{MacKay05} `symbol table',
Pethel, Corron and Bollt\rf{PetCorBol07} `symbol pattern', and,
in our notation\rf{GutOsi15,GHJSC16,CL18} symbol {\brick} or
winding number $\Mm$, see \refeq{pathBern}).
He focuses on the `destruction of chaos' as
one lowers the stretching parameter $a$, a much harder problem than what
we address here. Luckily for us, the strong coupling, strong local
stretching field theories' deterministic solutions are protected
by anti-integrability and live on horseshoes,
safely away from the regions of intermediate stretches, where dragons
live.

The reformulation of the lattice field theory 3-term recurrence
\refeqs{1dTemplatt}{1dPhi4} in terms of the 2-component field
\refeq{1stOrderDiffEqs} is a generalization of the passage from the
Lagrangian to the Hamiltonian formulation, also known as the `transfer
matrix' formulation of lattice field theories\rf{MonMun94,MunWal00} and
Ising models\rf{Onsager44,Kastening02}. We chose to prove it here using
only elementary linear algebra, not only because the Lagrangian
formalism\rf{BolTre10} is not needed for the problem at hand, but because
it actually obscures the generality of Hill's formula, which applies to
all systems, the dissipative ones, such as the Hill's formula
\refeq{PerViv} for the Bernoulli system, as well at the special cases,
such as the contraction parameter value $b=-1$ for which -in general
dissipative- {\HenonMap} \refeq{HenMap} happens to exhibit an additional
symplectic, time-reversal symmetry.

For \emph{forward-in-time evolution} \refeq{catMap}, the $[2\!\times\!2]$
\jacobianM\ $\jMat^\cl{}$ (the monodromy matrix of a \po) describes
the growth of an initial state perturbation in $\cl{}$ steps. For the
corresponding `Lagrangian' system, with action $\action$,
the first variation of the action $\delta\action=0$ yields the
{\ELe}s \refeq{eqMotion}, \refeq{catTempLatt}, while the
second variation, the $[\cl{}\!\times\!\cl{}]$ {\em \jacobianOrb}
\refeq{tempCatFix}, describes the stability of the \emph{entire} \po. In
the classical mechanics context, Bolotin and Treschev\rf{BolTre10} refer
to $\jMorb$ as the `Hessian operator', but, as it is clear from our
Bernoulli discussion, \refsect{s:JacobianOrb}, and the applications to
\KS\ and Navier-Stokes systems\rf{GuBuCv17}, the notion of global
(in)stability of orbits applies to all dynamical systems, not only the
Hamiltonian ones.

His 1878-1886 study\rf{Hill86} of the stability of planar motion of the
Moon around the Earth led  Hill to the {Hill's formula} \refeq{detDet}.
In Lagrangian setting, {\jacobianOrb} $\jMorb$ is the Hessian, the
second variation of the action functional. The discrete-time Hill's
formula for 1\dmn\ lattices with the nearest neighbor interactions that
we use here was derived  by Mackay and Meiss\rf{MacMei83} in 1983 (see
also Allroth\rf{Allroth83} eq.~(12)).
Why Hill deserves the credit, and why celestial mechanics and quantum
mechanics go hand-in-hand here, is explained in chapter~5 of Gutzwiller's
beautiful monograph\rf{gutbook}, as well as in Viswanath's masterly
calculation of Hill's lunar orbit\rf{DV02}.
Reader conversant with celestial literature might have hard time
recognizing the 3-term recurrences whose {\HillDet}s we compute here.
What took Hill from the spatial continuum to an integer lattice is the
fact that the Fourier modes for a compact orbit form a discrete set.
Hill's recurrence relations are made explicit in chapter~4 of Toda's 1967
theory of Toda lattices\rf{Toda89}, the classical mechanics of
one-dimensional lattices (chains) of particles with nearest neighbor
interaction, discrete and infinite in space, continuous in time
(their unexpected symmetries are discussed in chapter~4 of Gutzwiller
monograph\rf{gutbook}).
In his lunar application Hill was lucky: $\Det\jMorb_c$ that he computed
in a $[3\!\times\!3]$ Fourier modes matrix truncation turned out to be
a quite good approximation.
But his is a remarkable formula in the limit of $\cl{}\to\infty$
infinitesimal time steps, a formula that relates the $\infty$\dmn\
\emph{functional} {\HillDet} $\Det\jMorb_c$ to a determinant of the
finite $[d\!\times\!d]$ matrix $\jMat_c$, and it took
\Poincare\rf{Poinc1886} to prove that Hill's truncated Fourier modes
calculation is correct in the continuum limit.

While first discovered in a Lagrangian setting, Hill's formulas apply
equally well to dissipative dynamical systems, from the Bernoulli map of
\refsect{s:coinToss} to \NS\ and \KS\ systems\rf{GudorfThesis,GuBuCv17},
with the Lagrangian formalism of
\refrefs{MacMei83,TreZub09,BolTre10,kooknewt} mostly getting in the way
of understanding them.
Historically, dynamicists always compute $\jMat_c$.
However, in field theory it is the {\em \HillDet} $\Det\jMorb_c$ that
is the computationally robust quantity that one should evaluate.

The {\em fundamental fact} \refeq{fundFact} which equates the number of
periodic points in the {\fundPip} with its volume, \ie, its {\HillDet},
has a long history in the theory of integer lattices, and a key role in
cryptography\rf{MicG0l02,Barvinok02,DeLHTY04,BecRob07,Barvinok08}.
In two dimensions this formula is known since 1899 as
\HREF{https://en.wikipedia.org/wiki/Pick\%27s_theorem} {Pick's theorem},
in higher dimensions it was stated by Nielsen\rf{Nielsen1920,BBPT75} in
1920, and rederived several times since in different contexts, for
example in 1997 by Baake \etal\rf{BaHePl97}.
For the task at hand, Barvinok\rf{Barvinok04}
\HREF{http://www.math.lsa.umich.edu/~barvinok/lectures.pdf} {lectures}
offer a clear and simple introduction to integer lattices, and a proof of
the {`fundamental fact'} \refeq{fundFact}.

\Poincare\rf{poincare} was the first to  recognize the fundamental role
\emph{{\po}s} play in shaping ergodic dynamics. The first step
in this program is a census of {\po}s, addressed in
\refrefs{ArtMaz65,ChAnPi85,brucks,Lutzky88,Lutzky93,BLMS91,XH94,
CheLou97,BriPer05,CBcount}, starting with 1950's
\HREF{https://en.wikipedia.org/wiki/Pekka_Myrberg} {Myrberg}
investigations of {\po}s of quadratic maps, in what was arguably the
first application of computers to
dynamics\rf{Myrberg58a,Myrberg58b,Myrberg59,Myrberg62,Myrberg63}.
Such orbit counts are most elegantly encoded by \emph{topological zeta
functions} of \refsect{s:Lind1d}.
In 1966 Ihara\rf{Ihara66} defined the zeta function of an undirected
graph $\Gamma$ by analogy to Euler's product form of a zeta function,
\beq
\zeta_{\mbox{\footnotesize Ihara}}(z)_\Gamma =
        \prod_{[C]}\frac{1}{1-z^{|C|}}
\,,
\ee{Sato05Zeta}
where the product is over all equivalence classes of prime (non-self
retracing) loops $C$ of $\Gamma$, and $|C|$ denotes the length of $C$.
Ihara zeta functions%
\rf{Ihara66,Bass92,Pollicott01,Sato05,GuIsLa08,Terras10,RAEWH10,
Clair14,ZhXiHe15,Deitmar15,daCosta16,Saito18}
are ``graph-theoretic analogues of discrete Laplacians''\rf{Sunada13}
defined here in \refeq{LapOp}. Even though
`undirected' refers to no preferential
time direction, they do not appear related to
the time-reversal, group-theoretic Kim-Lee-Park zeta function
\refeq{KLPzetaFact} deployed here.
Still, as discussed in \refsect{sect:KiLePaEuler},
Ihara's idea that zeta function
can be written as a product \refeq{Sato05Zeta} over prime orbits holds.

In preparing this manuscript we have found expositions of Lagrangian
dynamics for discrete time systems by MacKay, Meiss and
Percival\rf{MacMei83,MKMP84,meiss92}, and Li and Tomsovic\rf{LiTom17b}
particulary helpful.
The \jacobianOrb\ \refeq{orbJprimeRpt} of a period-$(m\cl{})$ {\lst}
$\Xx$ was studied by Gade and Amritkar\rf{GadAmr93} in 1993. The
exposition of \refsect{s:latt1d} owes much to MacKay\rf{Bmack93} 1982 PhD
thesis' chapter on reversible maps, and Endler and Gallas Hamiltonian
\HenonMap\ orbit polynomials\rf{EG05a,EndGal06}.

\section{Hill's formula for a 2nd order difference equation}
\label{s:Hill2step}

Consider a map of form $\ssp_{\zeit+1}=g(\ssp_{\zeit-1},\ssp_{\zeit})$,
where $\ssp_{\zeit}$ is a scalar field (examples are the kicked rotor
\refeq{kittyMap} and the 3-term recurrence relations
(\ref{1dTemplatt}-\ref{1dPhi4})). Such a map can be replaced by a pair of 1st
order difference equations for the 2-component field
$\hat{\ssp}_{\zeit}=(\varphi_{\zeit},\ssp_{\zeit})$ at the temporal lattice
site ${\zeit}$,
 \beq
\hat{\ssp}_{\zeit+1}
  =
\hat{\map}(\hat{\ssp}_{\zeit})
  =
 \left(\begin{array}{c}
 \ssp_{\zeit}  \\
 g(\varphi_{\zeit},\ssp_{\zeit})
  \end{array} \right)
\,.
\ee{1stOrderDiffEqs}

As in \refsect{s:forwardHill}, the trace of the $\cl{}$th
iterate of the forward-in-time {\FPoper} can be evaluated in two ways.
First, using the
Dirac delta kernel of the operator $\Lop^\cl{}$,
\beq
\tr \Lop^\cl{} =  \int_\pS \!\!d^2\! \hat{\ssp}_{0} \,
        \delta(\hat{\ssp}_{0} - \hat{\map}^\cl{}(\hat{\ssp}_{0}))
\,.
\ee{ForwardInTimeTr2}
Restricting the integration to an infinitesimal open neighborhood of
the 2-component field $(\varphi_{c,0},\ssp_{c,0})$ at lattice site 0, the
period $\cl{}$ \lst\ $\Xx_c$ contribution to the trace is again
${1}/{\left|\det(\id-\jMat_c)\right|}$, 
with $\jMat_c$ the for\-ward-in-time $[2\times2]$ {\FloquetM} \refeq{FloqMat},
a product of the 1-time step \jacobianMs\ \refeq{d-1stepJac}
\beq
\jMat_\zeit =
            \left(\begin{array}{cc}
             0 & 1 \\
             \frac{\partial g(\varphi_{\zeit}, \ssp_{\zeit}) }
             {\partial \varphi_{\zeit}                } &
             \frac{\partial g(\varphi_{\zeit}, \ssp_{\zeit}) }
             {\partial \ssp_\zeit                }
            \end{array}\right)
\,,
\ee{GRRR}
where $(\varphi_{\zeit}, \ssp_{\zeit}) = \hat{\map}^{\zeit}(\varphi_{c,0}, \ssp_{c,0})$.

Alternatively, the trace can be evaluated as $2\cl{}$\dmn\ integral over
a product of one-time-step
{\FPoper}s \refeq{FPtrace},
\bea
\tr \Lop^\cl{} &=&
\int \prod_{\zeit=0}^{\cl{}-1} \left[ d^2\!\hat{\ssp}_\zeit \,
\delta(\hat{\ssp}_{\zeit+1} - \hat{\map}(\hat{\ssp}_{\zeit}))\right]
\continue
&=&
\int \prod_{\zeit=0}^{\cl{}-1} \left[
d\varphi_{\zeit}d\ssp_{\zeit}\,
\delta(\varphi_{\zeit+1} - \ssp_{\zeit})\,
\delta(\ssp_{\zeit+1} - g(\varphi_{\zeit},\ssp_{\zeit}))
                                \right]
\,,
\label{FPtrace2}
\eea
with a 1\dmn\ Dirac delta for each field component \refeq{1stOrderDiffEqs}.
With the periodic {\bcs} $\hat{\ssp}_{\zeit+\cl{}}=\hat{\ssp}_\zeit$, the
$d\varphi_{\zeit}$  integration eliminates the $\varphi_{\zeit}$ components,
resulting in the $\cl{}$\dmn\ scalar field integral
\bea
\tr \Lop^\cl{} &=&
\int d\Xx \prod_{\zeit=0}^{\cl{}-1}
           \delta(\ssp_{\zeit+1}-g(\ssp_{\zeit-1},\ssp_{\zeit}))
    \,,\quad
           d\Xx= \prod_{\zeit=0}^{\cl{}-1} d\ssp_{\zeit}
\,,
\label{FPtrace3}
\eea
or, in the \lst\ notation,
\beq
\tr \Lop^\cl{} = \int d\Xx \delta(F[\Xx])
    \,,\quad
F[\Xx] = \shift \Xx - g(\shift^{-1} \Xx, \Xx)
\,.
\label{FPtrace3}
\eeq
where $\Xx$ and $g(\shift^{-1}\Xx,\Xx)$ are $\cl{}$\dmn\ column vectors
with $\zeit$-th components $\ssp_{\zeit}$ and
$g((\shift^{-1}\Xx)_\zeit,\Xx_\zeit)$, respectively, $\shift$ is the cyclic
$[\cl{}\!\times\! \cl{}]$ time translation operator \refeq{hopMatrix}, and
the saddle-point condition $F[\Xx_c]=0$ is the {\ELe} \refeq{eqMotion} of the system.
The rest is as in \refeq{GlobalTr}; the trace is the
the deterministic partition sum \refeq{ClassPartitF} over all {\lsts},
\beq
\tr_c \Lop^\cl{} =
\int_{\pS_c}\!\!\!\!d\Xx\,\delta(F[\Xx])
                 = \frac{1}{\left|\Det\jMorb_c\right|}
\,,
\ee{globalRecur}
where $\jMorb_c$ is the $[\cl{}\!\times\!\cl{}]$ {\jacobianOrb} evaluated on
the period-$\cl{}$ {\lst} $\Xx_c$, enclosed by an infinitesimal open
neighborhood $\pS_c$.
Comparing the traces \refeq{ForwardInTimeTr} and \refeq{globalRecur}, we see
that we have
again proved the Hill's formula \refeq{detDet}.

Note that nowhere in the derivation have we assumed that the system has a
Lagrangian formulation: this version of Hill's formula applies to any  2nd
order difference equation, or 3-term recurrence of form
$\ssp_{\zeit+1}=g(\ssp_{\zeit-1},\ssp_{\zeit})$, for example, any
dissipative \HenonMap\ \refeq{HenMap} as well as its special $b=-1$
Hamiltonian case \refeq{Hen3term}.

\section{{\JacobianOrbs} of \Dn{\cl{}} prime orbits.}
\label{sect:SymmReducedJacobian}
As shown in \refeq{jacobianOrbD5}, the {\jacobianOrb} in the linear space of the prime
{\lst} with time reversal symmetry has a different form compared to the {\jacobianOrb}
 that acts on a Bravais {\lst}. It only acts on `half' of the Bravais cell and the
cyclic translational symmetry is broken.

Consider possible symmetries of periodic {\lsts} of the {\templatt}. If a {\lst} has
odd period \cl{},
reflection symmetry $(o)$ defined by \refeq{reflSymOdd}, the prime {\lst}
is a $m+1$-\dmn\ vector:
\beq
\transp{\tilde{\Xx}} = (\ssp_0,\ssp_1,\ssp_2,\ssp_3,\cdots,\ssp_{m}) \,.
\ee{primeLattStato}
And the $[(m+1)\times(m+1)]$ symmetry reduced {\jacobianOrb} is:
\bea
{\jMorb_o}
  =
\left(\begin{array}{cccccc}
 s& -2 & 0 & \dots &0& 0 \\
 -1 &  s& -1 & \dots &0&0 \\
0 & -1 &  s & \dots &0 & 0 \\
\vdots & \vdots & \vdots & \ddots &\vdots &\vdots\\
0 & 0 & \dots  &\dots  & s & -1 \\
0 & 0 & \dots  &\dots& -1 &  s-1
\end{array} \right)
\,.
\label{jacobianOrbO}
\eea
If {\lsts} have symmetries defined by \refeq{reflSymEvens0} or \refeq{reflSymEvens1},
the symmetry reduced {\jacobianOrbs} are $[(m+1)\times(m+1)]$ matrix:
\bea
{\jMorb_{ee}}
  =
\left(\begin{array}{cccccc}
 s& -2 & 0 & \dots &0& 0 \\
 -1 &  s& -1 & \dots &0&0 \\
0 & -1 &  s & \dots &0 & 0 \\
\vdots & \vdots & \vdots & \ddots &\vdots &\vdots\\
0 & 0 & \dots  &\dots  & s & -1 \\
0 & 0 & \dots  &\dots& -2 &  s
\end{array} \right)
\,,
\label{jacobianOrbEE}
\eea
or $[m\times m]$ matrix:
\bea
{\jMorb_{eo}}
  =
\left(\begin{array}{cccccc}
 s-1& -1 & 0 & \dots &0& 0 \\
 -1 &  s& -1 & \dots &0&0 \\
0 & -1 &  s & \dots &0 & 0 \\
\vdots & \vdots & \vdots & \ddots &\vdots &\vdots\\
0 & 0 & \dots  &\dots  & s & -1 \\
0 & 0 & \dots  &\dots& -1 &  s-1
\end{array} \right)
\,.
\label{jacobianOrbEO}
\eea
The determinants of these symmetry reduced {\jacobianOrbs} are the stabilities of the
{\lsts} in the time-reflection symmetric subspace. Using the fundamental fact,
one can compute the number of time-reversal invariant {\lsts} of {\templatt}.

Note that even though the symmetry reduced {\jacobianOrbs} are not circulant matrices,
they share eigenvalues with circulant {\jacobianOrbs} \refeq{Hessian}, because the
eigenvectors of the circulant {\jacobianOrbs} are Fourier modes and can be written as
symmetric and anti-symmetric vectors. The determinants of the symmetry reduced
{\jacobianOrbs} are:
\bea
\Det{\jMorb_o} &=&
\prod_{j=0}^{(\cl{}-1)/2} \left(s-2\cos\frac{2\pi j}{\cl{}}\right)
=\sqrt{\left(s-2\right)\Det\jMorb} \,,
\continue
\Det{\jMorb_{ee}} &=&
\prod_{j=0}^{(\cl{}-2)/2} \left(s-2\cos\frac{2\pi j}{\cl{}}\right)
=\sqrt{\left(s-2\right)\left(s+2\right)\Det\jMorb} \,,
\continue
\Det{\jMorb_{eo}} &=&
\prod_{j=0}^{\cl{}/2} \left(s-2\cos\frac{2\pi j}{\cl{}}\right)
=\sqrt{\frac{\left(s-2\right)}{\left(s+2\right)}\Det\jMorb}
\,,
\label{HillDetSymmReduced}
\eea
where $\Det\jMorb$ is given in \refeq{1stepDiffSolu}.
$\Det\jMorb$, $\Det\jMorb_{o}$, $\Det\jMorb_{ee}$ and $\Det\jMorb_{eo}$
count numbers of periodic {\lsts} that satisfy the symmetries defined by
\refeqs{reflSymNo}{reflSymEvens1} respectively.

MacKay\rf{Bmack93}
and Endler and Gallas\rf{EG05a}
give tables of {\lsts} and their orbits counts, together with
the counts of symmetric {\lsts} and orbits.

\section{Counting {\lsts} using zeta functions}
\label{sect:ZetaCounting}

\subsection{Counting {\lsts} using Artin-Mazur zeta function}
\label{sect:AMZetaCounting}

Let us first evaluate $\zeta_{\mbox{\footnotesize AM}}$ for scalar lattice
field theories studied here.

\paragraph{Bernoulli.}
The number of Bernoulli system period-\cl{} {\lsts} is given in
\refeq{noPerPtsBm}, so
\bea
\zetatop(z)
 =  \exp \Big(-\sum_{\cl{}=1}^\infty
\frac{{s}^\cl{} - 1}{\cl{}}z^\cl{}
         \Big)
 =
\frac{1 -  {s}z}{1 - z}
\,.
\label{BernZeta}
\eea
The numerator $(1 - {s}z)$ says that a Bernoulli system is a full
shift\rf{CBcount}: there are $s$ fundamental {\lsts}, in this case
fixed points $\{\ssp_0,\ssp_1,\cdots,\ssp_{s-1}\}$, and every other
{\lst} is built from their concatenations and repeats. The
denominator $(1 - z)$ compensates for the single overcounted
{\lst}, the fixed point $\ssp_{{s}-1}=\ssp_{0}$ $(\mbox{mod}\;1)$
of \reffig{fig:BernPart}, and its repeats. If the stretching factor
${s}=\beta$ is not an integer, the map \refeq{n-tuplingMap} is called the
`$\beta$-transformation'. For its Artin-Mazur zeta function see
\refref{FlLaPo94}.

\paragraph{Counting {Bernoulli} prime orbits.}
\label{s:bernPrime}

Substituting the Bernoulli map \tzeta\ \refeq{BernZeta}
into \refeq{zetatop-N}
we obtain
\bea
\sum_{n=1}N_n z^n
    &=&
 z+3 z^2+7 z^3+15 z^4+31 z^5+63 z^6+127 z^7
    \ceq
+255 z^8+511 z^9+ 1023z^{10} +2047 z^{11}
\cdots
\,,
\label{bernN_n-s=2}
\eea
in agreement with the {\lsts} count \refeq{detBern}.
The number of \emph{prime} orbits of period $\cl{}$ is given recursively by
subtracting repeats of shorter prime orbits \refeq{primeCount},
hence
\bea
\sum_{n=1}M_n z^n
    &=&
 z+  z^2+2 z^3+3 z^4+6 z^5+9 z^6+18 z^7
    \ceq
+30 z^8+56 z^9+99 z^{10} +186 z^{11}
\cdots
\,,
\label{bernM_n-s=2}
\eea
in agreement with the usual numbers of binary symbolic dynamics prime
cycles\rf{CBcount}.

\paragraph{\tempLatt.}
Substituting the number of \templatt\ period-\cl{} {\lsts} given in
\refeq{1stepDiffSolu} into the Artin-Mazur zeta \refeq{AMzeta},
Isola\rf{Isola90} obtains
\bea
\zetatop(z)
 =  \exp \Big(-\sum_{\cl{}=1}^\infty
\frac{\ExpaEig^\cl{} + \ExpaEig^{-\cl{}} - 2}{\cl{}} z^\cl{}
         \Big)
 =
\frac{1 - s z + z^2}
     {(1 - z)^2}
\,.
\label{Isola90-13}
\eea
Conversely, given the \tzeta, the generating function for the number of
temporal {\lsts} of period $\cl{}$ is given by the logarithmic
derivative \refeq{zetatop-N},
\bea
\sum_{{n}=0}^\infty N_{n} z^{n}
    & = & \frac{2-{s}z}{1 - s z + z^2}-\frac{2}{1 - z}
    \continue
& = & (s-2)\left[z + ({s}+2) z^2 + ({s}+1)^2 z^3 \right.
    \ceq
      \left.\qquad\quad
      +\,({s}+2)\,{s}^2 z^4 + (s^2+ s-1)^2 z^5  +  \cdots\right]
\,,
\label{1stChebGenF}
\eea
which is indeed the generating function for $T_{\cl{}}(s/2)$, the
{Chebyshev polynomial of the first kind} \refeq{POsChebyshev}.
Why Chebyshev? Essentially because $T_k(x)$ are also
defined by a 3-term recurrence:
\bea
&& T_0(x) = 1\,,\quad T_1(x) = x \,,
    \continue
&&  - T_{k+1}(x) +2x T_{k}(x) - T_{k-1}(x) = 0
    \quad \mbox{for } k \geq 2
\,.
\label{Cheb1stRecurr} 
\eea

\paragraph{\Henlatt.}
The problem of counting orbits for the {\HenonMap} \refeq{HenMap} was
first addressed in 1979 by Sim{\'o}\rf{Simo79}. For the complete
horseshoe, ${a}=6$ {\HenonMap} repeller there are $2^\cl{}$
period-$\cl{}$ {\lsts}, so
\bea
\zetatop(z)
 =  \exp \Big(-\sum_{\cl{}=1}^\infty
\frac{2^\cl{}}{\cl{}} z^\cl{}
         \Big)
\,=\,
1- 2z
\,.
\label{HenonZeta}
\eea
The numbers of the shortest period {\lsts} and prime orbits are
listed in \reftab{tab:HamHenon}.

\begin{table}
\begin{center}
\begin{tabular}{c|rrrrr|rrrrr|rrrrr}
$\cl{}$ &  1 &  2 &  3 &  4 &  5 &
       6 &  7 &  8 &  9 & 10 &
      11 & \\
\hline
$N_\cl{}$ &  2 &  4 &  8 & 16 &  32 &
       64 &  128 &  256 & 512 & 1024 &
      2048 & 
             \rule[-1ex]{0ex}{3.5ex} \\
$M_\cl{}$ &   2 &   1 &   2 &  3 &  6 &
         9 & 18 &  30 & 56  & 99 &
       186 &  
\end{tabular}
\bigskip
\caption{\label{tab:HamHenon}
$N_\cl{}$ and $M_\cl{}$ are the numbers of
the period-$\cl{}$ {\lsts} and
orbits, respectively,  for the ${a}=6$ {\HenonMap}.
}
\end{center}
\end{table}
%

\paragraph{{$\phi^4$} field theory.}
The same as the \henlatt, with $2\to3$ replacement in
\refeq{HenonZeta}.

\subsection{Counting \templatt\ {\lsts} using Kim-Lee-Park zeta function}
\label{sect:LC21catCounts}    

\begin{table}
\begin{center}
\begin{tabular}{c|rrrrr|rrrrr|rrrrr}
$n$ &  1 &  2 &  3 &  4 &  5 &
       6 &  7 &  8 &  9 & 10 \\
\hline
$M_\cl{}$ &   1 &   2 &   5 &  10 &   24 &
         50 & 120 & 270 & 640 & 1500 \\
$N^a_\cl{}$ &   1 &   5 &  16 &  45 &  121 &
        320 & 841 & 2205 &5776 &15125 \\
$N^s_{\cl{},0}$ &   1 &   5 &   4 &  15 &  11 &
         40 & 29 & 105 & 76 & 275 \\
$N^s_{\cl{},1}$ &   1 &   1 &   4 &  3 &  11 &
         8 & 29 & 21 & 76 & 55
\end{tabular}
\bigskip
\caption{\label{tab:lattstateCountCat}
Lattice state and {prime} orbit counts for the ${s}=3$ {\templatt}.
$M_\cl{}$ is the number of period-$\cl{}$ prime orbits
(see Bird and Vivaldi\rf{BirViv}).
$N^a_{\cl{}}$, $N^s_{\cl{},0}$ and $N^s_{\cl{},1}$ are numbers of {\lsts}
that are invariant under group actions of $H_{\cl{}}$, $H_{\cl{},0}$
and $H_{\cl{},1}$ respectively.
    }
\end{center}
\end{table}

As the symmetry of {\templatt} is $\Dn{\infty}$,
the number of {\lsts} for {\templatt} given by the $\Cn{\infty}$
{\HillDet} 
\refeq{1stepDiffSolu} miscounts states related by reflections.
In this case one uses the fundamental fact
to count each type  the time reversal invariant {\lsts} separately.
As an example, consider {\templatt} with $s=3$.
The {\HillDet}s of the symmetry
reduced {\jacobianOrbs}
\refeq{HillDetSymmReduced} count the corresponding {\lsts}:
\bea
N^a_\cl{} &=& \left(\ExpaEig^{\cl{}/2}-\ExpaEig^{-\cl{}/2}\right)^2 \,, \continue
N^s_{\cl{},0} &=& \ExpaEig^{\cl{}/2}-\ExpaEig^{-\cl{}/2} \, , \qquad \qquad\;\;
\cl{} \mbox{ odd,} \continue
N^s_{\cl{},0} &=& \sqrt{5}\left(\ExpaEig^{\cl{}/2}-\ExpaEig^{-\cl{}/2}\right) \, , \qquad
\cl{} \mbox{ even,} \continue
N^s_{\cl{},1} &=& \frac{1}{\sqrt{5}}\left(\ExpaEig^{\cl{}/2}-\ExpaEig^{-\cl{}/2}\right) \, , \qquad
\cl{} \mbox{ even.}
\label{symmNumbers}
\eea
Substitute into \refeq{KLPzetaExp} to count symmetric \lsts:
\bea
h(t) &=& \sum_{m=1}^{\infty} \left[
       N_{2m-1, 0}\,t^{2m-1}
       + \left(N_{2m,0}+N_{2m,1}\right)\,\frac{ t^{2m}}{2}
                               \right] \continue
&=&
\frac{\ExpaEig^{1/2} t}{1-\ExpaEig t^2}
-\frac{\ExpaEig^{-1/2}t}{1-\ExpaEig^{-1}t^2}
\label{HLsymmCatZetaExp}
+
\sqrt{\frac{9}{5}}\frac{\ExpaEig t^2}{1- \ExpaEig t^2}
-\sqrt{\frac{9}{5}}\frac{\ExpaEig^{-1} t^2}{1- \ExpaEig^{-1} t^2} \,.
\eea
Using the definition of the Kim-Lee-Park zeta function \refeqs{KLPzeta}{KLPzetaExp},
generating functions of the {\lst} counts are
\bea
t \frac{\partial}{\partial t}\ln\zeta_{\mbox{\footnotesize a}}(t^2)
&=& \sum_{\cl{}=1}^{\infty} N^a_{\cl{}} t^{2\cl{}} \continue
&=& t^2 + 5 t^4 + 16 t^6 + 45 t^8 + 121 t^{10} + 320 t^{12} + 841 t^{14} \continue
&&+ 2205 t^{16} + 5776 t^{18} + 15125 t^{20} + \dots
\,,
\eea
and
\bea
h(t) &=& \sum_{m=1}^{\infty} \left[
       N^s_{2m-1, 0}\,t^{2m-1}
       + \frac{\left(N^s_{2m,0}+N^s_{2m,1}\right)}{2}\, t^{2m}
                               \right] \continue
&=&
t + 3 t^2 + 4 t^3 + 9 t^4 + 11 t^5 + 24 t^6 +  \continue
&&29 t^7 + 63 t^8 + 76 t^9 +165 t^{10}
+ \dots
\,,
\eea
in agreement with the numbers of {\lsts} with period up to 10 listed in
the \reftab{tab:lattstateCountCat}.
Kim-Lee-Park zeta function counts as advertised.

    \ifsubmission
\section*{References}
\bibliographystyle{iopart-num}       
\bibliography{../bibtex/siminos}
    \else
\printbibliography[
heading=bibintoc,
title={References}
				  ] 
    \fi


    \ifboyscout
     {\color{blue}

     } 

    \clearpage
    \input{../spatiotemp/chapter/LC21blog}
    \fi 

\end{document}

%% file: biblatex.tex
\usepackage[
    backend=bibtex,
    sorting=nyt,
    style=numeric, 
    natbib=true,
    style=phys, 
    biblabel= brackets, 
    articletitle=true,  
    chaptertitle=true,  
    pageranges = true , 
    sortlocale=en_US,
    giveninits=true,
    url=false, 
    doi=false, 
    eprint=false
            ]{biblatex}

\DeclareFieldFormat
  [article,
   inbook,
   incollection,
   inproceedings,
   patent,
   thesis, 
   unpublished,
   report, 
   misc,
  ]{title}{\href{\thefield{url}}{#1}}

\newbibmacro{string+doiurlisbn}[1]{%
  \iffieldundef{doi}{%
    \iffieldundef{url}{%
      \iffieldundef{isbn}{%
        \iffieldundef{issn}{%
          #1%
        }{%
          \href{http://books.google.com/books?vid=ISSN\thefield{issn}}{#1}%
        }%
      }{%
        \href{http://books.google.com/books?vid=ISBN\thefield{isbn}}{#1}%
      }%
    }{%
      \href{\thefield{url}}{#1}%
    }%
  }{%
    \href{http://dx.doi.org/\thefield{doi}}{#1}%
  }%
}

\DeclareFieldFormat{title}{\usebibmacro{string+doiurlisbn}{\mkbibemph{#1}}}
\DeclareFieldFormat[article,incollection]{title}%
    {\usebibmacro{string+doiurlisbn}{\mkbibquote{#1}}}



%% file: defsReversal.tex
\ifsubmission
    \newcommand{\wwwcb}[1]{{\tt ChaosBook.org#1}}
    
    \newcommand{\HREF}[2]{{#2}}
    
    \newcommand{\toChaosBook}[2]{}{ChaosBook\, #2}
    \newcommand{\videoLink}[1]{}{}
    \newcommand{\toVideo}[1]{}
\else  
    \newcommand{\wwwcb}[1]{
                  {\tt \href{http://ChaosBook.org#1}
              {ChaosBook.org#1}}}
    
    \newcommand{\HREF}[2]{{\href{#1}{#2}}}
    
    \newcommand{\toChaosBook}[2] 
    {\href{http://ChaosBook.org/chapters/ChaosBook.pdf\##1}{ChaosBook\, #2}}
    \newcommand{\videoLink}[1]{    
        \raisebox{-0.4ex}[0pt][0pt]{$\!\!\!$
        \HREF{https://#1}{\includegraphics[height=1em]{YouTubeIcon}}
        $\!$             }}
    \newcommand{\toVideo}[1]    
        {\marginpar[\videoLink{#1}]{\videoLink{#1} \hfill}}
\fi 

\newcommand{\rf}     [1] {~\cite{#1}}
\newcommand{\refref} [1] {\cite{#1}}

\newcommand{\refrefs}[1] {\cite{#1}}

\newcommand{\refeq}  [1] {(\ref{#1})}
\newcommand{\refeqs} [2]{(\ref{#1}--\ref{#2})}
\newcommand{\reffig} [1] {figure~\ref{#1}}
\newcommand{\reffigs} [2] {figures~\ref{#1} and~\ref{#2}}
\newcommand{\refFig} [1] {Figure~\ref{#1}}
\newcommand{\refFigs} [2] {Figures~\ref{#1} and~\ref{#2}}
\newcommand{\reftab} [1] {table~\ref{#1}}

\newcommand{\refsect}[1] {section~\ref{#1}}
\newcommand{\refsects}[2] {sections~\ref{#1} and \ref{#2}}

\newcommand{\refappe}[1] {\ref{#1}} 
\newcommand{\refAppe}[1] {\ref{#1}} 

\newcommand{\lst}{lattice state}
\newcommand{\Lst}{Lattice state}
\newcommand{\lsts}{lattice states}

\newcommand{\templatt}{temporal cat}
\newcommand{\tempLatt}{Temporal cat}
\newcommand{\spt}{spatiotemporal}

\newcommand{\catlatt}{spatiotemporal cat}

\newcommand{\Henon}{H\'enon}
\newcommand{\HenonMap}{H\'enon map}
\newcommand{\henlatt}{temporal H{\'e}non}
\newcommand{\Henlatt}{Temporal H{\'e}non}
\newcommand{\bcs}{bc's}
\newcommand{\PV}{Percival-Vivaldi}

\newcommand{\sPe}{screened Poisson equation}

\newcommand{\zetatop}{\ensuremath{1/\zeta_{\mbox{\footnotesize AM}} }}
\newcommand{\msr}{\ensuremath{\rho}}                
\newcommand{\coord}{q}        
\newcommand{\jMps}{\ensuremath{J}}   
\newcommand{\jMorb}{{\ensuremath{\mathcal{\jMps}}}}
\newcommand{\jMat}{\mathbb{\jMps}}
\newcommand{\ssp}{\ensuremath{\phi}}             
\newcommand{\cssp}{\ensuremath{\tilde{\phi}}}    
\newcommand{\Source}{\mathsf{J}}
\newcommand{\source}{{j}}

\newcommand{\fundPip}{fundamental parallelepiped}

\newcommand{\lattice}{\ensuremath{\mathcal{L}}}
\newcommand{\sitebox}[1]{%
        {
   \setlength{\fboxsep}{-2\fboxrule}
   \fbox{\hspace{1.2pt}\strut$#1$\hspace{1.2pt}}
        }
                        } 

\newcommand{\Mm}{\ensuremath{\mathsf{M}}}
\newcommand{\Xx}{\ensuremath{\mathsf{\Phi}}}      
\newcommand{\Refl}{\ensuremath{\mathsf{\sigma}}}            
\newcommand{\shift}{\ensuremath{\mathsf{r}}}
\newcommand{\A}{\ensuremath{\mathcal{A}}}       
\newcommand{\reals}{\mathbb{R}}

\newcommand{\integers}{\ensuremath{\mathbb{Z}}}

\newcommand{\vel}{\ensuremath{v}}   
\newcommand{\jacobianOrb}{orbit Jacobian matrix}

\newcommand{\jacobianOrbs}{orbit Jacobian matrices}
\newcommand{\JacobianOrbs}{Orbit Jacobian matrices}
\newcommand{\HillDet}{Hill determinant}
\newcommand{\ecs}{exact coherent structure}

\ifboyscout 
  \newcommand{\PC}[2]
                     {\begin{quote}\PCedit{[#1 Predrag] \small #2}\end{quote}}
  \newcommand{\PCedit}[1]{{\color{magenta}#1}}
  \newcommand{\HL}[2]
                     {\begin{quote}{\color{green}[#1 Han] \small #2}\end{quote}}
  
  \newcommand{\Private}[1]{{\color{blue}#1}}
  \newcommand{\toCB}{\marginpar{\footnotesize 2CB}}  
  \newcommand{\CBlibrary}[1]
             {\href{http://ChaosBook.org/library/#1.pdf} { (click here)}}
\else 
  \newcommand{\PC}[2]{}{}
  \newcommand{\PCedit}[1]{#1}
  \newcommand{\HL}[2]{}
  
  \newcommand{\Private}[1]{}
  \newcommand{\toCB}{}
  
  \newcommand{\CBlibrary}[1]{}
\fi  

\newsavebox{\bartName}

{\hspace*{\fill}\nolinebreak[1]\usebox{\bartName}\vspace*{1ex}\end{minipage}}
\newcommand{\beq}{\begin{equation}}
\newcommand{\continue}{\nonumber \\ }
\newcommand{\nnu}{\nonumber}
\newcommand{\eeq}{\end{equation}}
\newcommand{\ee}[1] {\label{#1} \end{equation}}
\newcommand{\bea}{\begin{eqnarray}}
\newcommand{\ceq}{\nonumber \\ & & }
\newcommand{\eea}{\end{eqnarray}}
\newcommand{\barr}{\begin{array}}
\newcommand{\earr}{\end{array}}

\newcommand{\period}[1]{{\ensuremath{T_{#1}}}}         
\newcommand{\cl}[1]{{\ensuremath{n_{#1}}}}   
\newcommand{\speriod}[1]{{\ensuremath{L_{#1}}}}  

\newcommand{\statesp}{state space}
\newcommand{\Statesp}{State space}
\newcommand{\stateDsp}{state-space}
\newcommand{\StateDsp}{State-space}

\newcommand{\jacobianM}{Jacobian matrix}  
\newcommand{\jacobianMs}{Jacobian matrices}   
\newcommand{\JacobianM}{Jacobian matrix} %
\newcommand{\FloquetM}{Floquet matrix} 
\newcommand{\FPoper}{Perron-Frobenius oper\-ator} 
\newcommand{\FP}{Perron-Frobenius}
\newcommand{\dzeta}{dyn\-am\-ic\-al zeta func\-tion}

\newcommand{\tzeta}{top\-o\-lo\-gi\-cal zeta func\-tion}

\newcommand{\cycForm}{cycle averaging formula}

\newcommand{\admissible}{admissible}
\newcommand\map{f}                  
\newcommand\flow[2]{{f^{#1}(#2)}}

\newcommand{\ExpaEig}{\ensuremath{\Lambda}}
\newcommand{\Lyap}{\ensuremath{\lambda}}            

\newcommand{\Lop}{\ensuremath{\mathcal{L}}}       
\newcommand{\Det}{\mbox{\rm Det}\,}

\newcommand{\cycle}[1]{{\ensuremath{\overline{#1}}}}
\newcommand{\brick}{block}

\newcommand{\Ssym}[1]{{\ensuremath{m_{#1}}}}
\newcommand{\action}{S}
\newcommand{\ELe}{Euler–Lagrange equation}

\newcommand{\po}{periodic orbit}
\newcommand{\Po}{Periodic orbit}

\newcommand{\pS}{\ensuremath{\mathcal{M}}}          
\newcommand{\Poincare}{Poincar\'e }
\newcommand{\PoincSec}{Poincar\'e section}
\newcommand{\PoincMap}{return map} 
\newcommand{\zeit}{\ensuremath{t}}  
\newcommand{\transp}[1]{{#1}{}^\top}

\newcommand{\bra}[1]{\langle{#1}\vphantom{ }|}
\newcommand{\ket}[1]{|\vphantom{}{#1}\rangle}

\newcommand{\Group}{\ensuremath{G}}         
\newcommand{\LieEl}{\ensuremath{g}}  
\newcommand {\id}{{\ \hbox{{\rm 1}\kern-.62em\hbox{\rm 1}}}} 
\newcommand{\Cn}[1]{\ensuremath{\textrm{C}_{#1}}}
\newcommand{\Dn}[1]{\ensuremath{\textrm{D}_{#1}}}              
\newcommand{\KS}{Ku\-ra\-mo\-to-Siva\-shin\-sky}
\newcommand{\NS}{Navier-Stokes}
\newcommand{\dmn}{-dimensional}  

\newcommand{\etc}{{etc.}}       
\newcommand{\ie}{{i.e.}}        